\documentclass{JFM-FLM_Au}

\usepackage{psfrag}
\usepackage{tikz}
\usepackage{graphicx}
\usetikzlibrary{patterns}
\usepackage{multirow}

\newcommand{\redline}{\raisebox{2pt}{\tikz{\draw[-,red!40!red,solid,line width = 0.9pt](0,0) -- (5mm,0);}}}

\newcommand{\brightgreenline}{\textcolor[RGB]{0,204,0}{\rule{5mm}{1.5pt}}}
\newcommand{\brightredline}{\textcolor[RGB]{255,0,0}{\rule{5mm}{1.5pt}}}
\newcommand{\darkblueline}{\textcolor[RGB]{0,0,139}{\rule{5mm}{1.5pt}}}

\newcommand{\greenline}{\raisebox{2pt}{\tikz{\draw[-,green!40!green,solid,line width = 0.9pt](0,0) -- (5mm,0);}}}

\newcommand{\lightblueline}{\raisebox{2pt}{\tikz{\draw[-,cyan!60!white,solid,line width = 0.9pt](0,0) -- (5mm,0);}}}

\newcommand{\yellowline}{\raisebox{2pt}{\tikz{\draw[-,yellow!80!yellow,solid,line width = 0.9pt](0,0) -- (5mm,0);}}}
\newcommand{\pinkline}{\raisebox{2pt}{\tikz{\draw[-,magenta!60!white,solid,line width = 0.9pt](0,0) -- (5mm,0);}}}

\newcommand{\cyanline}{\textcolor{cyan!60!white}{\rule{1em}{1.5pt}}}
\newcommand{\redorangeline}{\textcolor{red!70!orange}{\rule{1em}{1.5pt}}}
\usepackage{siunitx}
\usepackage{subcaption}

\lefttitle{U. Cadambi Padmanaban, Midya. S, He. P, Ganapathisubramani. B, Symon. S}
\righttitle{Journal of Fluid Mechanics}

\title{Efficient three-dimensional variational data assimilation of multi-plane PIV data}

\author{Uttam Cadambi Padmanaban\aff{1}, Samaresh Midya\aff{1}, Ping He\aff{2} Bharathram Ganapathisubramani\aff{1} \and Sean Symon\aff{1}}

\affiliation{\aff{1}Department of Aeronautical and Astronautical Engineering, University of Southampton, University Road, Southampton, SO17 1BJ, UK
\aff{2}Department of Aerospace Engineering, Iowa State University, Ames, USA}

\corresau{Uttam Cadambi Padmanaban, \email{ucp1y23@soton.ac.uk}}

\begin{document}
\maketitle

\begin{abstract}
We perform three-dimensional variational data assimilation (3DVar) using a discrete adjoint approach to optimise the time-averaged momentum equations. The experimental data consist of sparse stereoscopic particle image velocimetry (PIV) measurements collected along $12$ cross-stream planes in the wake of a vehicle-like bluff body at a Reynolds number $Re_L = 5.64 \times 10^5$ based on the streamwise body length. Adjoint localisation is proposed and implemented to reduce the memory footprint of the discrete adjoint method for spatially-varying control variables in 3DVar by confining the control variable space to a user-defined subdomain. Restricting the control variable to $12$~\% of the full control space yields a maximum reduction in peak memory of $64$~\%, while producing assimilated fields of comparable fidelity with respect to mean velocity and the optimised momentum forcing field. The localised adjoint case improves upon the baseline Spalart--Allmaras turbulence model and recovers the correct asymmetric topology of the complex three-dimensional (3D) recirculation bubble. The assimilated Reynolds shear stress agrees well with the experiment, and the assimilated mean pressure is shown to be physically consistent when correlated with the in-plane vorticity fields. A data efficiency study is also performed, in which the number of planes provided for assimilation is progressively reduced, demonstrating that the data coverage must extend at least to the end of the primary recirculation bubble to adequately constrain the near-wake dynamics. The efficiency that adjoint localisation affords is crucial for assimilating sparse, experimental data for 3D separated flows on fine meshes that can tackle industrial problems of interest. 
\end{abstract}


\section{Introduction}
\label{section:Introduction}

The study of turbulent flows is challenging for engineering applications ranging from automotive and aerospace to wind engineering. One of the main reasons for this challenge is the lack of low-cost yet sufficiently accurate computational fluid dynamics (CFD) models at high Reynolds numbers for computing three-dimensional (3D) flow fields and obtaining time-averaged quantities such as lift and drag coefficients, skin friction, and mean pressure. The trade-off between computational cost and numerical fidelity is well established. Attempting scale-resolving simulations for a passenger aircraft is clearly not tractable \citep{pope2001turbulent, choi2012grid}, while the use of Reynolds-averaged Navier--Stokes (RANS) models may not be accurate. This significantly impairs our ability to understand complex 3D flow physics without performing experiments. The advancement of flow imaging techniques such as particle image velocimetry (PIV) \citep{adrian1991particle}, has allowed us to understand flow physics at high Reynolds numbers that remain inaccessible to computational models. Planar PIV yields two velocity components on a single plane, which may be insufficient to characterise 3D flows that arise in most engineering applications. Stereoscopic and tomographic PIV \citep{elsinga2006tomographic} make it possible to obtain all three velocity components on a plane and within a volume, respectively, but require significantly more demanding setups and yield limited spatial coverage. Quantities such as surface and volumetric pressure are not directly accessible from PIV and must be measured separately or inferred from the velocity data. 

Data assimilation (DA) is a method that can improve the predictions of computational models, de-noise experimental observations, and enhance their spatial resolution. DA emerged in the context of numerical weather prediction (NWP), where the need to initialise forecasts from sparse atmospheric measurements was most acute \citep{le1986variational}. One class of methods, known as variational DA, solves a minimisation problem in which the misfit between model predictions and observations (objective function) is reduced by tuning a control variable of choice. The minimisation of the objective function can be done using gradient-based techniques that require the total derivative of the objective function with respect to the control variable (hereafter referred to as sensitivity). The state of the art in computing the sensitivity is the adjoint method \citep{peter2010numerical}. The adjoint method can be applied using a continuous approach (linearising the partial differential equations to generate adjoint equations, followed by their discretisation) or a discrete approach (formulating the adjoint equations once the governing equations of fluid flow are discretised). Several studies provide a good comparison between these methods \citep{nadarajah2000comparison, giles2000introduction, peter2010numerical}.
\subsection{Assimilation of time-averaged flows}
The direct application of DA to turbulent flows beyond the geophysical setting of NWP emerged only recently. We review some of the works in the context of mean flow assimilation. \cite{foures2014data} assimilated direct numerical simulation (DNS) data of a flow past a circular cylinder at $Re = 150$ using two-dimensional (2D) mean velocity data by tuning a forcing term in the momentum equations. While a pioneering work, the study was restricted to a low Reynolds number and noted the impracticality of relying on DNS data for flows at high Reynolds numbers (of practical significance). \cite{symon2017data} performed DA on an idealised airfoil at a chord-based Reynolds number $Re_c = 13,500$ using planar PIV experimental data and the same framework as \cite{foures2014data}. The use of experimental data was shown to be possible (by implementing projection operators) as was the case in \cite{symon2019tale}, albeit for increasing spatial resolution and filling in the gaps that existed in the experimental data. The lack of a turbulence model in both these studies made the optimisation problem considerably harder to solve. This was overcome by \cite{franceschini2020mean}, who implemented a turbulence model to allow the initial state of the optimisation problem to be initialised with a baseline RANS model output. A crucial distinction was made between a turbulence model correction and a momentum source term in the context of data sparsity and quality of reconstruction. A Dirac-delta type pointwise forcing was observed when the momentum forcing control variable was used to assimilate sparse data. This was overcome by implementing regularisation to smooth out the forcing \citep{mons2021linear, franceschini2020mean}. These studies were, however, confined to 2D configurations at low Reynolds numbers.

Despite the successes of variational DA for canonical 2D flows, the assimilation of 3D flows with limited observational data remains challenging. \cite{he2018data} provided an early demonstration of 3D variational (3DVar) DA by introducing a spatially-varying $\beta$ correction to the SA turbulence production term within the field inversion framework of \cite{singh2017augmentation}, implemented in OpenFOAM \citep{weller1998tensorial}. The continuous adjoint equations were derived with simplifications to improve numerical stability, at the expense of gradient accuracy. For the 3D flow past a wall-mounted cube at $Re = 10^5$, \cite{he2018data} demonstrated that 3D variational DA with a turbulence model correction is feasible from limited observations. \cite{yan2022data} performed 3D DA for the flow past a canonical configuration of the Faith hill \citep{bell2012surface} model to derive data-augmented turbulence models. Due to the large number of design variables, the authors limited the correction region \textit{a priori} to ensure efficient optimisation. \cite{he2021data} used the continuous adjoint to assimilate measurements from PIV and large eddy simulation (LES) for jets and heated plates using an anisotropic eddy viscosity which was optimised using the variational method. The study also examined Reynolds stresses from the assimilated fields. 

A more recent class of techniques approaches assimilation through machine learning. Physics-informed neural networks (PINNs) \citep{raissi2019physics} integrate the governing equations of fluid flow into the training of neural networks. Unlike traditional neural networks, PINNs impose the governing equations as soft constraints by minimising their residuals, ensuring that the model adheres to physical laws while learning the solution. PINNs have been applied to fluid mechanics problems ranging from simple flows to demonstrate the method \citep{sliwinski2023mean} to problems that have integrated turbulence models into the network \citep{patel2024turbulence}. Most of these studies use DNS or LES reference data to train the model. Some studies have explored the use of experimental data for jets \citep{von2022mean}, magnetic resonance imaging to correct noisy data and displacement artefacts \citep{villie2025physics}, and a more complicated case of stalled flow past an airfoil at high Reynolds number \citep{toma2026mixed}. \cite{steinfurth2024assimilating} used PINNs to reconstruct the 3D flow field through a diffuser with a turbulent separation bubble (TSB). To achieve this, the study required three-component mean velocity fields along multiple planes demonstrating the steep data requirement and the challenge of 3D mean flow assimilation.

\subsection{Continuous versus discrete adjoint}
Returning to 3DVar, the continuous adjoint method benefits from reusing the primal solver for the adjoint equations without additional memory requirements. FreeFem++ \citep{hecht2012new} has been one tool suitable for DA. However, deriving and implementing adjoint equations for each turbulence model is a laborious task, and introducing new objective functions or design variables requires rigorous mathematical derivations. \cite{brenner2022efficient} applied a discrete adjoint approach to correct the eddy viscosity field in RANS simulations using a $k$-$\epsilon$ turbulence model, optimising a spatially-varying scalar multiplier with a frozen eddy viscosity. This approach is constrained by the Boussinesq approximation, and gradient accuracy remained a challenge. \cite{brenner2024variational} subsequently improved the algorithm and introduced a momentum source term correction, demonstrating that mean-velocity reconstruction accuracy was unaffected when coarse input data were considered. In a similar vein, \cite{mons2024data} used a corrective source term in the turbulence transport equation of the SA model for mean velocity assimilation of flow past a NACA 0012 airfoil. The near-stall flow was at an angle of attack $\alpha = 10^\circ$ and in the Reynolds number range $4.3 \times 10^4 \leq Re \leq 6.4 \times 10^4$. The reference mean velocity field was obtained from time-averaged PIV. DA was able to fill in the gaps that resulted from shadows and reflections during the data acquisition stage of PIV. The effects of three-dimensionality on reconstructed flow quantities, such as Reynolds stresses, were not investigated.

One of the challenges of performing 3D assimilation using a discrete adjoint approach is the lack of a modular tool that combines fluid solvers, turbulence models and adjoint solvers. \cite{thompson2024effect} demonstrated the use of DAFoam \citep{he2018aerodynamic, he2020dafoam} for performing variational DA of a flow through a periodic hill by providing synthetic PIV data and optimising a scalar multiplier to the production term of the Spalart-Allmaras (SA) \citep{spalart1992one} turbulence model. \cite{cadambi2026three} performed 3DVar for a NACA 0012 airfoil in deep stall at $Re_c \approx 7.5 \times 10^4$ using a single plane of two-component PIV reference data by optimising a forcing term in the momentum equations. The use of 3DVar by applying 3D constraints and solving the incompressible RANS equations was shown to minimise the 2D continuity errors in the experimental data. As such, it was posited that for separated flows, the use of 3DVar is imperative to ensure that quantities such as mean pressure and Reynolds stresses are physically consistent when using a single plane of planar PIV data as a reference for assimilation. An alternative is to use multiple planes of planar PIV data, as shown in \cite{stallcell2026}, where the full-field reconstruction of a flow past a stalled wing was performed using as few as two PIV data. The assimilation successfully reproduced the features of a stall cell \citep{moss1971two, winkleman1982, dell2016measurement, de2022effects}, particularly the mushroom-shaped structure on the surface. 

It can be concluded that the use of the discrete adjoint method is an attractive choice for performing 3DVar of high Reynolds number flows. One of the bottlenecks is its steep memory requirement. This is elaborated in \cite{kenway2019effective} and is a particular problem for field control variables that are defined on every cell of the mesh. In all the studies examined so far (with the exception of \cite{yan2022data}), the control variable space spans the entire computational grid, while in most cases, the corrections are applied only in specific physical locations. This is expected, since the corrections are targeted at regions in the flow that show maximum departure from the reference experimental data. In fact, methods that apply data-driven turbulence models locally have already been shown to be effective \citep{wu2025data, wu2025development, buchanan2025data, he2026field}. Restricting the optimisation of the control variables to a selective region of the flow directly reduces the memory footprint of the discrete adjoint method. Moreover, such a restriction ties in naturally with the manner in which data-driven turbulence models are conditioned to act selectively so that the baseline performance is unaltered.

The application of variational DA to full-field reconstruction of 3D separated flows from sparse experimental data is challenging for several reasons. One arises from the proliferation of turbulence models developed for a wide range of flow problems, either as extensions of existing ones or as entirely new closures. The variational DA framework must therefore remain agnostic to the choice of turbulence model so that it applies equally to existing and newly developed models. Another challenge is the computational difficulty of the discrete adjoint method. The memory required to construct and store the discrete adjoint grows rapidly with mesh size. For the fine 3D meshes that separated flows demand, this cost, rather than the optimisation itself, can become the bottleneck. Scaling variational DA to configurations of engineering relevance, therefore, demands efficiency as much as accuracy. Since sparse experimental data are used, the assimilated fields must demonstrate that the 3D flow physics is well reproduced. It is essential to examine secondary or derived quantities (those that are not provided as input to the assimilation) to justify the utility of DA in recovering unmeasured quantities. A final challenge is the lack of clarity on the density of measurements that are needed to correct the turbulence model to recover a 3D, incompressible mean flow field through variational DA.

\subsection{Contributions of this study}
To address these challenges, we assimilate stereoscopic PIV data taken from \cite{midya2025experimental} comprising the three components of velocity along $12$ cross-stream planes in the wake of a vehicle-like bluff body at a Reynolds number $Re_L = 5.64 \times 10^5$ based on the streamwise length of the model. We use 3DVar with a discrete adjoint method, using the SA model as the baseline, and optimise a forcing term in the momentum equations. Since the forcing acts on the momentum equations and the discrete adjoint requires no model-specific derivation, the framework applies equally to any turbulence model, although only the SA model is considered here. We propose adjoint localisation as a way to reduce the peak memory usage of the discrete adjoint method by restricting the control variable space to a user-defined region. The method is demonstrated to achieve significant memory savings without negatively impacting the assimilation quality. Using adjoint localisation, we demonstrate a full-field reconstruction, including an accurate reconstruction of the asymmetric topology of the primary recirculation bubble, with only cross-stream measurement plane data as input. We also examine quantities that are not provided as input to the assimilation, namely, the Reynolds shear stress and the mean pressure. Finally, we perform a data-efficiency study by restricting the number of data planes and assessing its impact on the Reynolds shear stress. 

The rest of the paper is organised as follows. Section~\ref{section:methods} describes the computational methods which include the governing equations and the variational method. Section~\ref{section:experimental_data} covers the experimental setup and the data processing pipeline and \S \ref{section:computational_setup} explains the computational setup. Section~\ref{section:baseline_model} assesses the baseline SA model, where its deficiencies relative to the experimental data are quantified. One of the important contributions of this study is adjoint localisation. This is developed in \S \ref{section:adjoint_localisation} together with the assimilation setup and an assessment of three candidate localisation regions. Section~\ref{section:mean_flow_assimilation} examines the assimilated mean flow, the 3D topology of the primary recirculation bubble, and derived quantities not directly provided as input to the assimilation, namely the Reynolds shear stress and the mean pressure field. Section~\ref{section:data_efficiency} examines the sensitivity of the reconstruction to the number of input measurement planes. Section~\ref{section:conclusion} provides conclusions and future work.  

\section{Methods}
\label{section:methods}
\subsection{Governing equations}
\label{subsection:governing_equations}
The experimental data used in this study consist of incompressible, time-averaged velocity fields. Since DA operates on the mean flow, the time-averaged continuity and momentum equations are solved. These equations are obtained by Reynolds averaging the governing equations after decomposing each quantity into its mean (ensemble- or time-average) and fluctuation. The resulting RANS equations are given by:
\begin{gather}
\label{equation:momentum}
\frac{\partial U_i}{\partial x_i} = 0, \\
U_j \frac{\partial U_i}{\partial x_j} + \frac{1}{\rho} \frac{\partial p}{\partial x_i} - \frac{\partial}{\partial x_j} \left( \nu \left(\frac{\partial U_i}{\partial x_j} + \frac{\partial U_j}{\partial x_i} \right)\right) - \frac{\partial \tau_{ij}}{\partial x_j} = 0,
\end{gather}
where $U_i$ is the mean velocity, $p$ is the pressure, $\rho$ is the density of the fluid, $\nu$ is the kinematic viscosity, $\tau_{ij}$ is the Reynolds stress tensor, and $x_i$ represents the spatial coordinates. The Reynolds stress tensor $\tau_{ij}$ is defined as the averaged outer product of the fluctuating velocity components, presenting the well-known problem of closure.

To close the system, the Reynolds stress tensor is modelled using the Boussinesq hypothesis, $\tau_{ij} = -\nu_t \left(\partial U_i/\partial x_j + \partial U_j / \partial x_i \right) + \frac{2}{3}k\delta_{ij}$, where the isotropic term is absorbed into a modified pressure $P^*$. Substituting this closure into equation~\ref{equation:momentum} and simplifying using the continuity equations results in
\begin{gather}
\label{equation:turb_model_rans}
U_j \frac{\partial U_i}{\partial x_j} + \frac{1}{\rho} \frac{\partial P^*}{\partial x_i} - \frac{\partial}{\partial x_j} \left( [\nu + \nu_t] \frac{\partial U_i}{\partial x_j} \right) = 0,
\end{gather}
where $\nu_t$ is the eddy viscosity. To obtain $\nu_t$, the Spalart--Allmaras (SA) turbulence model is employed, which solves a transport equation for a surrogate variable $\Tilde{\nu}$. The open-source, finite-volume method (FVM) CFD package OpenFOAM is used in solving the governing equations in \eqref{equation:turb_model_rans} and the turbulence transport equation. 

\subsection{Variational data assimilation}
\label{subsection:variational_da}
We expect a discrepancy in the mean velocity field between the SA turbulence model output and the experimental observations, necessitating the use of DA. Variational DA minimises this discrepancy by solving an optimisation problem in which an objective function is minimised, subject to constraints, by tuning a control variable. In this study, we use a forcing term added to the momentum equations, which modifies equation~\ref{equation:turb_model_rans} into
\begin{gather}
\label{equation:mom_correction}
U_j \frac{\partial U_i}{\partial x_j} +\frac{1}{\rho} \frac{\partial P^*}{\partial x_i} - \frac{\partial}{\partial x_j} \left( \nu \frac{\partial U_i}{\partial x_j} \right) - f_{R_i} - f_{c_i},
\end{gather}
where $f_{R_i} = \partial \tau_{ij}/\partial x_j$ is the modelled Reynolds forcing and $f_{c_i}$ is the corrective forcing that serves as the control variable. This choice of control variable is motivated by the greater flexibility (see \citet{franceschini2020mean, symon2017data, brenner2024variational}) that it provides in reconciling modelling deficiencies, particularly for high Reynolds number, separated flows. The corresponding objective function is
\begin{align}
\label{equation:mom_source_obj_func}
J(\mathbf{u}, \mathbf{f}_c) = \frac{1}{2} \|\mathcal{Q}(\mathbf{u}, \mathbf{f}_c) - \Tilde{\mathbf{Q}}\|_Q^2,
\end{align}
where $\Tilde{\mathbf{Q}}$ denotes the reference data, and the operator $\mathcal{Q}(\cdot)$ projects the computational mean velocity field onto the measurement space $Q$, such that $\mathcal{Q}(\mathbf{w}, \mathbf{x}) \in Q$. The norm $\|\cdot\|_Q$ is defined in the measurement space. We use the nearest-neighbour operator $\mathcal{Q}$, following \cite{franceschini2020mean}. This projects the computational mean velocity field onto the experimental grid by searching for the nearest computational grid point to each experimental data point within the field of view and sampling the velocity on those points. The objective function is then evaluated directly at these computational grid points, which eliminates the need for the adjoint of the measurement operator to transfer the discrepancy field back onto the computational mesh.

To clarify the role of each contribution, a combined forcing term $f_i$ is defined as
\begin{gather}
\label{equation:total_forcing_vector}
\mathbf{f} = \mathbf{f_{R}} + \mathbf{f_{c}}.
\end{gather}
The contribution of the Reynolds stress to the momentum equation, as represented by the Boussinesq hypothesis, is the deviatoric part $-2\nu_t S_{ij}$, with the isotropic part $\tfrac{2}{3}k\,\delta_{ij}$ absorbed into the modified pressure $P^*$. The corrective forcing $\mathbf{f}_c$ then accounts for the residual physics that this linear eddy-viscosity representation fails to capture, including any component of the true deviatoric Reynolds stress that is not aligned with the mean strain rate. The corrective forcing is initialised to zero at the start of the assimilation, recovering the baseline case.

Using a forcing term in the momentum equations as a control variable has its drawbacks. When sparse observations are used as reference data, the quality of the assimilation can degrade unless regularisation is applied \citep{brenner2022efficient, franceschini2020mean, symon2017data, cato2023comparison}. Without regularisation, the forcing is confined to locations where reference data is available, producing a spurious, oscillatory forcing field of the Dirac delta-type. To suppress these oscillations, a regularisation term is included in the objective function, giving the combined form:
\begin{align}
    \label{equation:combined_obj_func}
    J_{\text{tot}} = J^u + \lambda J^f,
\end{align}
where $J^u$ is the velocity objective function (equation~\ref{equation:mom_source_obj_func}), $J^f$ is the regularisation term, and $\lambda$ is a hyperparameter controlling the relative contribution of the two terms. The value of $\lambda$ is case-dependent and is determined through an L-curve analysis. This procedure is demonstrated in Appendix~\ref{appA}. A gradient-based regulariser is adopted:
\begin{align}
    \label{eqn:regulariser}
    J^f = \sum_{i \in \Omega_q} \left| \nabla \mathbf{f}_c \right|^2,
\end{align}
where $\Omega_q$ is the set of cells for which $\mathbf{f}_c \neq \mathbf{0}$, and $|\nabla \mathbf{f}_c|$ denotes the Frobenius norm of the gradient of the forcing field. This term penalises sharp gradients in the forcing field and encourages a smooth spatial distribution of the correction. 

\subsection{Discrete adjoint method}
\label{subsection:discrete_adjoint_method}
The minimisation of the objective function defined in equation~\ref{equation:mom_source_obj_func} requires the sensitivity. This is obtained using the discrete adjoint method. 
\begin{align}
\label{equation:sensitivity_equation}
\frac{dJ}{d \mathbf{f}_c} = \frac{\partial J}{\partial \mathbf{f}_c} - \psi^T \frac{\partial \mathbf{R}}{\partial \mathbf{f}_c},
\end{align}
where $\psi^T$ is the transpose of the adjoint vector, and $\mathbf{R}$ is the residual of the RANS equations (which also serves as an equality constraint). The adjoint vector is obtained by solving the adjoint equation
\begin{align}
\label{equation:adjoint_equation}
\left[\frac{\partial \mathbf{R}}{\partial \mathbf{w}}\right]^T \psi = \left[\frac{\partial J}{\partial \mathbf{w}}\right]^T,
\end{align}
where $\mathbf{w}$ is the vector of state variables ($\rho,~\mathbf{U}, ~p, ~\Tilde{\nu}$) and $\partial\mathbf{R}/\partial\mathbf{w}$ is the state Jacobian. Detailed derivation can be found in \citet{kenway2019effective}. We use the OpenFOAM-based package DAFoam for the discrete adjoint solver. DAFoam is an attractive choice since it uses reverse mode automatic differentiation (AD) for computing the partial derivatives which does not require the explicit assembly of the state Jacobian \citep{kenway2019effective}. Additionally, it is modular and open-source, which allows fast implementation of new capabilities without hampering the functionality of the existing solvers. Although it was mentioned earlier that DAFoam does not explicitly store the state Jacobian for the solution of equation~\ref{equation:adjoint_equation}, it must be emphasised that there is a requirement for the explicit assembly and storage of the state preconditioner matrix $[\partial\mathbf{R}/\partial\mathbf{w}^T]_{PC}$ which scales with the size of the mesh. This is a crucial point which will be revisited in \S \ref{section:adjoint_localisation}.

With the sensitivities computed, the minimisation of the objective function can be performed. The objective function is minimised subject to the equality constraints of the RANS equations and bound constraints on the control variable, forming a nonlinear, constrained optimisation problem that requires a robust optimiser. We use SNOPT \citep{gill2005snopt} for the optimisation which operates a quasi-Newton method in conjunction with a line-search filter that minimises a penalty function to obtain the search directions. SNOPT exploits the sparsity of the constraint Jacobian and employs a limited-memory  Broyden-Fletcher-Goldfarb-Shanno (L-BFGS) approximation of the Hessian, allowing efficient handling of large-scale problems. Convergence is governed by the default stopping criteria of SNOPT, which monitors a combination of projected gradient norms and step sizes to determine when the optimisation is complete.

\section{Experimental data}
\label{section:experimental_data}
The data used in this study are obtained from experiments performed in \cite{midya2025experimental}. This section provides details of the experimental setup used to generate these data and the post-processing steps required to prepare them as input for DA. The model geometry and the imaging experiments are described in \S \ref{subsection:experimental_setup}. The flow physics is characterised in \S \ref{subsection:flow_physics} through data presented on select planes. As discussed in \S \ref{subsection:data_processing}, a post-processing step is applied to the experimental data to make them compatible with the assimilation procedure. The resulting data density is representative of a general experimental campaign employing field-imaging techniques, allowing the robustness of the method under sparse data conditions to be demonstrated.

\subsection{Experimental setup and data acquisition}
\label{subsection:experimental_setup}
\begin{figure}
    \centering
    \psfrag{a}[cc][cc]{Bottom face}
    \psfrag{b}[cc][cc]{Top face}
    \psfrag{c}[cc][cc]{Rear face}
    \psfrag{d}[cc][cc]{Side face (R)}
    \psfrag{e}[cc][cc]{Side face (L)}
    \psfrag{f}[cc][cc]{Wind tunnel wall}
    \psfrag{x}{$x$}
    \psfrag{y}{$y$}
    \psfrag{z}{$z$}
    \psfrag{u}[cc][cc]{$U_\infty$}
    \includegraphics[width=0.8\textwidth]{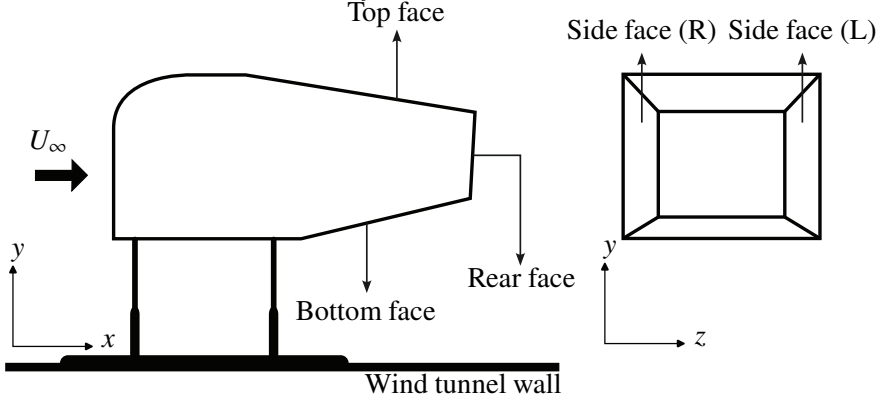}
    \caption{Front ($x$--$y$ plane) and back ($y$--$z$ plane) views of the multi-wake model.}
    \label{fig:multiWake}
\end{figure}

The multi-wake model (hereafter referred to as the model) shares geometric features with the Ahmed body and the DrivAer model. A schematic of the model is shown in figure~\ref{fig:multiWake}. It is a streamlined body with gentle inclinations on its sides and unequal inclinations on the top and bottom faces. The top and bottom face inclination angles are considerably lower than those of the Ahmed body, resulting in distinct flow features that are discussed in detail in \cite{midya2025experimental}. The exact dimensions of the model may also be found therein. The streamwise length of the model is $L_x = 0.864$~m, which is used as the length scale to non-dimensionalise all distances, defined as $x^* = x/L_x$, $y^* = y/L_x$, and $z^* = z/L_x$, where $x$, $y$, and $z$ are the streamwise, wall-normal, and spanwise coordinates, respectively. The model has a plane of symmetry at $z^* = 0$ (the $x$--$y$ plane), but lacks symmetry about the $y$--$z$ and $x$--$z$ planes. This symmetry is exploited in the computational setup, which is discussed in \S \ref{section:computational_setup}.

\begin{figure}
    \centering
    \psfrag{a}[cc][cc]{Camera $1$}
    \psfrag{b}[cc][cc]{Camera $2$}
    \psfrag{c}[cc][cc]{Fixed dual pulse laser}
    \psfrag{d}[cc][cc]{S$1$}
    \psfrag{e}[cc][cc]{S$2$}
    \psfrag{f}[cc][cc]{$100~\mathrm{mm}$}
    \psfrag{g}[cc][cc]{Plane$1$}
    \psfrag{p}[cc][cc]{Plane$2$}
    \psfrag{i}[cc][cc]{Plane$3$}
    \psfrag{j}[cc][cc]{Movable platform}
    \psfrag{k}[cc][cc]{Subplanes}
    \psfrag{x}{$x$}
    \psfrag{y}{$y$}
    \psfrag{z}{$z$}
    \psfrag{l}[cc][cc]{$U_\infty$}
    \includegraphics[width=1\textwidth]{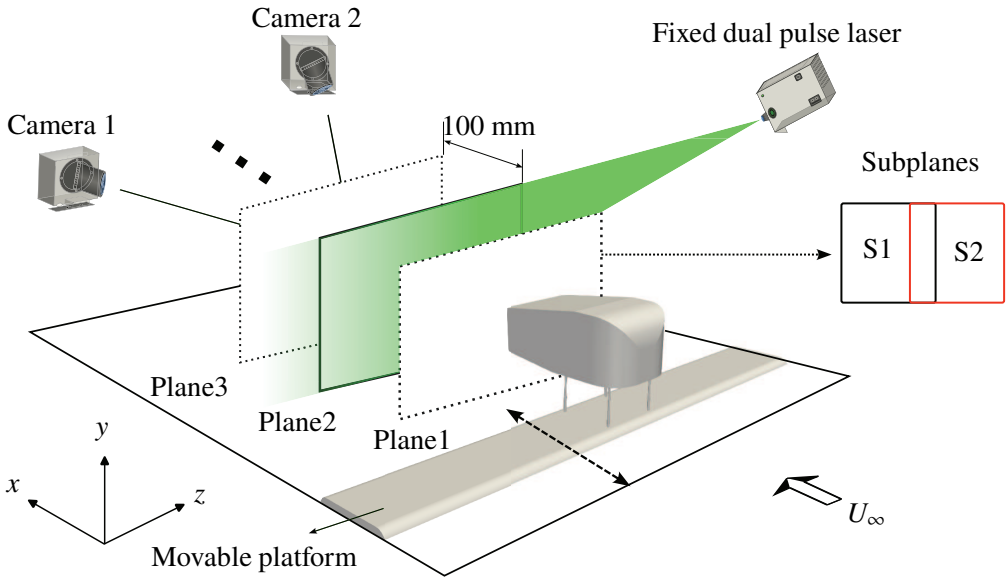}
    \caption{Experimental setup for stereo PIV of the multi-wake model.}
    \label{fig:chap_4_schematic}
\end{figure}

The model is mounted on four telescopic struts whose diameters are kept as small as feasible to minimise interference with the wake of the main body. The struts are placed on a movable platform that traverses the body along the streamwise direction, and the platform is attached to a cross rail fixed to the wind tunnel wall. The experiments are performed in the R.\ J.\ Mitchell Wind Tunnel at the University of Southampton, a closed-return facility with a working test section of $3.6~\mathrm{m} \times 2.4~\mathrm{m} \times 10~\mathrm{m}$, a speed range of $4$--$40~\mathrm{ms^{-1}}$, and a free-stream turbulence level of less than $0.2$~\%. The experimental setup is shown in figure~\ref{fig:chap_4_schematic}. Data are collected in the wall-normal--spanwise ($y$--$z$) plane at twelve streamwise stations separated by $100~\mathrm{mm}$, beginning $11~\mathrm{mm}$ downstream of the rear face of the body. The free-stream velocity is $U_\infty = 10~\mathrm{ms^{-1}}$, giving a Reynolds number based on the streamwise body length of $Re_L = 5.64 \times 10^5$. The normalised streamwise, wall-normal, and spanwise velocity components are defined as $U_x^* = U_x/U_\infty$, $U_y^* = U_y/U_\infty$, and $U_z^* = U_z/U_\infty$, respectively, where $U_x$, $U_y$, and $U_z$ are the dimensional mean velocity components and $(^*)$ denotes normalisation by $U_\infty$. The bottommost face of the body is positioned $290~\mathrm{mm}$ above the wind tunnel floor, well clear of the maximum boundary layer height in the tunnel, which is $\mathcal{O}(35~\mathrm{mm})$. For ease of reference, the normalised streamwise locations of the measurement planes are assigned plane numbers and are tabulated in table~\ref{tab:chap_4_planes}.

The wake is imaged using a stereo PIV system comprising two identical cameras, denoted Camera~1 and Camera~2 in figure~\ref{fig:chap_4_schematic}, inclined at $\approx 35^\circ$ and $\approx 40^\circ$ to the freestream, respectively. To achieve a large combined field of view (FOV), two separate experiments are performed to capture subplanes S1 and S2 (shown in figure~\ref{fig:chap_4_schematic}), which are subsequently stitched together using a linear combination in the overlap region (see \cite{midya2025experimental} for details). The identical camera specifications yield individual FOVs measuring $448~~\mathrm{mm}$ in $z$ and $405~\mathrm{mm}$ in $y$. The flow is seeded with smoke particles generated by a fog machine with a nominal particle diameter of $\mathcal{O}(\SI{1.5}{\micro m})$, and the wake cross-plane ($y$--$z$) is illuminated by a dual-pulse Neodymium-doped Yttrium Aluminium Garnet (\mbox{Nd:YAG}) laser with a sheet thickness of $2~\mathrm{mm}$. The laser operates at a wavelength of $532~\mathrm{nm}$, a pulse energy of $200~\mathrm{mJ}$, a repetition rate of $15~\mathrm{Hz}$, and a pulse duration of $6~\mathrm{ns}$. The PIV images are processed using PIVTools \citep{taylor6661397pivtools} with a multi-pass cross-correlation scheme. The interrogation window is initialised at $128$~px $\times$ $128$~px and progressively reduced to a final size of $48$~px $\times$ $48$~px over five passes (three at the final window size), with a $75$~\% overlap applied throughout. The resulting vector grid spacing in the time-averaged velocity field is $\Delta y \times \Delta z = \SI{0.857}{mm} \times \SI{0.939}{mm}$.

\begin{figure}
    \centering
    \begin{subfigure}{\textwidth}
            \caption{}
        \centering
        \psfrag{A}[cc][cc]{$U_x^*$}
        \psfrag{B}[cc][cc]{$U_y^*$}
        \psfrag{C}[cc][cc]{$U_z^*$}
        \psfrag{P}[cc][cc]{$y^*$}
        \psfrag{Q}[cc][cc]{$z^*$}
        \psfrag{a}[rc][rc]{$-0.15$}
        \psfrag{b}[rc][rc]{$0.0$}
        \psfrag{c}[rc][rc]{$0.15$}
        \psfrag{e}[tc][tc]{$-0.25$}
        \psfrag{f}[tc][tc]{$0.0$}
        \psfrag{g}[tc][tc]{$0.25$}
        \psfrag{I}[cc][cc][0.8]{$-0.3$}
        \psfrag{J}[cc][cc][0.8]{$0.5$}
        \psfrag{K}[cc][cc][0.8]{$1.2$}
        \psfrag{L}[cc][cc][0.8]{$-0.1$}
        \psfrag{M}[cc][cc][0.8]{$0.0$}
        \psfrag{N}[cc][cc][0.8]{$0.1$}
        \psfrag{O}[cc][cc][0.8]{$-0.1$}
        \psfrag{R}[cc][cc][0.8]{$0.0$}
        \psfrag{S}[cc][cc][0.8]{$0.1$}
        \includegraphics[width=\textwidth]{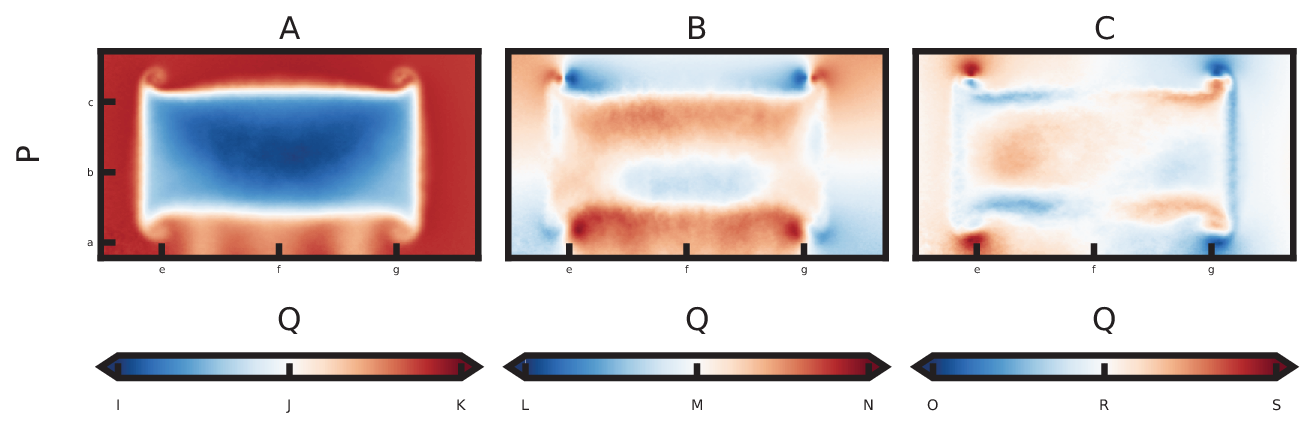}
        \label{fig:raw_plane2}
    \end{subfigure}

    \vspace{-1em}

    \begin{subfigure}{\textwidth}
        \centering
                \caption{}
        \psfrag{A}[cc][cc]{$U_x^*$}
        \psfrag{B}[cc][cc]{$U_y^*$}
        \psfrag{C}[cc][cc]{$U_z^*$}
        \psfrag{P}[cc][cc]{$y^*$}
        \psfrag{Q}[cc][cc]{$z^*$}
        \psfrag{a}[rc][rc]{$-0.15$}
        \psfrag{b}[rc][rc]{$0.0$}
        \psfrag{c}[rc][rc]{$0.15$}
        \psfrag{e}[tc][tc]{$-0.25$}
        \psfrag{f}[tc][tc]{$0.0$}
        \psfrag{g}[tc][tc]{$0.25$}
        \psfrag{I}[cc][cc][0.8]{$0.7$}
        \psfrag{J}[cc][cc][0.8]{$0.9$}
        \psfrag{K}[cc][cc][0.8]{$1.1$}
        \psfrag{L}[cc][cc][0.8]{$-0.1$}
        \psfrag{M}[cc][cc][0.8]{$0.0$}
        \psfrag{N}[cc][cc][0.8]{$0.1$}
        \psfrag{O}[cc][cc][0.8]{$-0.1$}
        \psfrag{R}[cc][cc][0.8]{$0.0$}
        \psfrag{S}[cc][cc][0.8]{$0.1$}
        \includegraphics[width=\textwidth]{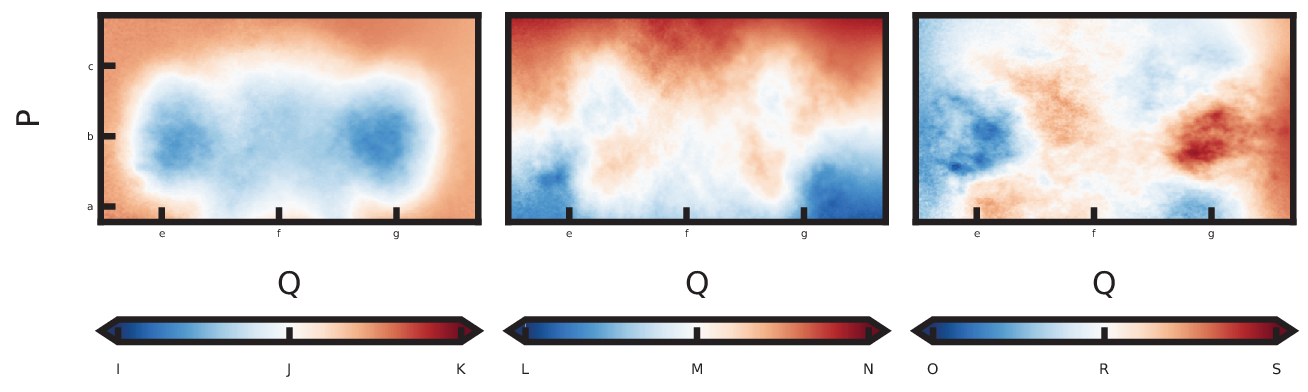}
        \label{fig:raw_plane10}
    \end{subfigure}

    \caption{(From left to right) Mean streamwise, wall-normal and spanwise components of velocity in the wake of the model along (a) Plane $2$ and (b) Plane $10$. See table~\ref{tab:chap_4_planes} for the streamwise locations of the measurement planes.}
    \label{fig:raw_uvw_all}
\end{figure}

\subsection{Mean flow features}
\label{subsection:flow_physics}
Due to the complexity of the flow and the large number of measurement planes, only a representative subset of the imaged flow is presented here. A thorough characterisation of the flow physics is beyond the scope of this paper and the reader is referred to \cite{midya2025experimental} for a detailed analysis. Two planes are selected to illustrate distinct dynamical regimes present in the wake. The mean streamwise, wall-normal, and spanwise velocity components on Plane~$2$ and $10$ are presented in figure~\ref{fig:raw_uvw_all}, covering the near and far wake regions, respectively. As will be shown later, these planes also serve as useful reference cases for demonstrating the limitations of baseline RANS models in capturing the flow to a satisfactory level of accuracy.

\begin{table}
    \centering
\begin{tabular}{ccccccccccccc}
        Plane & $1$ & $2$ & $3$ & $4$ & $5$ & $6$ & $7$ & $8$ & $9$ & $10$ & $11$ & $12$\\
        $x^*$ & $0.013$ & $0.127$ & $0.241$ & $0.356$ & $0.470$ & $0.584$ & $0.698$ & $0.813$ & $0.927$ & $1.041$ & $1.158$ & $1.272$ \\
    \end{tabular}
    \caption{Streamwise locations of the measurement planes.}
    \label{tab:chap_4_planes}
\end{table}

Figure~\ref{fig:raw_plane2} shows the mean streamwise, wall-normal and spanwise velocity components on Plane~$2$. It is observed that mean streamwise and wall-normal velocity components are symmetric about $z^* = 0$ (the $x$--$y$ plane), while the spanwise component is anti-symmetric, consistent with the plane of symmetry that the model possesses, as described in \S \ref{subsection:experimental_setup}. First, we describe the mean streamwise velocity field (left-most panel), which shows a large region of recirculation, evident from the negative velocity contours, which is distinct from the four corner vortices. \cite{midya2025experimental} identify this recirculation region as the primary bubble, which is the dominant structure in the near wake. The footprint of the telescopic struts can also be observed near the bottom of the FOV. Next, the wall-normal velocity field  (middle panel) is examined. We can clearly observe the complex, three-dimensionality and the corner vortices in the flow field. The spanwise mean velocity field (right-most panel) also demonstrates this three-dimensionality as well as the footprint of the corner vortices. The fluid is moving towards the symmetry line $z^* = 0$ from the two vertical edges. 

Figure~\ref{fig:raw_plane10} shows the mean streamwise, wall-normal and spanwise velocity fields on Plane~$10$, and they exhibit markedly different structures. The mean streamwise velocity field (left-most panel) shows two distinct lobes of momentum deficit, characteristic of the far wake, with the magnitude of this deficit considerably lower than that observed on Plane~$2$, as is evident from the colour bar scales. This two-lobed structure has been identified as the natural state of the wake for the given geometry and flow conditions \citep{midya2025experimental}. The mean wall-normal velocity field (middle panel) demonstrates exchange of fluid between the top and bottom halves of the F.O.V. The mean spanwise component (right-most panel) has a complex structure, with no clear coherent structure as in the case of the mean streamwise field. All three velocity fields show noisy behaviour, which will be smoothed after assimilation. 

\subsection{Data processing}
\label{subsection:data_processing}
\begin{figure}
    \centering
    \psfrag{G}[cc][cc]{Raw}
    \psfrag{H}[cc][cc]{Folded}
    \psfrag{V}[cc][cc]{Downsampled}
    \psfrag{P}[cc][cc]{$y^*$}
    \psfrag{Q}[cc][cc]{$z^*$}
    \psfrag{U}[cc][cc]{$U_x^*$}
    \psfrag{a}[rc][rc]{$-0.15$}
    \psfrag{b}[rc][rc]{$0.0$}
    \psfrag{c}[rc][rc]{$0.15$}
    \psfrag{d}[rc][rc]{$0.3$}
    \psfrag{e}[tc][tc]{$-0.25$}
    \psfrag{f}[tc][tc]{$0$}
    \psfrag{g}[tc][tc]{$0.25$}
    \psfrag{h}[tc][tc]{$0.5$}
    \psfrag{i}[tc][tc]{$0.1$}
    \psfrag{j}[tc][tc]{$0.2$}
    \psfrag{k}[tc][tc]{$0.3$}
    \psfrag{I}[cc][cc]{$-0.3$}
    \psfrag{J}[cc][cc]{$0.5$}
    \psfrag{K}[cc][cc]{$1.2$}
    \includegraphics[width=\textwidth]{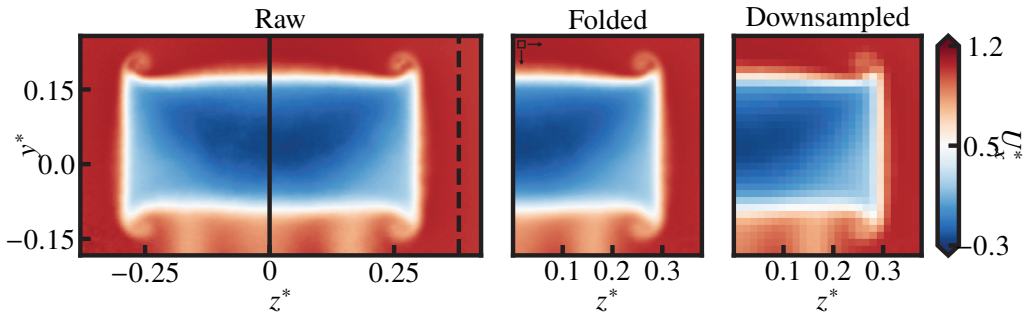}
\caption{Data processing pipeline applied as a post-processing step to the experimental mean velocity fields, illustrated using the mean streamwise velocity component on Plane~$2$. (From left to right) Raw: the mean velocity field at full resolution, with the solid black line indicating the symmetry plane and the dashed black line demarcating the boundary of the right half of the domain used for folding. Folded: the symmetry-averaged field, with the square box indicating the kernel size used for the moving average in the wall-normal and spanwise directions. Downsampled: the final processed velocity field at the reduced resolution used as input to the DA procedure.}
    \label{fig:pipeline_u}
\end{figure}

The experimental data obtained from stereo PIV are spatially well-resolved in the wall-normal and spanwise directions, owing to the subplane strategy described in \S \ref{subsection:experimental_setup}. While this is beneficial for identifying a wide range of flow structures in the wake, the spatial resolution is considerably higher than that encountered in a typical field-imaging campaign, where lower camera resolutions and practical constraints result in much sparser data. To ensure that the DA method has wider applicability beyond this special case of high-resolution data, the experimental fields are downsampled prior to being used as input to the DA procedure. The full data processing pipeline is illustrated in figure~\ref{fig:pipeline_u} and the vector counts at each stage are summarised in table~\ref{tab:chap_4_data_processing}.

\begin{table}
    \centering
    \begin{tabular}{lccccccc}
        Stage & $N_y$ & $N_z$ & Total vectors & $\Delta y$ (mm) & $\Delta z$ (mm) & Kernel size & Reduction factor \\
        Raw & $450$ & $750$ & $337{,}500$ & $0.857$ & $0.939$ & --- & --- \\
        Folded & $450$ & $354$ & $159{,}300$ & $0.857$ & $0.939$ & --- & $\approx 2$ \\
        Downsampled & $33$ & $28$ & $924$ & $\approx 12$ & $\approx 12$ & $14 \times 13$ & $\approx 172$ \\
    \end{tabular}
     \caption{Summary of the data processing pipeline applied to the experimental mean velocity fields. The reduction factor is computed relative to the preceding stage.}
    \label{tab:chap_4_data_processing}
\end{table}

The first step exploits the $x$--$y$ plane symmetry of the flow. The symmetry plane is identified (shown as the solid black line in the leftmost panel of figure~\ref{fig:pipeline_u}), and the data on either side are averaged. Since the right half of the measurement plane extends slightly further in the spanwise direction, vectors beyond the dashed black line are pruned prior to averaging. For the spanwise velocity component, which is anti-symmetric about this plane, the sign is inverted before averaging. This folding step reduces the number of vectors by approximately a factor of $2$ (the exact factor deviates slightly from $2$ due to the unequal number of vectors on either side of the symmetry plane, as discussed above).

The second step applies a moving average using a uniform kernel, whose size is determined based on a target vector spacing of $\Delta y \times \Delta z \approx \SI{12}{mm} \times \SI{12}{mm}$. This target resolution is also chosen to be as close as possible to the spacing of the computational grid, so that the experimental and CFD fields are compared on a consistent spatial basis. The resulting kernel size is $14 \times 13$ in the wall-normal and spanwise directions, respectively. The data are then sampled at every $14$th and $13$th vector in the wall-normal and spanwise directions, respectively, to yield the final downsampled field, shown in the rightmost panel of figure~\ref{fig:pipeline_u}. This resolution is found to be approximately the lowest at which the corner vortices remain discernible. Any coarser sampling would result in these features being averaged out. As can be seen, the primary recirculation bubble is largely unaffected by the downsampling, while the corner vortices remain visible, albeit marginally. The three mean velocity components from the downsampled data are used as reference data for assimilation.

\section{Computational setup}
\label{section:computational_setup}
\begin{figure}
    \centering
    \psfrag{a}{$16L_x$}
    \psfrag{b}{$3L_x$}
    \psfrag{c}{$2L_x$}
    \psfrag{x}[cc][cc]{$x$}
    \psfrag{y}[cc][cc]{$y$}
    \psfrag{z}[cc][cc]{$z$}
        \psfrag{l}[cc][cc]{$U_\infty$}
    \includegraphics[width=\textwidth]{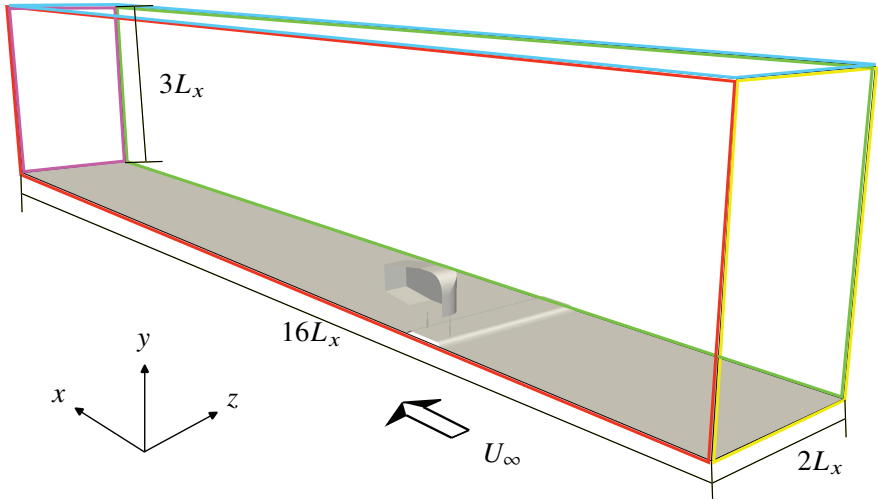}
\caption{Computational domain for the half-body configuration, with dimensions shown along the streamwise, wall-normal, and spanwise directions. The domain boundaries are indicated as follows: \protect\redline~symmetry plane, \protect\yellowline~inlet, \protect\pinkline~outlet, \protect\lightblueline~top face, \protect\greenline~side face.}
    \label{fig:domain}
\end{figure}

The details of the computational setup to perform DA are presented here. The symmetry of the model along the $x$--$y$ plane at $z^* = 0$ is exploited to designate it as a symmetry boundary along the spanwise mid-plane, and only the right half of the model is used to construct the computational domain, shown along with its boundaries in figure~\ref{fig:domain}. The domain is rectangular and measures $16L_x \times 3L_x \times 2L_x$ along the streamwise, wall-normal, and spanwise directions, respectively, with the domain length chosen to ensure sufficient wake recovery length downstream of the model ($\approx 10L_x$). The origin of the coordinate system is matched with that of the experiment and lies $y^* = 0.16$ above the bottom wall, along $z^* = 0$ and on the rear face along the streamwise direction. The computational mesh consists mostly of hexahedral cells with a region of refinement close to the model. Since the scope of this study is full-field reconstruction rather than resolving the near-wall region, no inflation layers are used to capture the viscous sublayer. Care is taken to ensure that the overall shape of the telescopic struts is captured with as few cells as possible, which naturally results in $y^+ > 100$ on all solid surfaces (the model and the tunnel floor). The discretisation yields a total of $564{,}163$ cells throughout the domain. This is a relatively coarse mesh, motivated by the need to demonstrate adjoint localisation and data sparsity tests. 

\begin{table}
    \centering
    \begin{tabular}{lccccc}
        \toprule
        & Inlet & Outlet & Symmetry, top \& side & Wall (model \& floor) \\
        $U$ & $U_{\infty}$ & $\dfrac{\partial U}{\partial n} = 0$ & $\hat{n} \cdot \vec{U} = 0$, $\dfrac{\partial U_\parallel}{\partial n} = 0$ & No slip \\[6pt]
        $p$ & $\dfrac{\partial p}{\partial n} = 0$ & $0$ & $\dfrac{\partial p}{\partial n} = 0$ & $\dfrac{\partial p}{\partial n} = 0$ \\[6pt]
        $\tilde{\nu}$ & $\tilde{\nu}_{\infty}$ & $\dfrac{\partial \tilde{\nu}}{\partial n} = 0$ & $\dfrac{\partial \tilde{\nu}}{\partial n} = 0$ & $\tilde{\nu} = 0$ \\[6pt]
        $\nu_t$ & \texttt{calculated} & \texttt{calculated} & $\dfrac{\partial \nu_t}{\partial n} = 0$ & \texttt{nutUSpaldingWallFunction} \\
        \bottomrule
    \end{tabular}
      \caption{Boundary conditions for the baseline SA model.}
    \label{tab:chap_4_bc_bsl}
\end{table}

The boundary conditions applied to velocity, pressure, $\nu_t$, and $\tilde{\nu}$ are listed in table~\ref{tab:chap_4_bc_bsl}. A velocity-inlet, pressure-outlet boundary condition is prescribed at the inlet and outlet, respectively, following the standard approach for incompressible flows. At the inlet, $\tilde{\nu}_{\infty}$ is prescribed in the range $3\nu_{\infty}$ to $5\nu_{\infty}$ as recommended by \cite{spalart1992one}, with a zero-gradient condition imposed at the outlet. The eddy viscosity $\nu_t$ at the inlet and outlet is designated as \texttt{calculated}, such that it is derived from the local value of $\tilde{\nu}$ via the relation $\nu_t = \tilde{\nu} f_{v1}$ rather than being independently prescribed. The model surface and the tunnel floor are treated as no-slip walls. Since the viscous sublayer is not resolved, the \texttt{nutUSpaldingWallFunction} is applied to $\nu_t$ at these boundaries, based on the continuous law of the wall formulation of \cite{spalding1961single}. This provides a smooth blending between the viscous sublayer and the log-law region without requiring an explicit switch, and the modified eddy viscosity is set to $\tilde{\nu} = 0$ at all wall boundaries. The spanwise mid-plane, top face, and side face of the domain are all assigned symmetry boundary conditions, consistent with the half-domain configuration described in \S \ref{section:computational_setup}, imposing zero normal velocity and zero normal gradients of all remaining flow quantities at these boundaries.

\section{Baseline model}
\label{section:baseline_model}
The baseline computations are performed using the SA turbulence model, which is equivalent to setting $\mathbf{f}_{c} = \mathbf{0}$ throughout the domain in \eqref{equation:mom_correction}. The baseline solution serves as the first optimisation iteration, providing the optimiser with a significantly better initial guess compared to random initialisation \citep{franceschini2020mean}. The momentum and turbulence transport equations are discretised using a second-order upwind scheme to ensure that the solution is free from the numerical diffusion associated with first-order schemes. The residual tolerance is set to $5~\times~10^{-7}$ for all flow variables, as a tight tolerance on the primal residuals is essential for ensuring efficient adjoint solver convergence. To allow for some flexibility in the solution process, a residual tolerance of up to $10^{-6}$ on any flow variable is accepted as a converged state. This has proven to be a reliable criterion that balances a sufficiently converged primal solution with the convergence requirements of the adjoint solver. Additionally, the preconditioner matrix used in the solution of the linear system in equation~\eqref{equation:adjoint_equation} is recomputed every $3$ optimisation iterations, ensuring faster convergence of the adjoint solver while adding negligible overhead.

\begin{figure}
    \centering
    \begin{subfigure}{\textwidth}
        \centering
        \caption{}
        \psfrag{X}[cc][cc]{Baseline}
        \psfrag{Y}[cc][cc]{Experiment}
        \psfrag{P}[cc][cc]{$y^*$}
        \psfrag{Q}[cc][cc]{$z^*$}
        \psfrag{D}[cc][cc]{$U_x^*$}
        \psfrag{E}[cc][cc]{$U_y^*$}
        \psfrag{F}[cc][cc]{$U_z^*$}
        \psfrag{a}[rc][rc]{$-0.15$}
        \psfrag{b}[rc][rc]{$0.0$}
        \psfrag{c}[rc][rc]{$0.15$}
        \psfrag{e}[tc][tc]{$-0.25$}
        \psfrag{f}[tc][tc]{$0.0$}
        \psfrag{g}[tc][tc]{$0.25$}
        \psfrag{I}[cc][cc][0.8]{$-0.28$}
        \psfrag{J}[cc][cc][0.8]{$0.48$}
        \psfrag{K}[cc][cc][0.8]{$1.23$}
        \psfrag{L}[cc][cc][0.8]{$-0.24$}
        \psfrag{M}[cc][cc][0.8]{$-0.03$}
        \psfrag{N}[cc][cc][0.8]{$0.17$}
        \psfrag{O}[cc][cc][0.8]{$-0.23$}
        \psfrag{R}[cc][cc][0.8]{$-0.06$}
        \psfrag{S}[cc][cc][0.8]{$0.11$}
        \includegraphics[width=\textwidth]{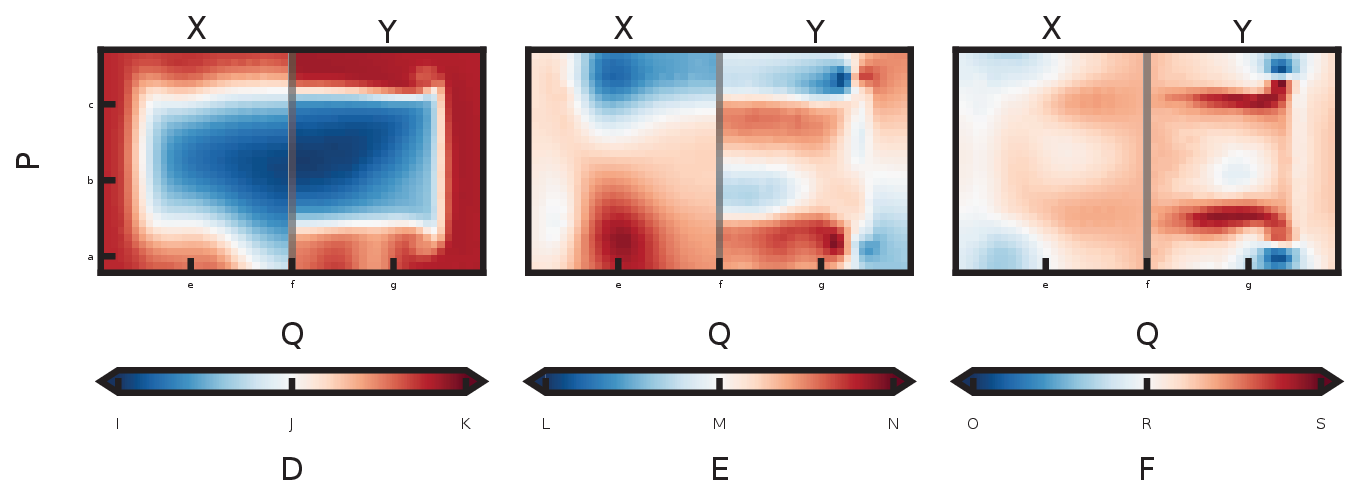}
        \label{fig:bsl_vs_exp_plane1}
    \end{subfigure}

    \vspace{-1em}

    \begin{subfigure}{\textwidth}
        \centering
        \caption{}
        \psfrag{X}[cc][cc]{Baseline}
        \psfrag{Y}[cc][cc]{Experiment}
        \psfrag{P}[cc][cc]{$y^*$}
        \psfrag{Q}[cc][cc]{$z^*$}
        \psfrag{D}[cc][cc]{$U_x^*$}
        \psfrag{E}[cc][cc]{$U_y^*$}
        \psfrag{F}[cc][cc]{$U_z^*$}
        \psfrag{a}[rc][rc]{$-0.15$}
        \psfrag{b}[rc][rc]{$0.0$}
        \psfrag{c}[rc][rc]{$0.15$}
        \psfrag{e}[tc][tc]{$-0.25$}
        \psfrag{f}[tc][tc]{$0.0$}
        \psfrag{g}[tc][tc]{$0.25$}
        \psfrag{I}[cc][cc][0.8]{$0.70$}
        \psfrag{J}[cc][cc][0.8]{$0.90$}
        \psfrag{K}[cc][cc][0.8]{$1.13$}
        \psfrag{L}[cc][cc][0.8]{$-0.13$}
        \psfrag{M}[cc][cc][0.8]{$-0.02$}
        \psfrag{N}[cc][cc][0.8]{$0.09$}
        \psfrag{O}[cc][cc][0.8]{$-0.04$}
        \psfrag{R}[cc][cc][0.8]{$0.01$}
        \psfrag{S}[cc][cc][0.8]{$0.07$}
        \includegraphics[width=\textwidth]{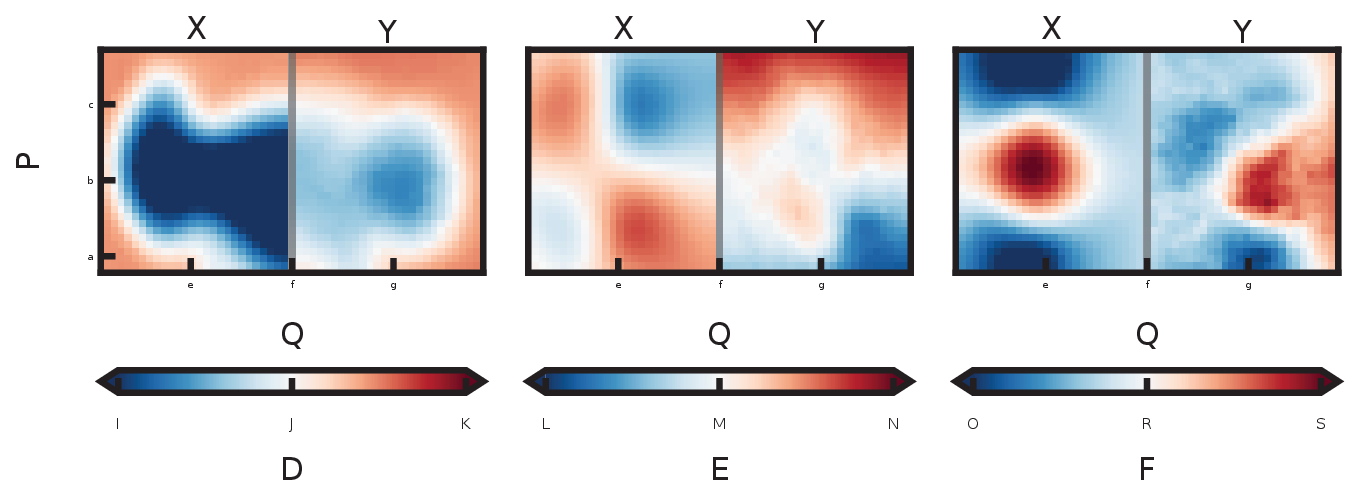}
        \label{fig:bsl_vs_exp_plane9}
    \end{subfigure}
\caption{Comparison of the mean streamwise, wall-normal, and spanwise velocity components between the baseline SA model and the experiment on (a) Plane~$2$ and (b) Plane~$10$. Results are presented as butterfly plots, where the left half shows the baseline fields reflected about the symmetry plane and interpolated onto the experimental grid using linear interpolation, and the right half shows the corresponding experimental data. The sign of the anti-symmetric spanwise component $U_z^*$ is retained when mirroring.}
    \label{fig:bsl_vs_exp}
\end{figure}

Figure~\ref{fig:bsl_vs_exp} presents butterfly plots which compare the mean streamwise, wall-normal, and spanwise velocity components from the baseline SA model with the corresponding experimental fields on Plane~$2$ (figure~\ref{fig:bsl_vs_exp_plane1}) and Plane~$10$ (figure~\ref{fig:bsl_vs_exp_plane9}). The right half of each panel shows the experimental data and the left half shows the baseline SA model field, mirrored about the spanwise symmetry plane ($z^*=0$) and interpolated onto the experimental grid using linear interpolation. The baseline is mirrored along the symmetry plane to allow direct comparison with the experimental data. The sign of the anti-symmetric spanwise component $U_z^*$ is retained when mirroring, since inverting it would render the two halves anti-symmetric and prevent direct comparison. It is emphasised that the left half does not represent independent data. It is a spatial reflection of the right-half baseline field, shown solely to facilitate comparison.

The near-wake comparison on Plane~2 (figure~\ref{fig:bsl_vs_exp_plane1}) reveals significant discrepancies between the baseline model and the experiment. In the streamwise component, the region of momentum deficit is not as clearly demarcated as in the experimental field, and the footprint of the telescopic struts is absent in the baseline. Instead, the baseline exhibits a large region of momentum deficit below the recirculation region. The wall-normal component has a complex 3D structure. The agreement with the experimental data is poor and the streamwise vortices are missing. The rightmost set of panels in figure~\ref{fig:bsl_vs_exp} shows the spanwise component. The corner vortices are missing and the baseline spatially-averages the structures due to the low mesh resolution. The experimental structures are much more concentrated in regions of the flow with high mean shear. These structures that are completely absent in the baseline, fail to capture the key aspect of the flow's three-dimensionality.

In the far wake, the comparison on Plane~10 (figure~\ref{fig:bsl_vs_exp_plane9}) shows deficiencies of a similar nature. First, consider the streamwise velocity. The baseline predicts a considerably weaker momentum recovery than that observed experimentally. Furthermore, the distinct single-lobe structure of the momentum deficit (seen as two lobes in the full field), clearly visible in the experimental data, is not reproduced by the baseline model. The wall-normal velocity is compared next. By this streamwise location, the vortices originating from the top and bottom corners of the model have grown in size. The exchange of momentum between them, however, cannot be clearly discerned from the wall-normal velocity of the baseline, pointing to a further deficiency of the prediction. Finally, consider the spanwise velocity. The baseline reproduces some essential features of the experimental field, with a positive $U_z^*$ region surrounded by a negative $U_z^*$ region. However, the predicted structure is considerably more compact and symmetric than the diffused, asymmetric structure observed in the experiment, and the magnitudes of both the positive and negative regions are larger in the baseline than in the experiment. These discrepancies are attributable to the inherent limitations of the SA model in this flow regime. In particular, an underprediction of the eddy viscosity in the wake region would result in insufficient turbulent momentum mixing, which could account for the weaker wake recovery observed in the far field.

\begin{figure}
    \centering
 \psfrag{n}[tc][tc]{$1$}
\psfrag{o}[tc][tc]{$2$}
\psfrag{p}[tc][tc]{$3$}
\psfrag{q}[tc][tc]{$4$}
\psfrag{r}[tc][tc]{$5$}
\psfrag{s}[tc][tc]{$6$}
\psfrag{t}[tc][tc]{$7$}
\psfrag{u}[tc][tc]{$8$}
\psfrag{v}[tc][tc]{$9$}
\psfrag{w}[tc][tc]{$10$}
\psfrag{x}[tc][tc]{$11$}
\psfrag{y}[tc][tc]{$12$}
\psfrag{P}[cc][cc]{$L_1$}
\psfrag{Q}[cc][cc]{Plane}
\psfrag{a}[rc][rc]{$0.08$}
\psfrag{b}[rc][rc]{$0.16$}
\psfrag{c}[rc][rc]{$0.24$}
\psfrag{A}[lc][lc]{$U_x$}
\psfrag{B}[lc][lc]{$U_y$}
\psfrag{C}[lc][lc]{$U_z$}
\psfrag{Z}[lc][lc]{Baseline}
    \includegraphics[width=\textwidth]{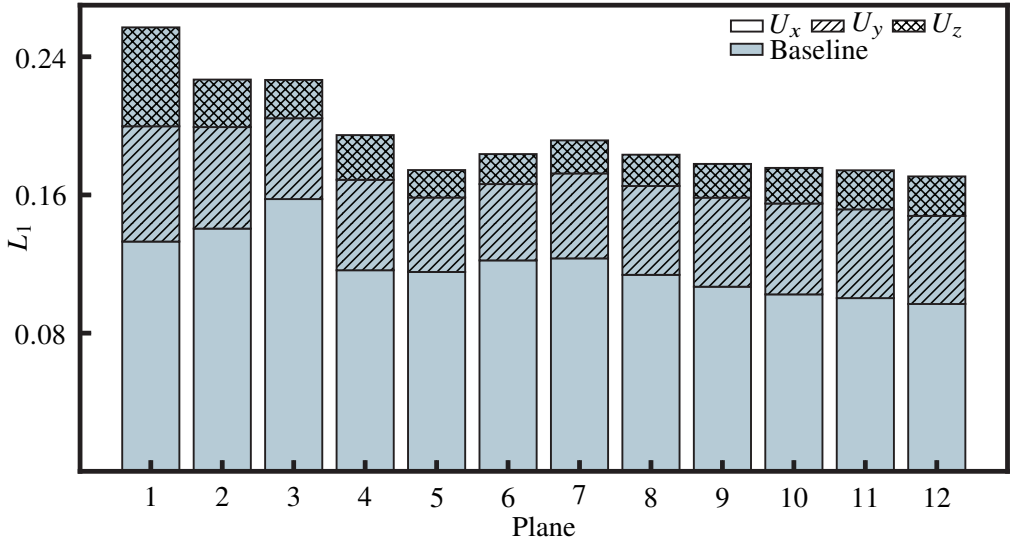}
\caption{$L_1$ norm per measurement plane, computed between the baseline SA model and the experimental data using \eqref{eqn:L1NormPaper}, evaluated separately for each of the three mean velocity components ($U_x^*$, $U_y^*$, $U_z^*$). The norm is evaluated on the downsampled experimental grid.}
\label{fig:l1_barchart}
\end{figure}

While the two planes shown in figure~\ref{fig:bsl_vs_exp} clearly demonstrate the deviation of the baseline SA model from the experiment, it is useful to quantify this deviation across all measurement planes and understand how it is distributed among the three velocity components. This deviation is quantified using an $L_1$ norm, defined as
\begin{align}
\label{eqn:L1NormPaper}
    L_1 = \frac{1}{{N_p U_\infty}}\sum_{p=1}^{N_p} |\mathcal{Q}(U_i^p) - \hat{U}_i^p|,
\end{align}
where $\hat{U}_i^p$ refers to the $i^{th}$ component of the mean velocity of the reference data, and the operator $\mathcal{Q}(\cdot)$ is a linear interpolator used to sample the computational data on the experimental grid. The $L_1$ norm is computed per plane and, on each plane, averaged over the number of data points $N_p$. This norm is selected for comparison due to its resistance to outliers, making it a better indicator of overall error reduction in the domain. Figure~\ref{fig:l1_barchart} shows the $L_1$ norm decomposed into the three velocity components across all $12$ measurement planes. As seen in figure~\ref{fig:raw_uvw_all}, the norm is significantly higher for $U_x^*$ in comparison to both $U_y^*$ and $U_z^*$, which is expected since $U_x^*$ is the dominant velocity component, with a maximum value $\mathcal{O}(10)$ times greater than that of $U_y^*$ and $U_z^*$. Among the remaining components, $U_y^*$ shows the next highest overall magnitude across all planes, followed by $U_z^*$. The maximum $L_1$ norm for $U_x^*$ occurs on Plane~$3$, which lies in the near wake close to the edge of the primary recirculation bubble ($x^* \approx 0.4$), whereas both $U_y^*$ and $U_z^*$ attain their maximum errors on Plane~$1$, immediately behind the rear face of the body.

The $L_1$ norm is now discussed region by region. In the near wake (Plane~$1$--$4$), the errors in $U_x^*$ are considerably higher and differ substantially from plane to plane within this region, reflecting the complexity of the primary recirculation bubble and the inability of the baseline SA model to capture its 3D structure. The wall-normal component, by contrast, shows comparable error magnitudes across all twelve planes, suggesting that the discrepancy in $U_y^*$ is distributed relatively uniformly throughout the wake rather than being concentrated in any particular region. The large error in $U_z^*$ immediately behind the rear face of the body is notable and is likely attributable to the strong flow gradients in this region owing to abrupt separation. In the intermediate region (Plane~$5$--$8$), where the corner vortices grow by diffusing in the spanwise-wall-normal plane and the wake recovers streamwise momentum, the $L_1$ norms are lower and weakly dependent on $x$, suggesting that the baseline model performs comparatively better once the flow is no longer dominated by the primary bubble. In the far wake (Plane~$8$--$12$), the errors are broadly consistent across planes. This is not unexpected. As noted in \S\ref{subsection:flow_physics}, the far wake corresponds to the natural state of the flow for the given geometry and flow conditions, and it is reasonable to expect that the baseline model also settles into a persistent but incorrect natural state in this region. Such complex discrepancies motivate the use of DA to correct the predictions of the baseline model, as will be discussed in the following section.
    
\section{Adjoint localisation}
\label{section:adjoint_localisation}
Before discussing the assimilation results, it must be noted that the DA studies presented in the literature consider the full computational mesh as the control variable space, despite the objective function being evaluated over a much smaller FOV owing to the sparse nature of the experimental data. While this affords maximum flexibility in the corrections that can be applied, it places a significant burden on the peak memory consumption of the adjoint solve. As described in \S~\ref{subsection:discrete_adjoint_method}, the discrete adjoint requires the explicit assembly and storage of the preconditioner of the state Jacobian $\partial\mathbf{R}/\partial\mathbf{w}^T$ of size $n_w \times n_w$, where $n_w$ is the number of state variables and scales with the mesh size. These are the dominant memory costs and grow with the mesh independently of the choice of control variables. In addition, the reverse-mode AD sweep that computes the matrix-vector product $\partial\mathbf{R}/\partial\mathbf{f}_c^T\boldsymbol{\psi}$ requires registering the control variables as inputs on the AD tape (a tape stores intermediate variables in reverse-mode AD). In the global case, this involves $3N_\Omega$ scalar inputs corresponding to the three components of $\mathbf{f}_c$ at every cell in the computational domain $\Omega$. This is partly circumvented by the introduction of adjoint localisation, which is developed and implemented within the existing DAFoam framework. Such an implementation is made possible by the modular architecture of DAFoam, as discussed in \S~\ref{subsection:discrete_adjoint_method}.

\begin{figure}
    \centering
    \psfrag{a}{Full grid}
    \psfrag{b}{Localised grid}
    \psfrag{x}{$x$}
    \psfrag{y}{$y$}
    \psfrag{z}{$z$}
    \psfrag{e}[cc][cc]{$(x_{min}, y_{min}, z_{min})$}
    \psfrag{f}[cc][cc]{$(x_{max}, y_{max}, z_{max})$}
    \psfrag{d}[cc][cc]{$U_\infty$}
    \includegraphics[width=0.7\textwidth]{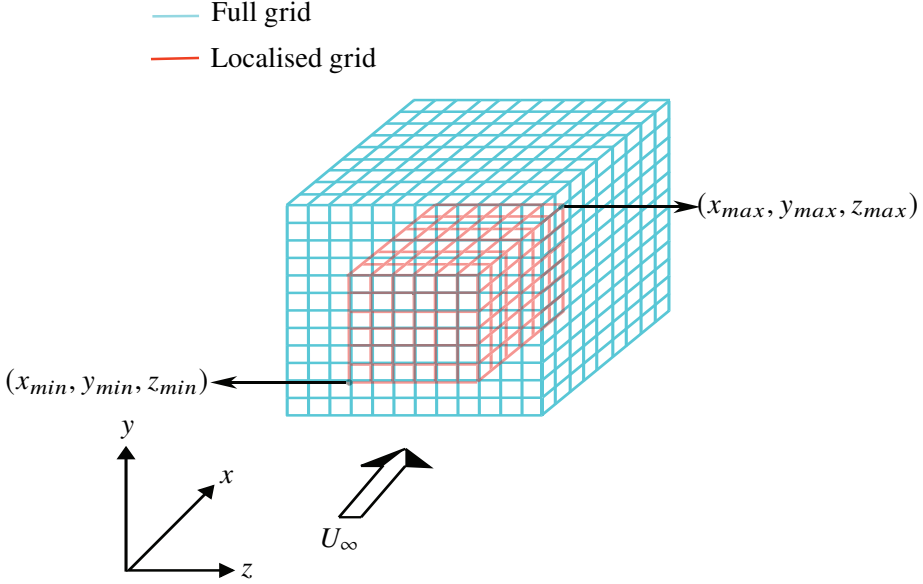}
\caption{Schematic of the adjoint localisation procedure. The full computational grid $\Omega$ and the localised subdomain $\Omega_p$, defined by a rectangular bounding box with corners $(x_{\min}, y_{\min}, z_{\min})$ and $(x_{\max}, y_{\max}, z_{\max})$, are shown. The momentum forcing $\mathbf{f}_c$ is restricted to cells within $\Omega_p$.}
\label{fig:localise}
\end{figure}

The objective of adjoint localisation is to restrict the control variables to a subdomain $\Omega_p \subset \Omega$, prescribed \textit{a priori} by the user, rather than optimising over the full computational domain $\Omega$. The control variable vector $\mathbf{x}$ is thus defined as:
\begin{align}
    \mathbf{x} = \{\mathbf{f}_c(\mathbf{r}) : \mathbf{r} \in \Omega_p\}, \quad \Omega_p \subset \Omega,
\end{align}
where $\mathbf{r}$ is the position vector. In the conventional formulation, $\Omega_p \equiv \Omega$, recovering the standard (hereafter referred to as the global) case. A simplified schematic of adjoint localisation is shown in figure~\ref{fig:localise}. For a regular full grid $\Omega$, consisting of hexahedral cells, the localised grid $\Omega_p$ is a subset of the full grid. In a general case, neither the localised nor the full grid need be rectangular or consist of purely hexahedral cells. In the current implementation, $\Omega_p$ is specified by the corners of a rectangular bounding box with coordinates $(x_{\min}, y_{\min}, z_{\min})$ and $(x_{\max}, y_{\max}, z_{\max})$. A crucial point worth noting is that $\Omega_p$ is static throughout the assimilation process. The bounds are chosen at the start and remain fixed throughout the entire process.

The localisation procedure operates as follows. Prior to the start of the optimisation, all mesh cells whose centres lie within $\Omega_p$ are identified and their indices are stored. At each subsequent optimisation iteration, the matrix-vector product $\partial\mathbf{R}/\partial\mathbf{f}_c^T\boldsymbol{\psi}$ and the objective gradient $\partial f/\partial\mathbf{f}_c$ are computed using reverse-mode AD. In this approach, the control variables $\mathbf{f}_c$ are designated as inputs to the AD computation. The residual evaluation is executed while recording the operations on a computational tape, and the chain rule is subsequently applied in reverse to propagate sensitivity information back to those inputs. The memory cost associated with the control variables scales with the number of registered inputs. In the global case, all $3N_\Omega$ components of $\mathbf{f}_c$ are registered, whereas with localisation only the $3N_p$ components associated with cells in $\Omega_p$ are registered, where $N_p = |\Omega_p|$. The reverse sweep therefore operates over a smaller input space, reducing the seed and gradient storage associated with the control variables. The optimiser consequently updates $\mathbf{f}_c$ only within $\Omega_p$, and the forcing field remains zero throughout the remainder of $\Omega$.

It is important to note what localisation does and does not affect. In the Jacobian-free approach adopted here, the state Jacobian $\partial\mathbf{R}/\partial\mathbf{w}^T$ is never explicitly assembled. Its transpose matrix-vector products are computed on the fly by reverse-mode AD \citep{kenway2019effective}. Instead, the dominant stored quantities are the incomplete lower upper (ILU) preconditioner $[\partial\mathbf{R}/\partial\mathbf{w}^T]_{PC}$, the generalised minimal residual method (GMRES) Krylov basis, and the reverse-mode AD tape of the full-domain residual evaluation. These are all governed by the state vector $\mathbf{w}$ and the full computational mesh, so they are unaffected by the choice of $\Omega_p$. They therefore constitute an irreducible memory floor that localisation cannot reduce. The memory saving achieved by localisation arises from the reduced registration of control variable inputs, and is consequently bounded. As $N_p \to 0$, the total memory asymptotes to the floor set by the state-space machinery, and no choice of localisation region, however small, can reduce it below this value. The memory reduction achieved in practice and a quantitative comparison between the localised and global reconstructions are presented in \S\ref{subsection:convergence_selection}, where it is shown that restricting the control variable space to $\Omega_p$ does not degrade the quality of the assimilated fields.

\subsection{Choice of localisation region}
\label{subsection:localisation_cases}
The prescription of $\Omega_p$ can be approached in several ways. In a manual approach, the user specifies the region \textit{a priori} based on knowledge of the flow physics, targeting the subdomain most likely to benefit the optimisation while preserving reconstruction quality relative to the global case. An automatic approach instead selects $\Omega_p$ on the basis of flow features deemed critical for accurate reconstruction such as vorticity, strain rate, or eddy viscosity magnitude, without requiring prior knowledge of the computational setup. However, such feature-based selection requires an implementation capable of handling non-box geometries. A further extension is a dynamic approach, in which $\Omega_p$ varies between iterations and adopts a greedy strategy to progressively refine the active region. In the present work, the manual approach is used, with $\Omega_p$ chosen on the basis of physical arguments.

\begin{figure*}
    \centering
    \psfrag{x}[cc][cc]{$x$}
    \psfrag{y}[cc][cc]{$y$}
    \psfrag{z}[cc][cc]{$z$}
    \psfrag{l}[cc][cc]{$U_{\infty}$}

    \begin{subfigure}{0.48\textwidth}
            \caption{}
        \centering
        \includegraphics[width=\textwidth]{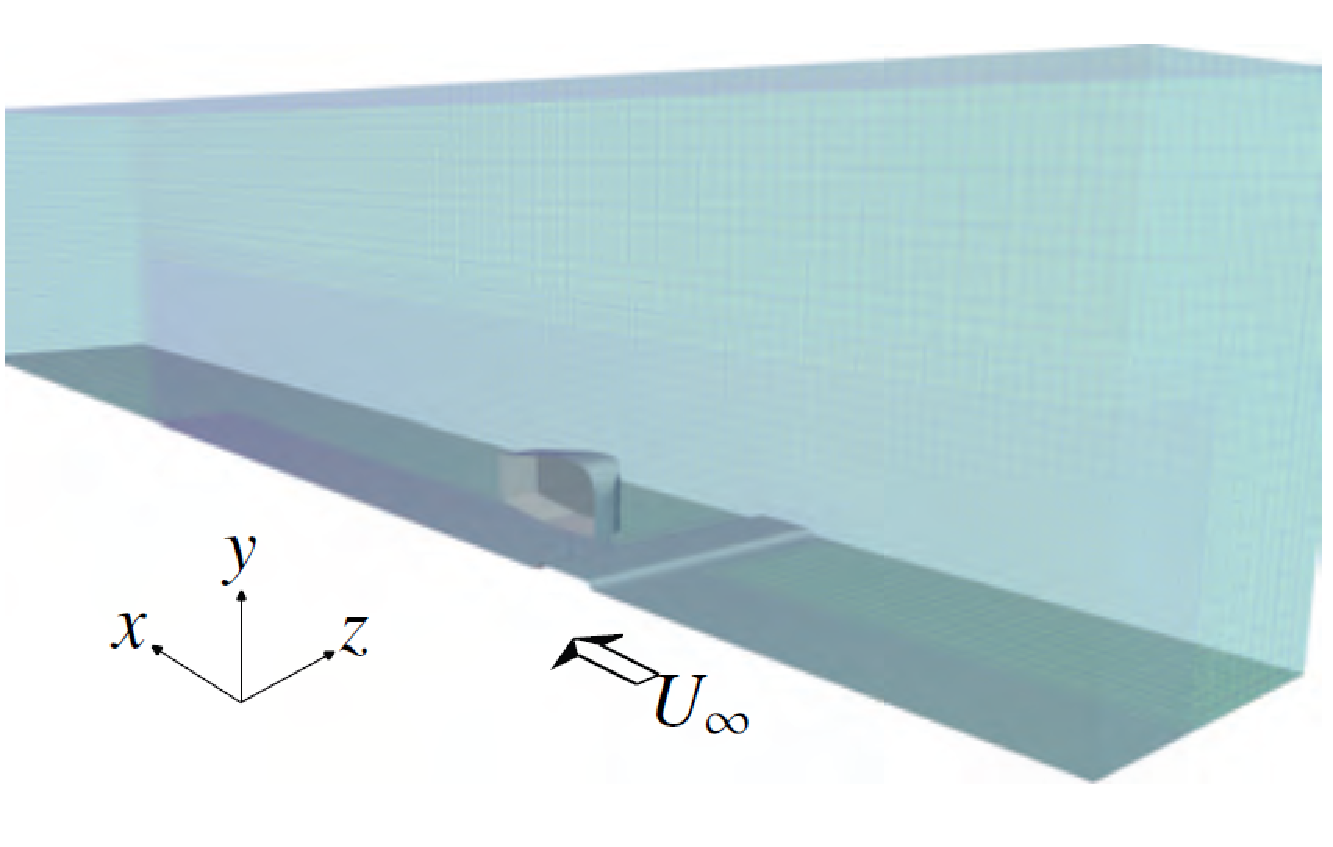}
        \label{fig:global}
    \end{subfigure}
    \hfill
    \begin{subfigure}{0.48\textwidth}
            \caption{}
        \centering
        \includegraphics[width=\textwidth]{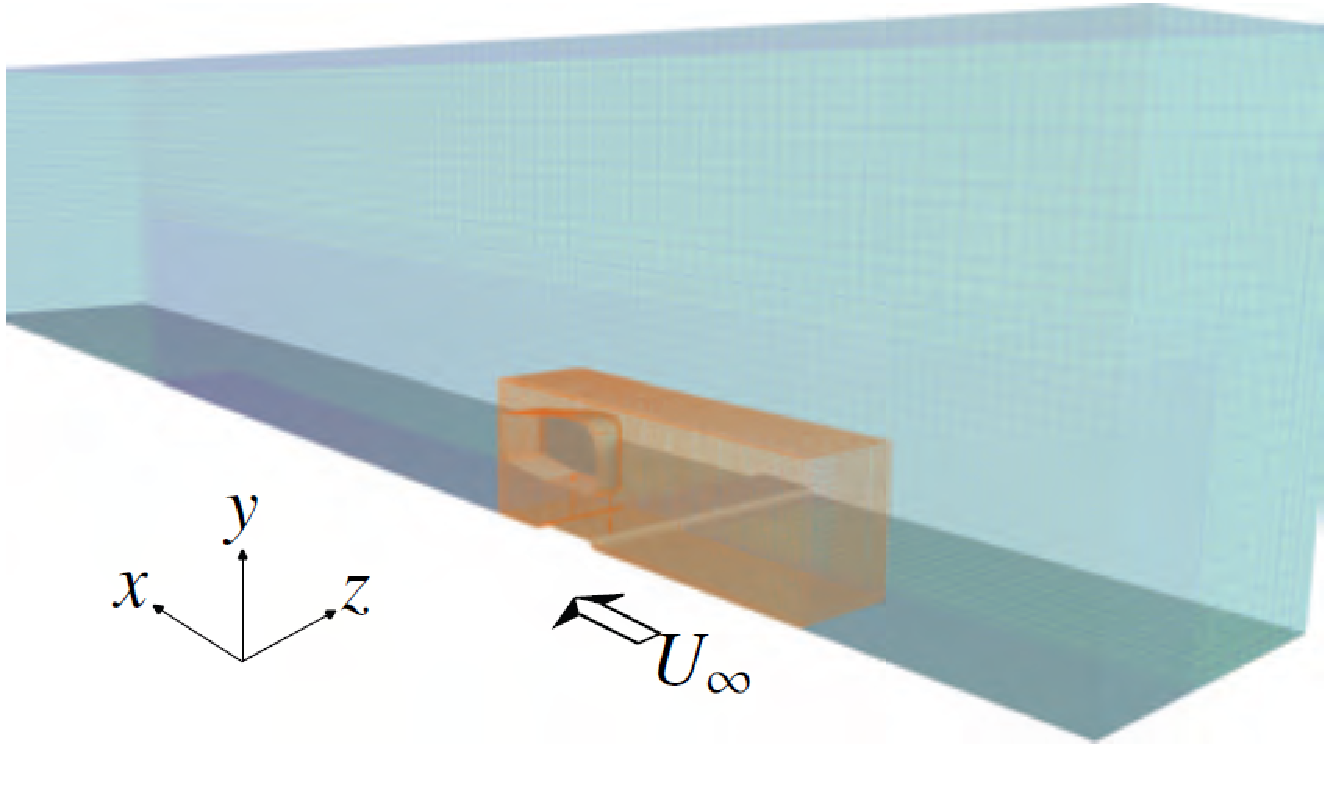}
        \label{fig:local_upstream}
    \end{subfigure}

    \vspace{0.5em}

    \begin{subfigure}{0.48\textwidth}
            \caption{}
        \centering
        \includegraphics[width=\textwidth]{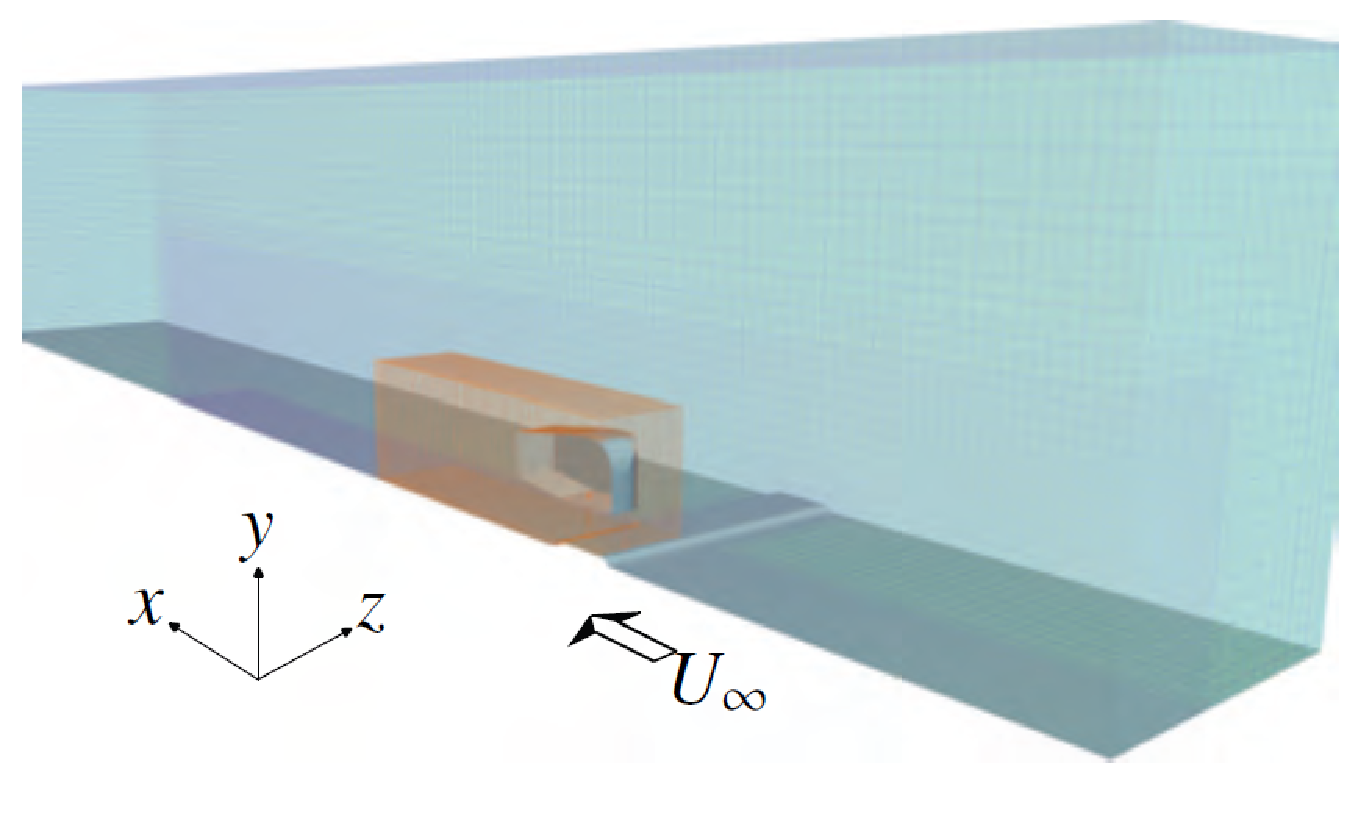}
        \label{fig:local}
    \end{subfigure}
    \hfill
    \begin{subfigure}{0.48\textwidth}
            \caption{}
        \centering
        \includegraphics[width=\textwidth]{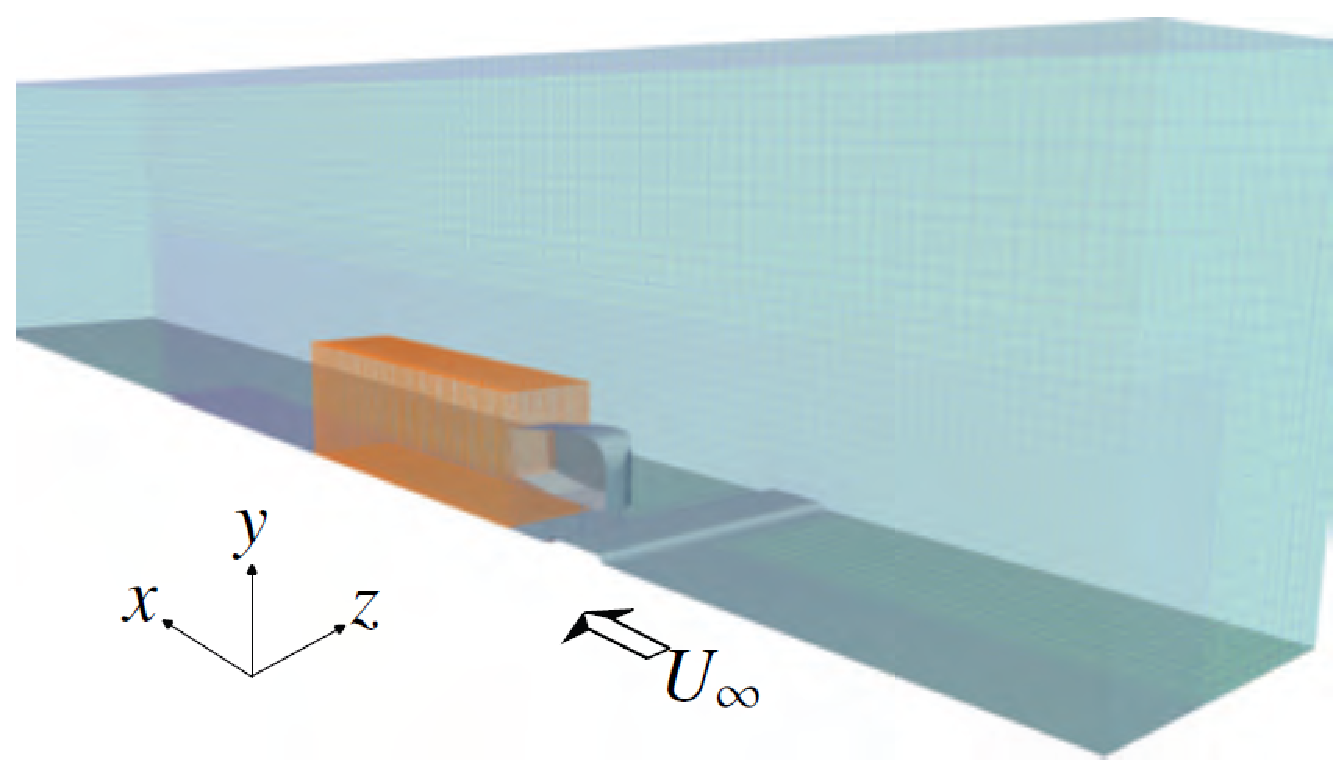}
        \label{fig:local_downstream}
    \end{subfigure}
\caption{Schematic of the localisation cases considered in this chapter, shown as wireframe representations of the computational mesh. The full grid $\Omega$ (\protect\cyanline) and the localised grid $\Omega_p$ (\protect\redorangeline) are shown in each panel. \textit{(a)} Global case, where $\Omega_p \equiv \Omega$. \textit{(b)} Upstream localised case, where $\Omega_p$ covers the upstream portion of the domain. \textit{(c)} Body-wake localised case, where $\Omega_p$ is confined to contain the model and its wake. \textit{(d)} Downstream localised case, where $\Omega_p$ covers the downstream portion of the domain.}
        \label{fig:local_grids_schematic}
\end{figure*}

The wireframe meshes of the three cases, clearly delineating the localised grid from the full grid, are shown in figure~\ref{fig:local_grids_schematic}, and the bounding box coordinates with the corresponding localisation ratios are given in table~\ref{tab:chap_4_localisation}. The cases are motivated by the physics of the problem and are designed such that $N_{\Omega_p}$ is approximately equal across all three, ensuring a consistent localisation ratio $N_{\Omega_p}/N_\Omega \approx 0.13$ for a fair comparison. The global case (figure~\ref{fig:global}) uses the full computational domain without any localisation, i.e.\ $\Omega_p \equiv \Omega$. In the upstream local case (figure~\ref{fig:local_upstream}), $\Omega_p$ encompasses the region upstream of and around the model, but does not extend into the measurement region downstream of the rear face. The motivation is to investigate whether corrections applied upstream and around the body can propagate convectively into the region where the experimental reference data are located. The body-wake case (figure~\ref{fig:local}) selects $\Omega_p$ to partly encompass the model and to extend into the measurement region, with $x^*_{\max} = 1.832$, corresponding to $0.56L_x$ beyond Plane~$12$ (see table~\ref{tab:chap_4_planes}), providing a balance between optimising the control variables upstream of the data planes and directly at the measurement locations. In the downstream case (figure~\ref{fig:local_downstream}), $\Omega_p$ is confined entirely to the region downstream of the model, excluding the body itself from the control variable space.

\begin{table}
    \centering
    \setlength{\tabcolsep}{6pt}
    \begin{tabular}{lcccccccc}
        & \multicolumn{2}{c}{$x^*$} & \multicolumn{2}{c}{$y^*$} & \multicolumn{2}{c}{$z^*$} & & \\
        \cmidrule[0.4pt](l{2pt}r{2pt}){2-3}
        \cmidrule[0.4pt](l{2pt}r{2pt}){4-5}
        \cmidrule[0.4pt](l{2pt}r{2pt}){6-7}
        Case & Min & Max & Min & Max & Min & Max & Cells & $N_{\Omega_p}/N_{\Omega}$ \\
        \midrule[0.4pt]
        Global     & ---      & ---     & ---      & ---     & --- & ---     & 564163 & 1.0 \\
        Upstream   & $-2.748$ & $0$     & $-0.491$ & $0.401$ & $0$ & $0.687$ & 70970  & 0.126 \\
        Body-wake  & $-0.916$ & $1.832$ & $-0.491$ & $0.401$ & $0$ & $0.687$ & 70307  & 0.125 \\
        Downstream & $0$      & $2.748$ & $-0.491$ & $0.401$ & $0$ & $0.687$ & 73728  & 0.131 \\
    \end{tabular}
    \caption{Comparison of localisation cases showing bounding box
             extents and cell count reduction.}
    \label{tab:chap_4_localisation}
\end{table}

\subsection{Comparison with the global case}
\label{subsection:convergence_selection}
Variational DA is performed with the SA model as the baseline and the momentum forcing $\mathbf{f}_c$ as the control variable. The velocity objective function is augmented with the gradient-based regularisation of the forcing field (equation~\ref{equation:combined_obj_func}), as described in \S\ref{subsection:variational_da}. The regularisation weight is set to $\lambda = 5 \times 10^{-5}$, determined through the L-curve analysis presented in Appendix~\ref{appA}, and is adopted for all assimilation cases in this chapter. Unless stated otherwise, all three downsampled, mean velocity components measured on the $12$ planes serve as reference data. The objective function computation is unaffected by localisation and is evaluated only at those points in the wake where experimental data are available. The three localised configurations are now compared with the global case in terms of the convergence of the optimisation, the plane-by-plane misfit, the recovered forcing fields, and the peak memory consumption, to determine the configuration adopted for the remainder of this chapter.

\begin{figure}[tbp]
    \centering
    \psfrag{a}[cc][cc]{$100$}
    \psfrag{b}[cc][cc]{$200$}
    \psfrag{c}[cc][cc]{$300$}
    \psfrag{d}[cc][cc]{$400$}
    \psfrag{e}[cc][cc]{$500$}
    \psfrag{f}[rc][rc]{$10^{-4}$}
    \psfrag{g}[rc][rc]{$10^{-3}$}
    \psfrag{h}[rc][rc]{$10^{-2}$}
    \psfrag{i}[rc][rc]{$10^{-1}$}
    \psfrag{j}[rc][rc]{$10^{0}$}
    \psfrag{Q}[cc][cc]{$N$}
    \psfrag{P}[cc][cc]{$J^u/J_0, J^f/J_0$}
    \psfrag{A}{Global}
    \psfrag{C}{Upstream}
    \psfrag{E}{Body-wake}
    \psfrag{G}{Downstream}
    \includegraphics[width=\textwidth]{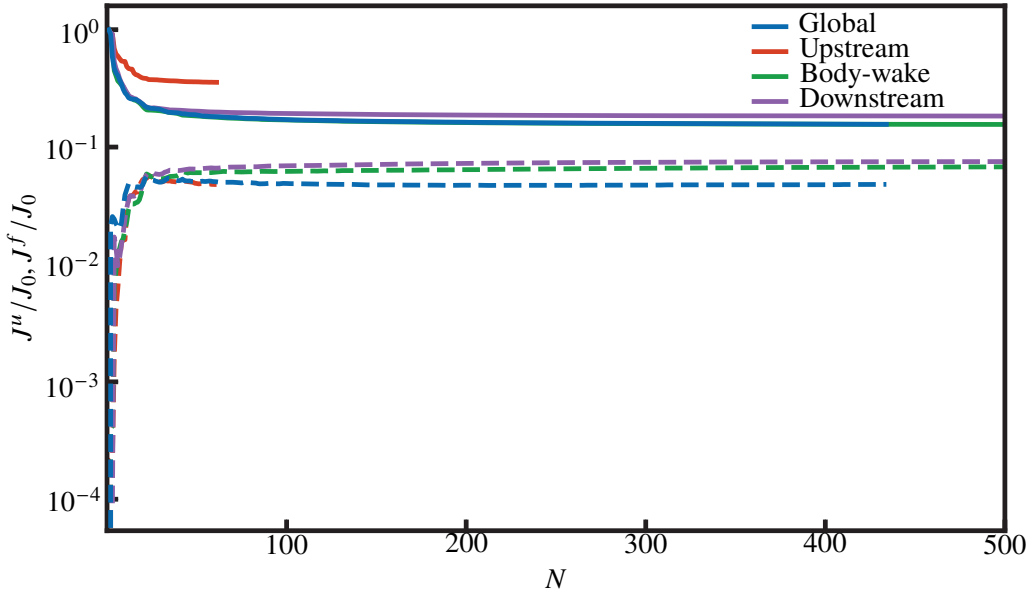}
    \caption{Convergence of the normalised objective function $J^u$ (solid) and regularisation term $J^f$ (dashed) for the global and three localisation cases. Values are normalised by the first recorded iterate $J_0$ of each case.}
    \label{fig:chap_4_convergence}
\end{figure}

The convergence of the optimisation for the four cases is presented in figure~\ref{fig:chap_4_convergence}, which shows the objective function $J^u$ and the regularisation term $J^f$, each normalised by the first recorded iterate $J_0$ of the respective case. The global case, in which $\Omega_p \equiv \Omega$, serves as the reference against which the localised cases are assessed. The upstream case converges to a substantially higher value of $J^u/J_0$ relative to the other cases and also terminates early. This indicates that restricting the control variable to a region that does not overlap with the reference data is ineffective and the optimised forcing field is unable to make modifications downstream of the body that reduce the misfit at the measurement planes. This is physically expected given the convective nature of the flow, and suggests that the upstream placement of $\Omega_p$ is not well-suited to this class of flow problem.

The body-wake case follows the global one almost exactly throughout the optimisation. The close agreement between the two, despite an approximately eightfold difference in the size of the control variable space, suggests that the two cases share a similar optimisation landscape and that the additional degrees of freedom in the global case do not improve the outcome. The convergence of $J^f/J_0$ further supports the similarity between the body-wake and global cases. The downstream case converges to a marginally higher value of $J^u/J_0$ than both the global and body-wake cases, while exhibiting a smoother forcing field than the global one, as indicated by its lower $J^f/J_0$. This suggests that restricting the control variable purely to the wake region precludes any corrective effect on the flow developing over the top and rear faces of the model, which contributes to the misfit at the measurement planes. The analysis of the convergence trends of the four cases indicates that the control variable must span both the model surface and the near-wake region to achieve the best reconstruction outcome. While the body-wake case appears to mirror the global reconstruction closely, the converged objective function is an integrated measure that carries no spatial information. A plane-by-plane comparison of the $L_1$ norm is therefore required before any conclusion is drawn.

\begin{figure}[tbp]
    \centering
    \psfrag{a}[rc][rc]{$0.08$}
    \psfrag{b}[rc][rc]{$0.16$}
    \psfrag{c}[rc][rc]{$0.24$}
    \psfrag{n}{$1$}
    \psfrag{o}{$2$}
    \psfrag{p}{$3$}
    \psfrag{q}{$4$}
    \psfrag{r}{$5$}
    \psfrag{s}{$6$}
    \psfrag{t}{$7$}
    \psfrag{u}{$8$}
    \psfrag{v}{$9$}
    \psfrag{w}{$10$}
    \psfrag{x}{$11$}
    \psfrag{y}{$12$}
    \psfrag{j}{$10^0$}
    \psfrag{i}{$10^{-1}$}
    \psfrag{h}{$10{-2}$}
    \psfrag{g}{$10^{-3}$}
    \psfrag{f}{$10^{-4}$}
    \psfrag{Q}[cc][cc]{Plane}
    \psfrag{P}[rc][rc]{$L_1$}
    \psfrag{A}{$U_x$}
    \psfrag{B}{$U_y$}
    \psfrag{C}{$U_z$}
    \psfrag{Z}{Baseline}
    \psfrag{X}{Global}
    \psfrag{W}{Upstream}
    \psfrag{V}{Body-wake}
    \psfrag{U}{Downstream}
    \includegraphics[width=\textwidth]{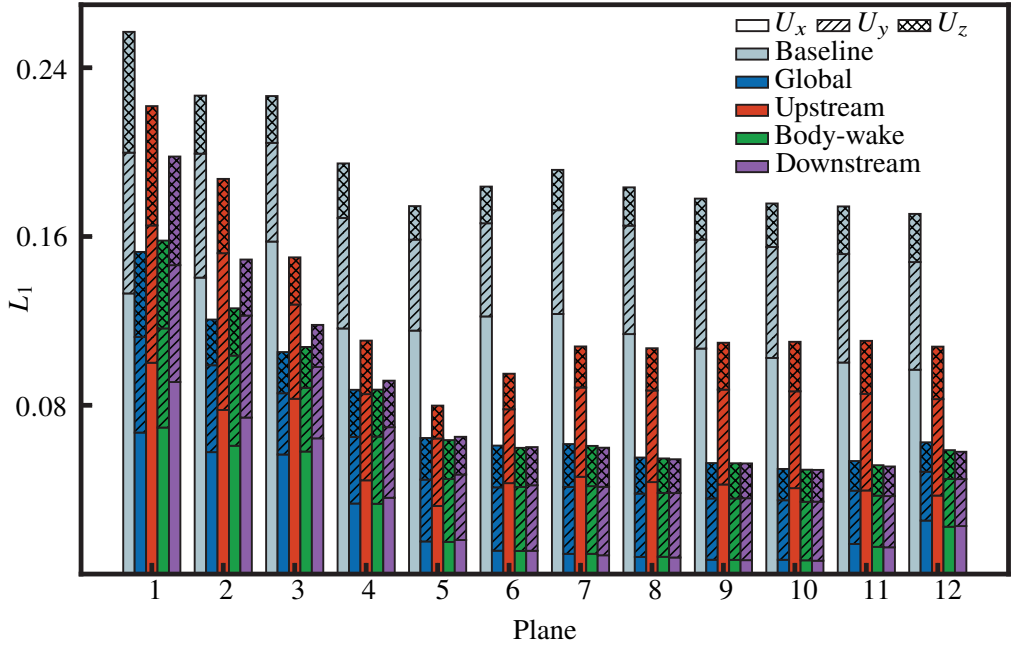}
    \caption{$L_1$ norm of the velocity misfit between the assimilated fields and the experimental reference data, decomposed by velocity component and shown for each measurement plane. Results are shown for the baseline and the four cases: global, upstream localised, body-wake localised, and downstream localised.}
    \label{fig:chap_4_l1_localisation}
\end{figure}

Figure~\ref{fig:chap_4_l1_localisation} presents the $L_1$ norm decomposed into the three velocity components along each measurement plane for the assimilated global and three localised cases, with the baseline $L_1$ norm included for reference. All assimilated cases show a reduction in the $L_1$ norm relative to the baseline across every plane, which confirms that the optimisation is effective regardless of the choice of $\Omega_p$. The global case again serves as the reference for the comparison. The upstream case shows consistently higher misfit than the global case across all planes, with a decreasing trend up to Plane~$5$ followed by a sudden increase and a plateau. The body-wake case shows near-perfect agreement with the global case across all planes, confirming that this configuration captures the essential physics of the optimisation landscape despite the significantly reduced control variable space. The downstream case shows $L_1$ norms comparable to the global case beyond Plane~$5$, but elevated values upstream of it. It can be concluded from the downstream case that the misfit within the primary recirculation bubble requires control variables to be active along the model surface and that optimising purely within the wake is insufficient. 

\begin{figure}[tbp]
    \centering
    \psfrag{P}[cc][cc]{$y^*$}
    \psfrag{Q}[cc][cc]{$x^*$}
    \psfrag{A}[lc][lc]{Global}
    \psfrag{B}[lc][lc]{Upstream}
    \psfrag{C}[lc][lc]{Body-wake}
    \psfrag{D}[lc][lc]{Downstream}
    \psfrag{a}[rc][rc]{$-0.25$}
    \psfrag{b}[rc][rc]{$0$}
    \psfrag{c}[rc][rc]{$0.25$}
    \psfrag{d}[rc][rc]{$0.5$}
    \psfrag{e}[cc][cc]{$-0.8$}
    \psfrag{f}[cc][cc]{$0$}
    \psfrag{g}[cc][cc]{$0.8$}
    \psfrag{h}[cc][cc]{$1.6$}
    \psfrag{I}[cc][cc]{$-0.6$}
    \psfrag{J}[cc][cc]{$0.0$}
    \psfrag{K}[cc][cc]{$0.6$}
    \psfrag{R}[cc][cc]{$f_{c,x}L_x/U_\infty^2$}
    \includegraphics[width=\textwidth]{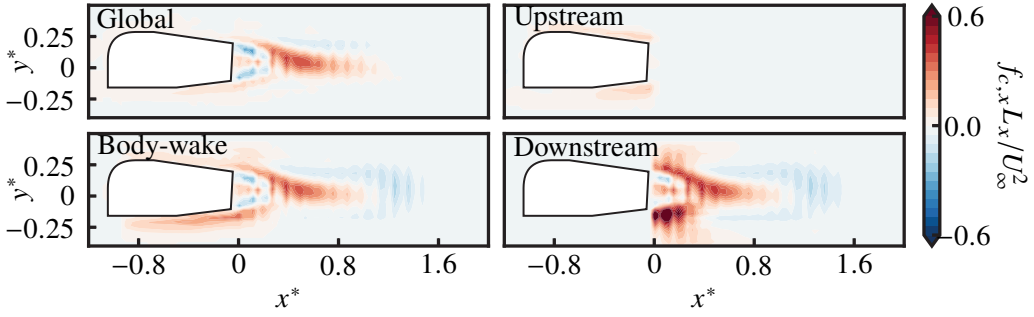}
    \caption{Streamwise component of the corrective forcing $f_{c,x}$ on the $x$--$y$ symmetry plane for the global and three localised cases. The fields are interpolated onto a common regular grid, and a single colour scale, symmetric about zero, is shared across all four panels. The localised cases recover a non-zero forcing only within their respective subdomains $\Omega_p$ (table~\ref{tab:chap_4_localisation}).}
    \label{fig:chap_4_xy_fvx_localisation}
\end{figure}

The $L_1$ norm, however, is an aggregate measure. It confirms that the body-wake case matches the global reconstruction in terms of the integrated velocity misfit, but carries no information about the spatial structure of the correction that produced it. To establish whether the body-wake case recovers the same physical correction as the global one, the forcing fields must be examined directly. The forcing fields are examined on the $x$--$y$ symmetry plane, where the assimilated data are linearly interpolated onto a structured grid of comparable resolution to the CFD mesh to facilitate visualisation. It is noted that the solution to an inverse problem is not generally unique, but the gradient-based regularisation applied here promotes smoothness of the forcing field and reduces the degeneracy of the solution, making the comparison between cases meaningful. Figure~\ref{fig:chap_4_xy_fvx_localisation} presents the streamwise component $f_{c,x}L_x/U_\infty^2$ for the global and three localised cases. The effect of the regularisation can be appreciated by examining figure~\ref{fig:forcing_reg_vs_unreg_global} in Appendix~\ref{appA} where a comparison between the regularised and plain cases is offered. The smoothness of the forcing field demonstrates that the regularisation is working well. 

In the global case, the forcing exhibits regions of momentum deficit in the immediate vicinity of the body and a compensating momentum increase further into the wake, collectively acting to reshape the recirculation bubble to better match the experimental data. A small but non-zero forcing is also present along the top and bottom faces of the model. The upstream case produces forcing exclusively in the upstream and surface regions, with no correction in the wake, consistent with the deliberate restriction of $\Omega_p$ to the region upstream of and around the body. In the body-wake case, the forcing in the wake closely matches that of the global case, and the correction along the top and bottom faces of the model is marginally stronger, providing additional streamwise acceleration in these regions. The downstream case confines the forcing entirely to the wake, and the magnitude of the correction is visibly higher immediately downstream of the rear face. This is consistent with the optimiser compensating for the absence of a gradual momentum build-up along the model surfaces, which the active surface correction provides in the global and body-wake cases. Taken together, the forcing fields indicate that the body-wake case recovers a correction that closely matches the global one, with minor differences attributable to the finite extent of $\Omega_p$, and the control variable is therefore well-matched between the two configurations.

\begin{figure}[tbp]
    \centering
    \psfrag{P}[cc][cc]{$y^*$}
    \psfrag{Q}[cc][cc]{$x^*$}
    \psfrag{A}[lc][lc]{Global}
    \psfrag{B}[lc][lc]{Upstream}
    \psfrag{C}[lc][lc]{Body-wake}
    \psfrag{D}[lc][lc]{Downstream}
    \psfrag{a}[rc][rc]{$-0.25$}
    \psfrag{b}[rc][rc]{$0$}
    \psfrag{c}[rc][rc]{$0.25$}
    \psfrag{d}[rc][rc]{$0.50$}
    \psfrag{e}[cc][cc]{$-0.8$}
    \psfrag{f}[cc][cc]{$0$}
    \psfrag{g}[cc][cc]{$0.8$}
    \psfrag{h}[cc][cc]{$1.6$}
    \psfrag{I}[cc][cc]{$-1.5$}
    \psfrag{J}[cc][cc]{$0.0$}
    \psfrag{K}[cc][cc]{$1.5$}
    \psfrag{R}[cc][cc]{$U_x^*$}
    \includegraphics[width=\textwidth]{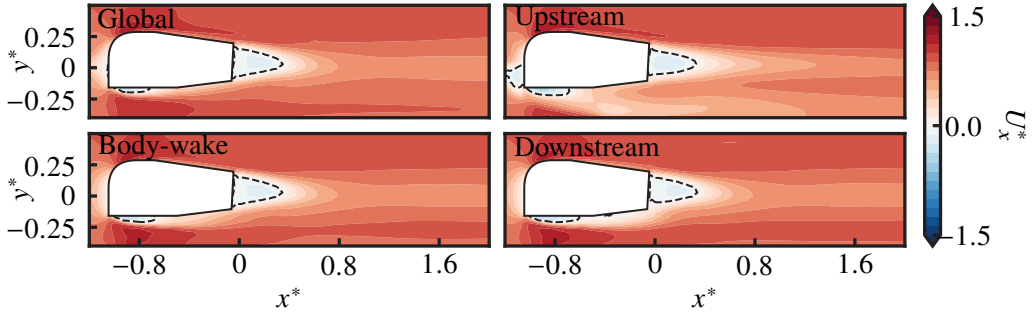}
    \caption{Mean streamwise velocity $U_x/U_\infty$ on the $x$--$y$ symmetry plane for the global and three localised cases. The dashed line denotes the $U_x = 0$ isoline of the respective case, marking the boundary of the recirculation bubble. The body outline is shown in white.}
    \label{fig:chap_4_xy_ux_localisation}
\end{figure}

The effect of the forcing on the reconstructed mean flow is now examined through the streamwise velocity field on the $x$--$y$ symmetry plane. Figure~\ref{fig:chap_4_xy_ux_localisation} presents the mean streamwise velocity $U^*$ for the four cases, with the $U^*=0$ contour shown as a dashed black line to demarcate the regions of recirculating flow. The global case exhibits two such regions. The first is a small separation bubble on the bottom face of the model, immediately downstream of the leading edge, where the abrupt change in geometry causes the boundary layer to separate and reattach. No separation is observed along the top and bottom slant surfaces. The second, and primary, recirculation region is the large wake bubble downstream of the rear face. The upstream case reconstructs both bubbles faithfully, but introduces a small region of spurious recirculating fluid near the stagnation point of the model. This is not in agreement with the global case at the location, and is likely a consequence of the control variable acting in the upstream region while the reference data are located further downstream, creating an underconstrained correction in the vicinity of the stagnation point. The body-wake case shows close agreement with the global case for both recirculation regions, with no unphysical features. The downstream case reproduces the wake bubble well, but exhibits a slightly elongated separation region along the bottom face at the leading-edge separation point. Despite the absence of control variables in this region, no unphysical flow features are introduced. 

One of the principal benefits of adjoint localisation is the reduction in peak memory consumption achieved by registering fewer control variable inputs on the AD tape. The memory saving factor $M_{\text{local}}/M_{\text{global}}$ and the localisation ratio $N_{\Omega_p}/N_\Omega$ for each case are presented in table~\ref{tab:chap_4_memory}. The peak memory reported here is the whole-job peak resident set size as recorded by the scheduler, which encompasses all memory allocated during the adjoint solve. The dominant contributors to this footprint are the explicitly stored ILU preconditioner $[\partial\mathbf{R}/\partial\mathbf{w}^T]_{PC}$, the GMRES Krylov basis, and the reverse-mode AD tape of the full-domain residual evaluation, all of which are governed by the state vector $\mathbf{w}$ and are independent of the choice of $\Omega_p$. The memory saving achieved by localisation arises solely from the reduction in the number of registered AD inputs from $3N_\Omega$ to $3N_{\Omega_p}$, and is consequently bounded below by the floor set by these state-governed terms, as discussed in \S\ref{section:adjoint_localisation}.

\begin{table}
\centering
\begin{tabular}{lccc}
\toprule
Case & $N_{\Omega_p}/N_\Omega$ & Peak memory (GB) & $M_{\text{local}}/M_{\text{global}}$ \\
\midrule
Global     & $1.000$ & $441.87$ & $1.000$ \\
Upstream   & $0.126$ & $160.18$ & $0.362$ \\
Body-wake  & $0.125$ & $254.89$ & $0.577$ \\
Downstream & $0.131$ & $227.40^\dagger$ & $0.515^\dagger$ \\
\bottomrule
\multicolumn{4}{l}{\footnotesize $^\dagger$ Run on Iridis~6 with 64 active cores} \\
\end{tabular}
\caption{Peak memory utilisation for the global and localisation cases. All cases except downstream were run on a single node with 64 cores on Iridis~5. The memory saving factor is defined as $M_{\text{local}}/M_{\text{global}}$. The downstream case was run on Iridis~6 and is included for indicative purposes only; see footnote.}
\label{tab:chap_4_memory}
\end{table}

The global, upstream, and body-wake cases were run on Iridis~5, equipped with dual $2.0$~GHz Intel Xeon Gold~$6130$ processors and $1.44$~TB of node memory, using 64 cores. The downstream case was run on Iridis~6, equipped with dual AMD EPYC~$9654$ processors, also using 64 active cores. While the two clusters differ in processor architecture, the memory footprint of the adjoint solve is governed primarily by the mesh size and the localisation fraction rather than by processor type, since the dominant stored objects (the preconditioner, Krylov basis, and AD tape) are determined by $n_w$ and $N_{\Omega_p}$ alone. The downstream figure is therefore considered representative and is included for completeness.

Considering the results in table~\ref{tab:chap_4_memory}, the upstream case achieves the largest memory saving ($M_{\text{local}}/M_{\text{global}} = 0.362$, a reduction of approximately $64$~\%) despite having a localisation ratio nearly identical to the body-wake case ($N_{\Omega_p}/N_\Omega \approx 0.126$ versus $0.125$). This is consistent with the upstream region lying in a lower mesh-density part of the domain, where the AD input registration and associated seed storage are reduced more aggressively than in the body-wake case, which covers the denser near-wake region. The body-wake case achieves a saving of approximately $42$~\%, and the downstream case approximately $48$~\%. All three localisation cases achieve meaningful reductions in peak memory relative to the global case. The body-wake configuration matches the global reconstruction in the integrated $L_1$ norm across the data planes, in the spatial distribution of the control variable, and in the structure of the mean flow on the $x$--$y$ plane, while requiring only $58$~\% of the global memory footprint. The combination of equivalent reconstruction fidelity and substantially reduced computational cost justifies the adoption of the body-wake localisation for the remainder of this chapter.

\section{Mean flow assimilation}
\label{section:mean_flow_assimilation}
While the body-wake case has been shown to closely match the global reconstruction at a fraction of the memory cost, the $L_1$ norm is an integrated quantity that provides a useful but incomplete picture of the reconstruction quality. It is important to examine the mean fields of flow variables and compare it to the experiment to distinguish between regions that have a good mean field reconstruction in comparison to those regions that do not. We discuss the mean flow variables in two parts. Section~\ref{subsection:assimilated_variables} presents the assimilated mean velocity fields and compares it to the experiment, while \S \ref{subsection:mean_flow_assimilation_derived_variables} presents the Reynolds shear stress and the mean pressure which are variables that are not provided as input to the assimilation.

\subsection{Assimilated variables}
\label{subsection:assimilated_variables}
\begin{figure}
    \centering
    \begin{subfigure}{\textwidth}
        \centering
        \caption{}
        \psfrag{X}[cc][cc]{Assimilated}
        \psfrag{Y}[cc][cc]{Experiment}
        \psfrag{P}[cc][cc]{$y^*$}
        \psfrag{Q}[cc][cc]{$z^*$}
        \psfrag{D}[cc][cc]{$U_x^*$}
        \psfrag{E}[cc][cc]{$U_y^*$}
        \psfrag{F}[cc][cc]{$U_z^*$}
        \psfrag{a}[rc][rc]{$-0.15$}
        \psfrag{b}[rc][rc]{$0.0$}
        \psfrag{c}[rc][rc]{$0.15$}
        \psfrag{e}[tc][tc]{$-0.25$}
        \psfrag{f}[tc][tc]{$0.0$}
        \psfrag{g}[tc][tc]{$0.25$}
        \psfrag{I}[cc][cc][0.8]{$-0.28$}
        \psfrag{J}[cc][cc][0.8]{$0.48$}
        \psfrag{K}[cc][cc][0.8]{$1.23$}
        \psfrag{L}[cc][cc][0.8]{$-0.24$}
        \psfrag{M}[cc][cc][0.8]{$-0.03$}
        \psfrag{N}[cc][cc][0.8]{$0.17$}
        \psfrag{O}[cc][cc][0.8]{$-0.23$}
        \psfrag{R}[cc][cc][0.8]{$-0.06$}
        \psfrag{S}[cc][cc][0.8]{$0.11$}
        \includegraphics[width=\textwidth]{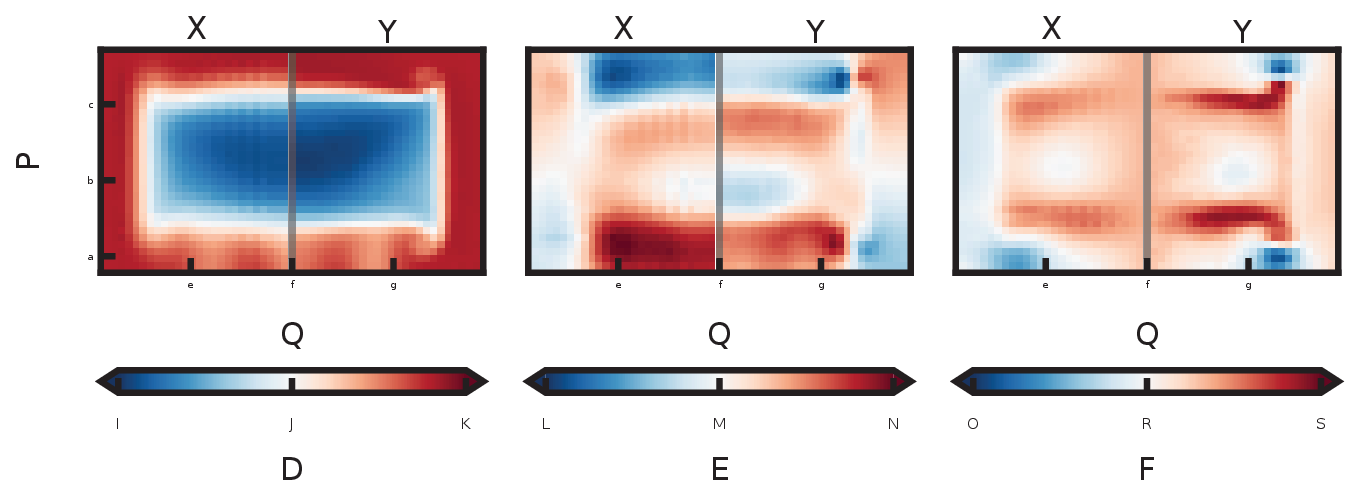}
        \label{fig:assm_vs_exp_plane1}
    \end{subfigure}

    \vspace{-1em}

    \begin{subfigure}{\textwidth}
        \centering
        \caption{}
        \psfrag{X}[cc][cc]{Assimilated}
        \psfrag{Y}[cc][cc]{Experiment}
        \psfrag{P}[cc][cc]{$y^*$}
        \psfrag{Q}[cc][cc]{$z^*$}
        \psfrag{D}[cc][cc]{$U_x^*$}
        \psfrag{E}[cc][cc]{$U_y^*$}
        \psfrag{F}[cc][cc]{$U_z^*$}
        \psfrag{a}[rc][rc]{$-0.15$}
        \psfrag{b}[rc][rc]{$0.0$}
        \psfrag{c}[rc][rc]{$0.15$}
        \psfrag{e}[tc][tc]{$-0.25$}
        \psfrag{f}[tc][tc]{$0.0$}
        \psfrag{g}[tc][tc]{$0.25$}
        \psfrag{I}[cc][cc][0.8]{$0.68$}
        \psfrag{J}[cc][cc][0.8]{$0.90$}
        \psfrag{K}[cc][cc][0.8]{$1.13$}
        \psfrag{L}[cc][cc][0.8]{$-0.13$}
        \psfrag{M}[cc][cc][0.8]{$-0.02$}
        \psfrag{N}[cc][cc][0.8]{$0.09$}
        \psfrag{O}[cc][cc][0.8]{$-0.04$}
        \psfrag{R}[cc][cc][0.8]{$0.01$}
        \psfrag{S}[cc][cc][0.8]{$0.07$}
        \includegraphics[width=\textwidth]{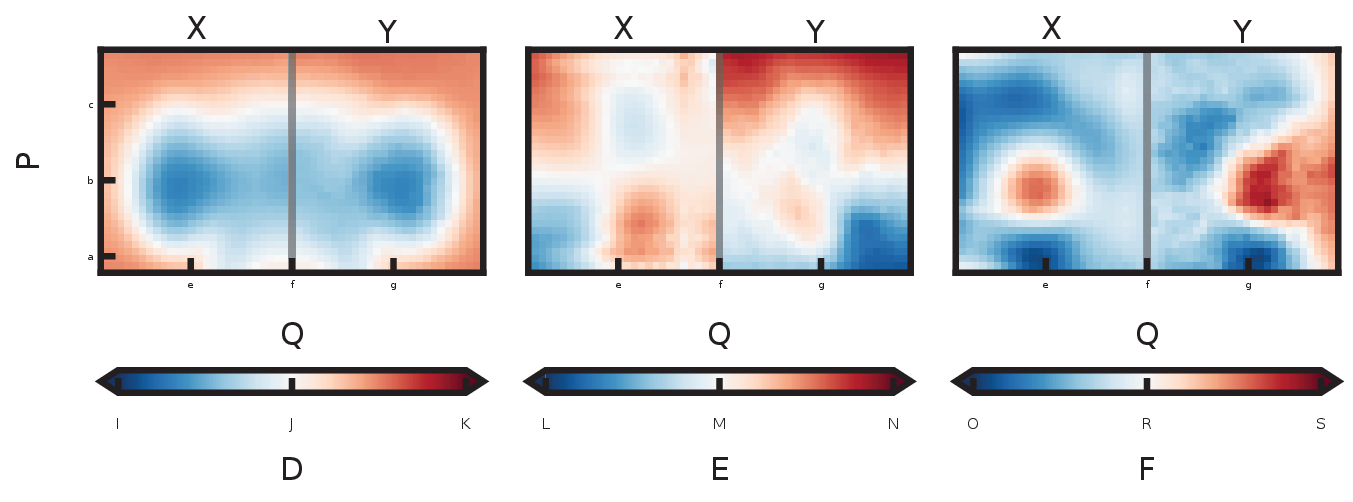}
        \label{fig:assm_vs_exp_plane9}
    \end{subfigure}
\caption{Comparison of the mean streamwise, wall-normal, and spanwise velocity components between the body-wake assimilated and the experiment on (a) Plane~$2$ and (b) Plane~$10$. Plotting format similar to figure~\ref{fig:bsl_vs_exp}.}
    \label{fig:assm_vs_exp}
\end{figure}

 Figure~\ref{fig:assm_vs_exp} presents the mean streamwise, wall-normal and spanwise assimilated fields for Plane~$2$ and~$10$ directly, in the same format as figure~\ref{fig:bsl_vs_exp}. Let us first discuss the assimilated field On Plane~$2$, which is shown in figure~\ref{fig:assm_vs_exp_plane1}. The assimilated mean streamwise velocity (left-most panel) shows a substantially improved match with the experiment relative to the baseline, with the wake deficit better captured and a modest recovery of the corner vortices. Next, the wall-normal component (middle panel) is also reproduced well, though with slightly elevated magnitudes in the corner vortex region. Finally, we examine the mean spanwise velocity field (right-most panel). The assimilated mean spanwise velocity field broadly reproduces most of the features of its experimental counterpart. The magnitude of the assimilated spanwise mean velocity is lower in comparison to the experiment. The residual discrepancy thats exist in the mean streamwise, wall-normal and spanwise velocity components of the assimilated field is reflected in the relatively greater magnitude of the $L_1$ norm along Plane~$2$ in comparison to the other planes, as can be seen in figure~\ref{fig:chap_4_l1_localisation}. 
 
 We discuss the the mean velocity fields along Plane~$10$. Figure~\ref{fig:assm_vs_exp_plane9} shows the mean streamwise, wall-normal and spanwise velocity fields from assimilation, comparing them to the experiment. The assimilated mean streamwise velocity field (left-most panel) shows a markedly better agreement with the experimental field with the magnitude of momentum recovery captured almost exactly. The single-lobe structure shows the correct level of streamwise velocity recovery consistent with the experiment. The assimilated wall-normal velocity field (middle panel) shows good agreement with the experiment, both in its magnitude and in its spatial distribution. The spanwise velocity field from assimilated (right-most panel) similarly agrees well with the experiment. In all three cases, it is noted that assimilation is able to smoothen the mean velocity field, which is a consequence of the regularisation that was applied. The close agreement between the assimilated field and the experiment along Plane~$10$, which was observed in the $L_1$ norm in figure~\ref{fig:chap_4_l1_localisation}, is also confirmed spatially for all three velocity components in figure~\ref{fig:assm_vs_exp_plane9}. These results confirm that the body-wake assimilation captures the dominant flow physics across the measurement domain, with the largest residual errors confined to the near-wake region closest to the model.

The results presented in figure~\ref{fig:assm_vs_exp} confirm that the assimilation reproduces the measured velocity components across the data planes. However, the sparse cross-stream measurements used as reference data provide no direct information about the streamwise organisation of the flow, and it is not \textit{a priori} evident that the assimilated field recovers the correct 3D structure of the recirculation bubble in unmeasured regions. The following discussion examines whether the assimilation achieves this, using the mean velocity streamlines and the streamwise velocity field on the $x$--$y$ symmetry plane.

\begin{figure}[tbp]
    \centering
\includegraphics[width=\textwidth]{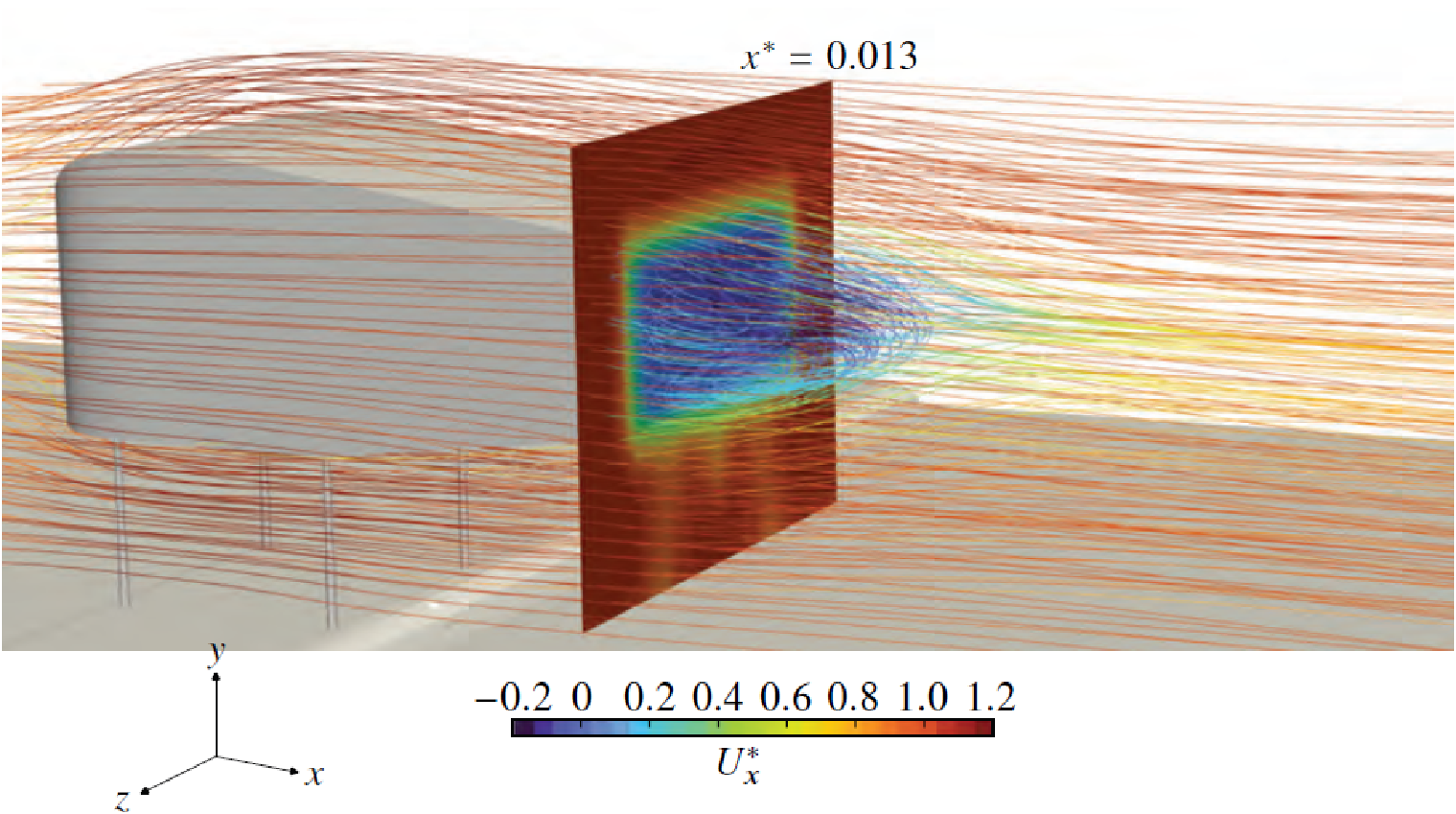}
\caption{Isometric view of mean velocity streamlines in the wake of the multi-wake model, coloured by the normalised streamwise velocity $U_x^*$. The streamlines are seeded upstream of the model and trace the 3D structure of the recirculating flow. The mean streamwise velocity field on Plane~$1$ ($x^* = 0.013$) is shown for reference.}
\label{fig:chap_4_streamlines_3d} 
\end{figure}

The 3D structure of the primary recirculation bubble is first examined through the mean velocity streamlines shown in figure~\ref{fig:chap_4_streamlines_3d}, coloured by $U_x^*$. The streamlines are seeded upstream of the model and computed using ParaView's Stream Tracer filter with the Runge-Kutta~4-5 integration method, following \cite{midya2025experimental}. The bubble is complex. The streamlines entering the wake from the top and bottom of the rear face recirculate and exchange fluid across the centreline, revealing a distinctly 3D topology. The mean streamwise velocity field on Plane~$1$ ($x^* = 0.013$) is shown for spatial reference. The overall topology closely resembles the experimental streamline patterns reported in \cite{midya2025experimental}, providing initial evidence that the assimilation recovers the correct 3D structure of the wake from sparse cross-stream data alone.

\begin{figure}[tbp]
    \centering
    \psfrag{a}[cc][cc]{$-0.2$}
    \psfrag{b}[cc][cc]{$0$}
    \psfrag{c}[cc][cc]{$0.2$}
    \psfrag{d}[cc][cc]{$0.4$}
    \psfrag{e}[cc][cc]{$0.6$}
    \psfrag{f}[cc][cc]{$0.8$}
    \psfrag{g}[cc][cc]{$1.0$}
    \psfrag{x}[cc][cc]{$x$}
    \psfrag{y}[cc][cc]{$y$}
    \psfrag{z}[cc][cc]{$z$}
    \psfrag{w}[cc][cc]{$U_x^*$}
    \psfrag{h}[cc][cc]{Model}
    \psfrag{i}[cc][cc]{Mean vorticity centres}
    \includegraphics[width=\textwidth]{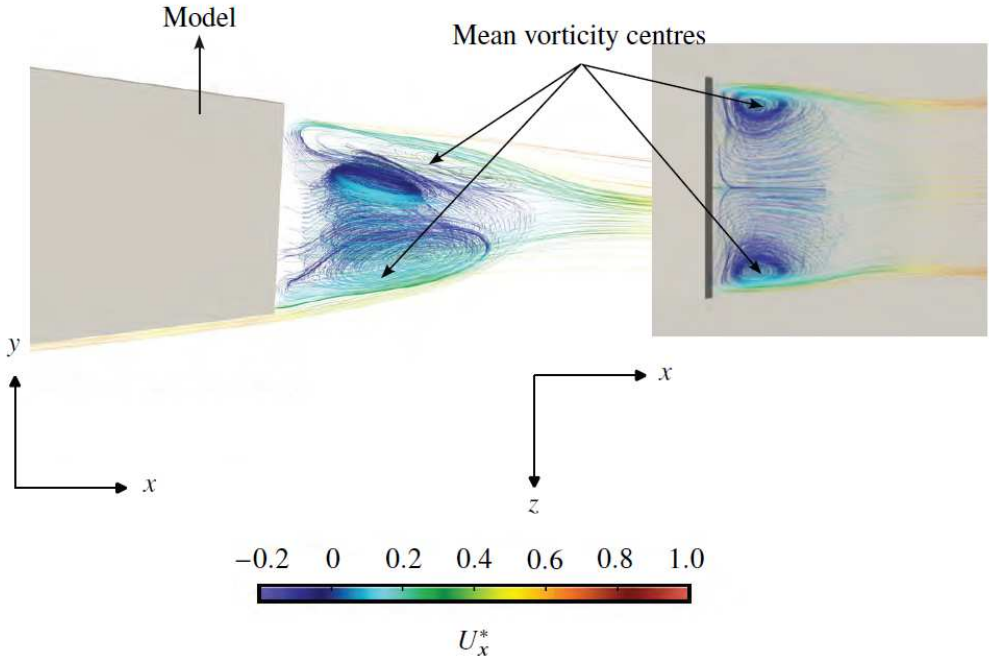}
\caption{Mean velocity streamlines coloured by the normalised streamwise velocity $U_x^*$ in the front view ($x$--$y$ plane, left) and top view ($x$--$z$ plane, right). The vorticity centres, identified from the mean streamline topology, are marked and the model geometry is shown for reference.}
\label{fig:chap_4_streamlines_2d}
\end{figure}

The internal structure of the bubble is examined more closely through projections of the mean velocity streamlines onto the $x$--$y$ and $x$--$z$ planes, shown in figure~\ref{fig:chap_4_streamlines_2d}. The front view ($x$--$y$ plane) reveals two distinct vortex cores within the primary bubble, identified from the mean streamline topology and annotated in the figure. The upper and lower cores correspond to the top and bottom recirculating fluid regions, respectively. The line connecting them is inclined with respect to the vertical, tilted away from the rear face of the model. This is a direct consequence of the geometry of the rear face itself, which imposes its asymmetry on the structure of the bubble along the $y$--$z$ plane. The top view ($x$--$z$ plane) shows that the bubble is more compact in the streamwise direction than in the spanwise direction, indicating that the recirculating fluid diffuses more rapidly in the streamwise direction compared to the spanwise one. This behaviour is consistent with the experimental observations of \cite{midya2025experimental}. Together, the two views establish that the assimilated field reproduces the complex 3D topology of the bubble, including features that were not directly measured.

\begin{figure}
    \centering
    \psfrag{A}[tl][tl]{Experiment}
    \psfrag{B}[tl][tl]{Baseline}
    \psfrag{C}[tl][tl]{Body-wake}
    \psfrag{Q}[cc][cc]{$x^*$}
    \psfrag{P}[cc][cc]{$y^*$}
    \psfrag{e}[cc][cc]{$-0.8$}
    \psfrag{f}[cc][cc]{$0.0$}
    \psfrag{g}[cc][cc]{$0.8$}
    \psfrag{h}[cc][cc]{$1.6$}
    \psfrag{a}[rc][rc]{$-0.25$}
    \psfrag{b}[rc][rc]{$0.0$}
    \psfrag{c}[rc][rc]{$0.25$}
    \psfrag{I}[lc][lc]{$-1.4$}
    \psfrag{J}[lc][lc]{$0.0$}
    \psfrag{K}[lc][lc]{$1.4$}
    \psfrag{R}[cc][cc]{$U_x^*$}
    \includegraphics[width=\textwidth]{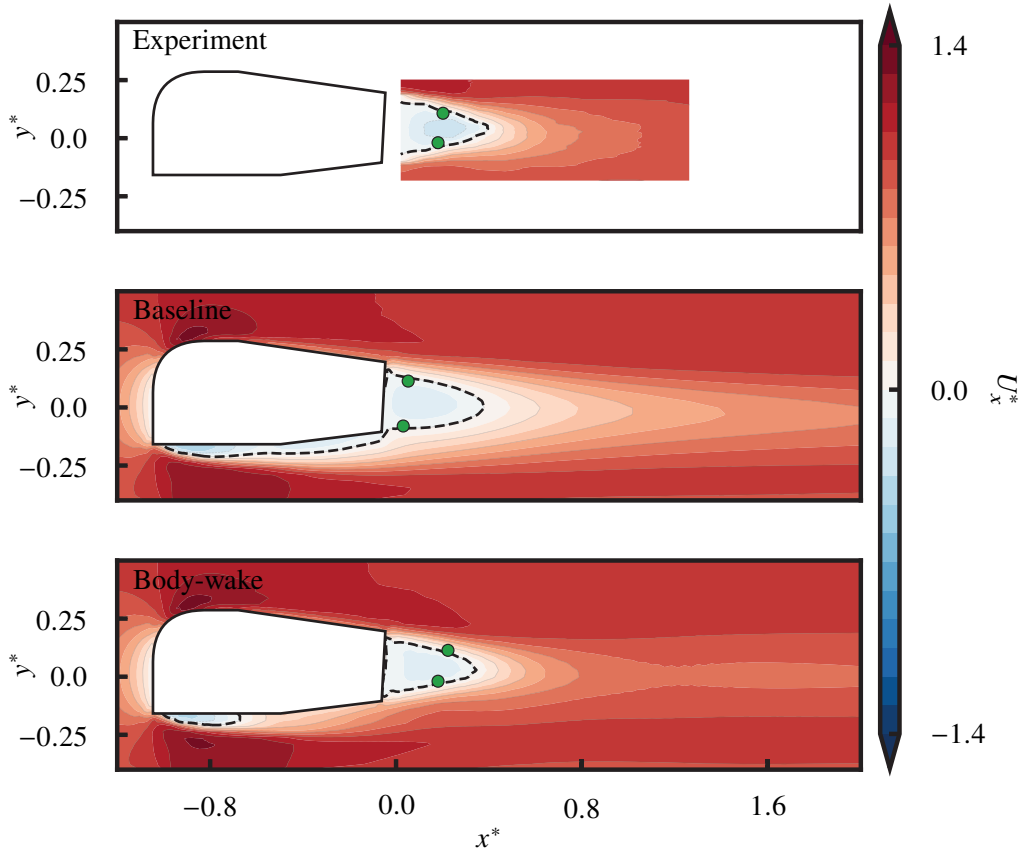}
    \caption{Mean streamwise velocity $U_x^*$ on the $x$--$y$ symmetry plane at $z^*=0$ for the experiment \textit{(top)}, baseline SA model \textit{(middle)} and the body-wake assimilated case \textit{(bottom)}. The dashed line denotes the $U_x^*=0$ isoline delineating the primary recirculation bubble, and the filled green circles indicate the dominant eddy centres identified by the $\Gamma_1$ criterion \citep{graftieaux2001combining}.}
\label{fig:chap_4_xy_streamlines}
\end{figure}

\textcolor{black}{A further assessment of the bubble topology is provided by the mean streamwise velocity on the $z^* = 0$ symmetry plane, shown in figure~\ref{fig:chap_4_xy_streamlines} for the experiment (top panel), baseline (middle panel), and body-wake assimilated (bottom panel) cases. It must be noted that the experimental data have low resolution along the streamwise direction (recall that the planes of data are separated by $100~\mathrm{mm}$). The data are interpolated onto a uniform grid with $300$ points along $x$ and $150$ points along $y$. The effect of such upsampling is visible in the top panel of figure~\ref{fig:chap_4_xy_streamlines}, where the contour lines are not smooth. The primary recirculation bubble is delineated by the $U_x^*=0$ isoline (dashed), and the dominant eddy centres within it are identified using the $\Gamma_1$ criterion \citep{graftieaux2001combining}. The criterion is evaluated on the interpolation grid using a circular neighbourhood of radius $0.08L_x$, and the eddy centres are taken as the local extrema of the smoothed $|\Gamma_1|$ field exceeding a threshold of approximately $0.8$. For the experimental field, points whose neighbourhood is truncated by the edge of the measurement domain are excluded to avoid spurious extrema. The identified centres are consequently accurate only to within the spacing of the interpolation grid.} 

\textcolor{black}{In figure \ref{fig:chap_4_xy_streamlines}, it is clear that the dominant eddy centres for the experiment are detached from the body, and the upper core is further downstream in comparison to the lower core. The baseline (middle panel) exhibits a secondary separation bubble at the leading edge of the bottom face of the body that fails to reattach and merges directly with the primary bubble, behaviour that is physically inconsistent with the expected reattachment downstream of the local separation on the bottom face, as described in \cite{midya2025experimental}. In the baseline, the two dominant eddy centres are positioned symmetrically about the horizontal centreline, yielding a near-zero inclination angle between them, inconsistent with the expected asymmetry arising from the geometry of the rear face. The assimilated field (bottom panel) shows similarities with the experiment. Firstly, the secondary bubble reattaches well upstream of the rear face, consistent with the physical description of \cite{midya2025experimental}. Secondly, the mean streamwise velocity field shows two eddy centres within the primary bubble, where the upper core is displaced further downstream than the lower one, similar to what is observed in the experiment. The agreement in the locations of the dominant eddy centres on the spanwise mid-plane between the experiment and the assimilation provides strong evidence that the assimilation is able to recover the correct asymmetric topology of the recirculation bubble despite the low data resolution along the streamwise direction.}

\subsection{Derived variables}
\label{subsection:mean_flow_assimilation_derived_variables}
Beyond the mean velocity field, variational DA offers the opportunity to recover quantities that are not directly provided as input to the assimilation. Two such quantities are examined here: the Reynolds stresses and the mean pressure field. 

The Reynolds shear stress $\overline{u'v'}$ is selected for comparison as it is the dominant off-diagonal component of the Reynolds stress tensor and represents the best-converged quantity from the stereo PIV measurement. Under the Boussinesq hypothesis, the deviatoric part is modelled as $\mathbf{R}_\text{dev} = -2\nu_t\mathbf{S}$, where $\mathbf{S} = \frac{1}{2}\left(\nabla\mathbf{U} + \nabla\mathbf{U}^T\right)$ is the mean strain rate tensor. Before presenting the comparison, it is important to note that the Reynolds shear stress $\overline{u'v'}$ computed from the optimised eddy viscosity field $\nu_t$ is not a direct equivalent of the experimentally measured quantity. The corrective forcing $\mathbf{f}_c$ enters the momentum equation alongside the divergence of the Reynolds stress tensor $\nabla\cdot\mathbf{R}$, and an ideal comparison would involve matching the total forcing vector $\mathbf{f} = \mathbf{f}_R + \mathbf{f}_c$ (see \eqref{equation:total_forcing_vector}) from the assimilation against $\nabla\cdot\mathbf{R}$ computed from the experimental Reynolds stresses. However, this is not feasible here for two reasons: the experimental Reynolds stresses are not fully converged, and the sparse measurement coverage in the streamwise direction precludes a reliable computation of $\nabla\cdot\mathbf{R}$ from the experimental data. The comparison of $\overline{u'v'}$ presented here should therefore be interpreted with these caveats in mind, and as an indicative rather than exact validation of the turbulence representation recovered by the assimilation.

\begin{figure}
    \centering
    \begin{subfigure}{\textwidth}
            \caption{}
        \psfrag{A}[cc][cc]{Experiment}
        \psfrag{B}[cc][cc]{Baseline}
        \psfrag{C}[cc][cc]{Assimilated}
        \psfrag{P}[cc][cc]{$y^*$}
        \psfrag{Q}[cc][cc]{$z^*$}
        \psfrag{a}[rc][rc]{$0.15$}
        \psfrag{b}[rc][rc]{$0.0$}
        \psfrag{c}[rc][rc]{$-0.15$}
        \psfrag{e}[cc][cc]{$-0.2$}
        \psfrag{f}[cc][cc]{$0.0$}
        \psfrag{g}[cc][cc]{$0.2$}
        \psfrag{I}[lc][lc]{$-0.023$}
        \psfrag{J}[lc][lc]{$0.0$}
        \psfrag{K}[lc][lc]{$0.023$}
        \psfrag{R}[cc][cc]{$\overline{u'v'}/U_\infty^2$}
        \includegraphics[width=\textwidth]{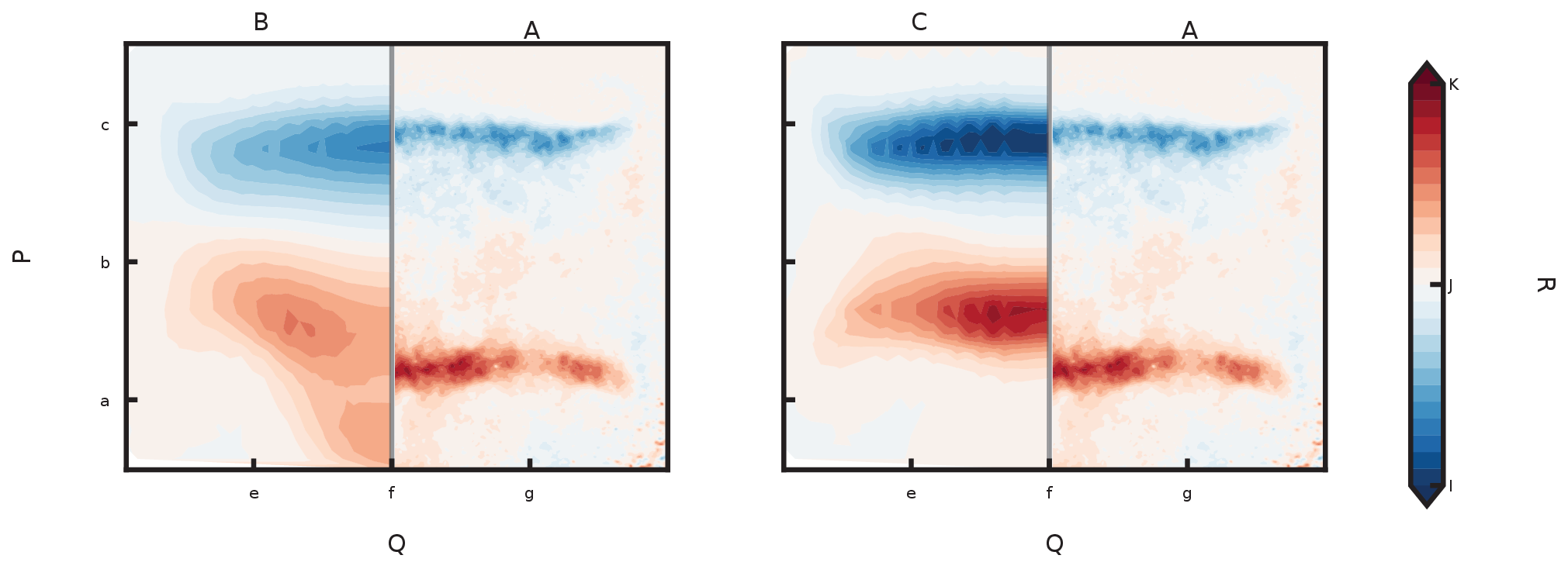}
        \label{fig:chap_4_uv_plane2}
    \end{subfigure}

    \vspace{-1em}

    \begin{subfigure}{\textwidth}
            \caption{}
        \psfrag{A}[cc][cc]{Experiment}
        \psfrag{B}[cc][cc]{Baseline}
        \psfrag{C}[cc][cc]{Assimilated}
        \psfrag{P}[cc][cc]{$y^*$}
        \psfrag{Q}[cc][cc]{$z^*$}
        \psfrag{a}[rc][rc]{$0.15$}
        \psfrag{b}[rc][rc]{$0.0$}
        \psfrag{c}[rc][rc]{$-0.15$}
        \psfrag{e}[cc][cc]{$-0.2$}
        \psfrag{f}[cc][cc]{$0.0$}
        \psfrag{g}[cc][cc]{$0.2$}
        \psfrag{I}[lc][lc]{$-0.035$}
        \psfrag{J}[lc][lc]{$0.0$}
        \psfrag{K}[lc][lc]{$0.035$}
        \psfrag{R}[cc][cc]{$\overline{u'v'}/U_\infty^2$}
        \includegraphics[width=\textwidth]{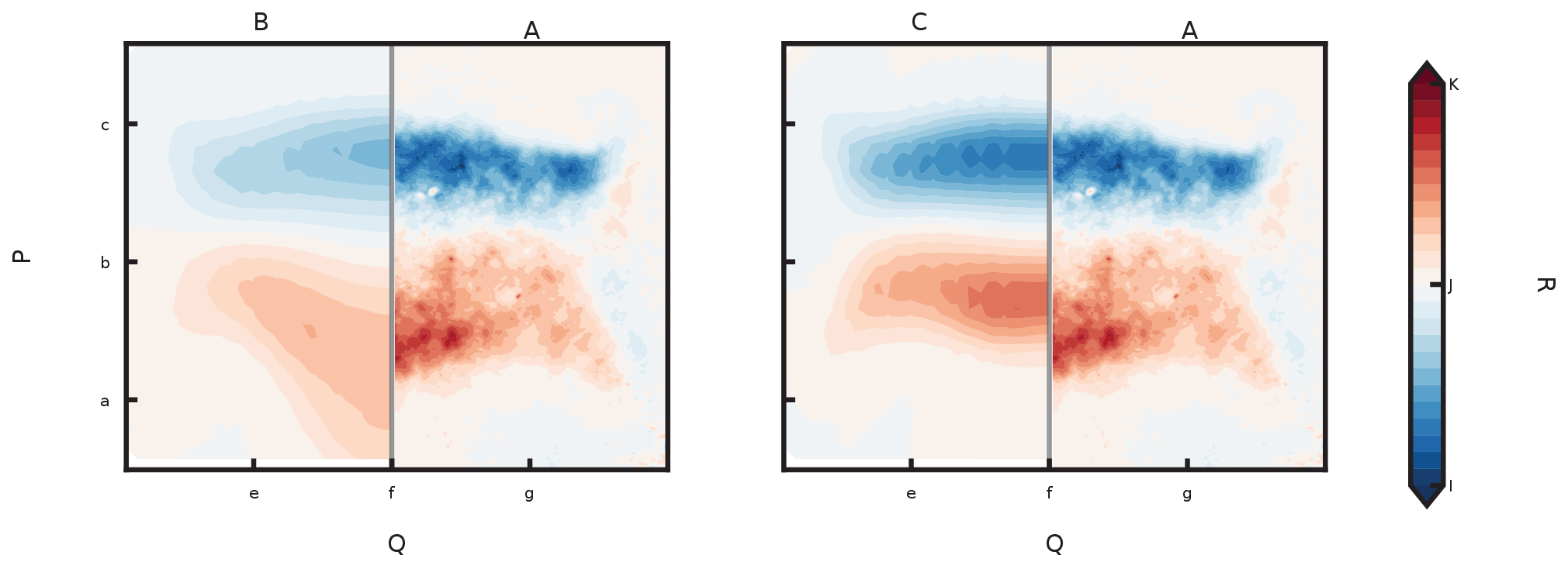}
        \label{fig:chap_4_uv_plane3}
    \end{subfigure}

    \caption{Comparison of the Reynolds shear stress $\overline{u'v'}/U_\infty^2$ between the baseline SA model, the body-wake assimilated field, and the experiment on (a) Plane~$2$ and (b) Plane~$3$. In both panels, the right half of the butterfly plot corresponds to the experiment and the left half to the baseline/assimilated fields.}
    \label{fig:chap_4_uv_butterfly}
\end{figure}

The normalised Reynolds shear stress $\overline{u'v'}/U_\infty^2$ from the experiment is compared with the baseline and the assimilated fields on Plane~$2$ and~$3$ in figure~\ref{fig:chap_4_uv_butterfly}, with the corresponding peak values summarised in table~\ref{tab:chap_4_uv_minmax}. The left column of images in figure~\ref{fig:chap_4_uv_butterfly} compares the baseline with the experiment and the right column, the body-wake assimilated with the experiment. The stress fields exhibit a plane of symmetry about $z^*=0$, as expected from the spanwise symmetry of the geometry. Two distinct regions of opposite sign are visible on each plane, both elongated in the spanwise direction and relatively narrow in the wall-normal direction. This horizontal elongation is consistent with the geometry of the model. The inclination angles on the top and bottom faces are substantially larger than those on the side faces, so the dominant turbulent fluctuations originate from the shear layers detached from the top and bottom trailing edges. The positive $\overline{u'v'}$ region is located below the centreline and the negative region above it, which is consistent with the sign convention for a shear layer where the mean velocity gradient $\partial U/\partial y$ changes sign across the wake centreline — the upper shear layer produces negative $\overline{u'v'}$ and the lower positive.

On Plane~$2$, which lies closer to the rear face of the model, the two regions of opposite sign are clearly separated by a gap near $y^*=0$, indicating limited interaction between the upper and lower shear layers at this streamwise location. This is consistent with the description in \cite{midya2025experimental}, where the two shear layers are shown to remain relatively distinct in the near wake. By Plane~$3$, the two regions have broadened in the wall-normal direction and the gap between them has closed, indicating the onset of interaction between the upper and lower shear layers as they develop downstream. The assimilated field reproduces this behaviour qualitatively on both planes. The separation between the two regions on Plane~$2$ is captured, though less sharply than in the experiment, and the broadening on Plane~$3$ is reproduced with close agreement in both structure and magnitude.

\begin{table}
\centering

\begin{tabular}{llcc}
\toprule
Plane & Case & $\min(\overline{u'v'}/U_\infty^2)$ & $\max(\overline{u'v'}/U_\infty^2)$ \\
\midrule
\multirow{3}{*}{2} & Experiment  & $-0.0164$ & $0.0203$ \\
                   & Baseline    & $-0.0160$ & $0.0120$ \\
                   & Assimilated   & $-0.0230$ & $0.0205$ \\
\midrule
\multirow{3}{*}{3} & Experiment  & $-0.0349$ & $0.0295$ \\
                   & Baseline    & $-0.0164$ & $0.0120$ \\
                   & Assimilated   & $-0.0257$ & $0.0205$ \\
\bottomrule
\end{tabular}
\caption{Maximum and minimum values of the Reynolds shear stress $\overline{u'v'}/U_\infty^2$ for the experiment, baseline SA model, and body-wake assimilated field on Plane~$2$ and~$3$.}
\label{tab:chap_4_uv_minmax}
\end{table}

Examining the peak values in table~\ref{tab:chap_4_uv_minmax}, the assimilated field shows a consistent enhancement of the Reynolds shear stress relative to the baseline on both planes, with the peak magnitudes moving substantially closer to the experimental values. The baseline significantly underpredicts the stress on both planes, with peak magnitudes roughly half those of the experiment on Plane~$3$. This enhancement in the assimilation is a direct consequence of the increase in eddy viscosity driven by the optimised velocity field, confirming that the corrective forcing acts primarily by modifying the turbulence representation through the SA transport equation in addition to directly supplying momentum. \textcolor{black}{This is in stark contrast to what was observed in \cite{cadambi2026three}, where two-component mean velocity was used to assimilate a stalled airfoil using 2D variational data assimilation. The absence of the spanwise velocity component and 2D constraints was shown to reduce the magnitude of eddy viscosity, and thereby the Reynolds shear stress. We speculate that the presence of all three mean velocity components for assimilation allows the momentum forcing to exclusively target model discrepancies, enhancing the eddy viscosity and subsequently the Reynolds stress.}

The mean pressure field is examined next. Since $\mathbf{f}_c$ is divergence-free, it acts as a solenoidal body force on the momentum equation and does not contribute to the pressure Poisson solve. This means that the recovered pressure field $P^*$ is determined entirely by the corrected velocity field and the modified turbulent pressure $\frac{2}{3}k$. As the velocity field is well-matched to the experiment and $\mathbf{f}_c$ is divergence-free, the recovered pressure is expected to be physically consistent, even in the absence of experimental pressure measurements for direct validation.

\begin{figure}
    \centering
        \psfrag{X}[bc][bc]{Pressure coefficient}
    \psfrag{Y}[bc][bc]{Vorticity: $\omega_y$}
        \psfrag{Z}[bc][bc]{Vorticity: $\omega_z$}
    \psfrag{P}[cc][cc]{$y^*$}
    \psfrag{Q}[cc][cc]{$z^*$}
    \psfrag{a}[rc][rc]{$0.15$}
    \psfrag{b}[rc][rc]{$0.00$}
    \psfrag{c}[rc][rc][0.8]{$-0.15$}
    \psfrag{e}[cc][cc]{$-0.25$}
    \psfrag{f}[cc][cc]{$0.00$}
    \psfrag{g}[cc][cc]{$0.25$}
    \psfrag{I}[cc][cc][0.8]{$-0.15$}
    \psfrag{J}[cc][cc][0.8]{$-0.09$}
    \psfrag{K}[cc][cc][0.8]{$-0.04$}
    \psfrag{L}[cc][cc][0.8]{$-18$}
    \psfrag{M}[cc][cc][0.8]{$0$}
    \psfrag{N}[cc][cc][0.8]{$18$}
    \psfrag{R}[cc][cc]{$C_p$}
    \psfrag{S}[cc][cc]{$\omega_y^*$}
    \psfrag{A}[cc][cc]{$\omega_z^*$}
    
    \includegraphics[width=\textwidth]{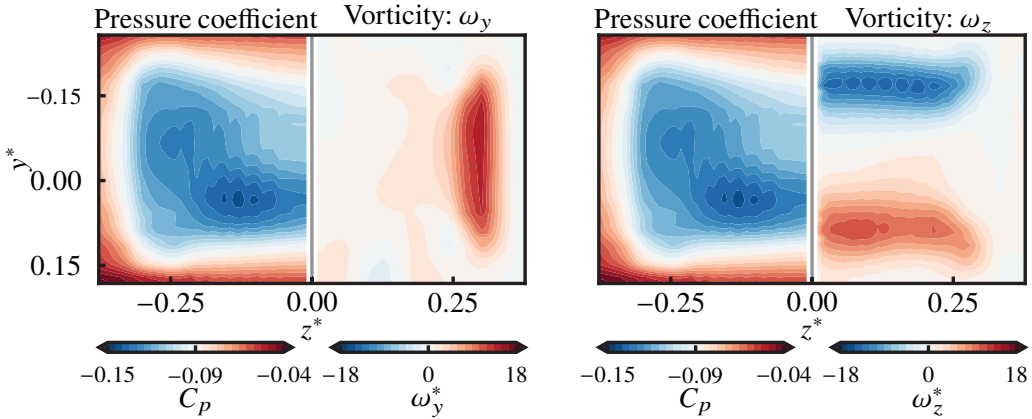}
    \caption{Assimilated fields on Plane~$2$ in the near wake. Each subplot presents the mean pressure coefficient $C_p$ on the left half and a vorticity component on the right half: wall-normal vorticity $\omega_y^*$ in the left subplot, and spanwise vorticity $\omega_z^*$ in the right subplot. The $C_p$ field is reflected about the spanwise symmetry plane $z^* = 0$ for display. The dividing line at $z^* = 0$ marks the symmetry plane of the model.}
    \label{fig:chap_4_cp_vort_plane2}
\end{figure}

\begin{figure}
    \centering
    \psfrag{X}[bc][bc]{Pressure coefficient}
    \psfrag{Y}[bc][bc]{Vorticity: $\omega_y$}
        \psfrag{Z}[bc][bc]{Vorticity: $\omega_z$}
    \psfrag{P}[cc][cc]{$y^*$}
    \psfrag{Q}[cc][cc]{$z^*$}
    \psfrag{a}[rc][rc]{$0.15$}
    \psfrag{b}[rc][rc]{$0.00$}
    \psfrag{c}[rc][rc]{$-0.15$}
    \psfrag{e}[cc][cc]{$-0.25$}
    \psfrag{f}[cc][cc]{$0.00$}
    \psfrag{g}[cc][cc]{$0.25$}
    \psfrag{I}[cc][cc][0.8]{$-0.06$}
    \psfrag{J}[cc][cc][0.8]{$-0.03$}
    \psfrag{K}[cc][cc][0.8]{$0.003$}
    \psfrag{L}[cc][cc][0.8]{$-15.1$}
    \psfrag{M}[cc][cc][0.8]{$0.0$}
    \psfrag{N}[cc][cc][0.8]{$15.1$}
    \psfrag{R}[cc][cc]{$C_p$}
    \psfrag{S}[cc][cc]{$\omega_y^*$}
    \psfrag{A}[cc][cc]{$\omega_z^*$}    \includegraphics[width=\textwidth]{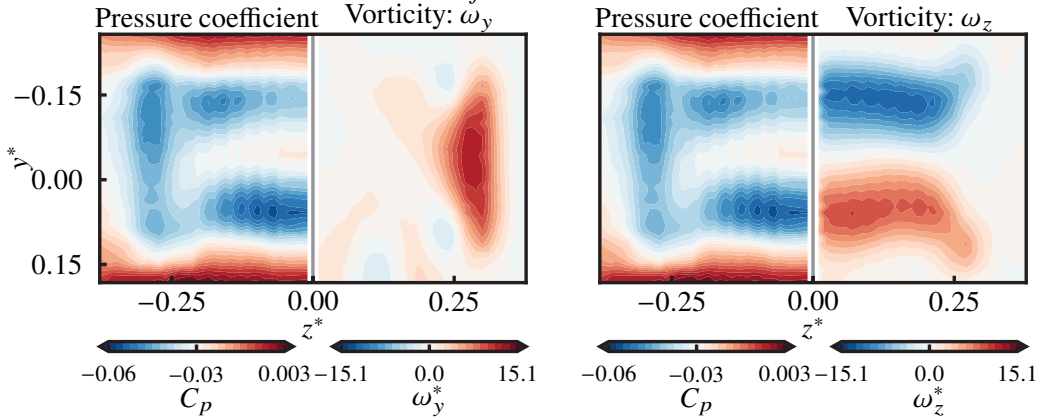}
    \caption{Assimilated fields on Plane~$3$ in the near wake. Each subplot presents the mean pressure coefficient $C_p$ on the left half and a vorticity component on the right half: wall-normal vorticity $\omega_y^*$ in the left subplot, and spanwise vorticity $\omega_z^*$ in the right subplot. The $C_p$ field is reflected about the spanwise symmetry plane $z^* = 0$ for display. The dividing line at $z^* = 0$ marks the symmetry plane of the model.}
    \label{fig:chap_4_cp_vort_plane3}
\end{figure}

Figures~\ref{fig:chap_4_cp_vort_plane2} and~\ref{fig:chap_4_cp_vort_plane3} show the mean pressure coefficient and the two in-plane vorticity components from the body-wake assimilated field on Planes~$2$ and~$3$, respectively. Each figure contains two subplots. In each subplot, the left half displays the assimilated $C_p$ and the right half displays one of the in-plane vorticity components, with $\omega_y^* = \omega_y L_x / U_\infty$ shown in the left subplot and $\omega_z^* = \omega_z L_x / U_\infty$ in the right subplot. It must be noted that these two vorticity components ($\omega_y^*$ and $\omega_z^*$) cannot be computed for the experimental data due to the poor resolution of the mean velocity along the streamwise direction. The term \textit{in-plane} refers here to the direction of the vorticity vector, which lies in this plane, even though the rotation each component describes takes place out of plane. The pressure field has been interpolated onto the downsampled experimental grid and reflected about the symmetry plane so that it can be displayed alongside the vorticity. In the absence of experimental pressure measurements, the vorticity fields provide a qualitative reference, and the spatial correspondence between the flow features they reveal and the gradients of the recovered pressure field offers indirect evidence for the physical consistency of the assimilated pressure.

On Plane~$2$ (figure~\ref{fig:chap_4_cp_vort_plane2}), the assimilated $C_p$ field is negative throughout the plane, with values more negative in the central region than at the periphery. The resulting pressure gradient points inwards from the periphery, drawing fluid into the wake from the surrounding freestream. The negative-pressure region is not symmetric about the $x$-$z$ plane. The lower portion of the plane exhibits more strongly negative $C_p$ than the upper portion. This asymmetry is consistent with the inclined rear face of the model, which warps and tilts the bound vortex behind the body, as also observed by \citet{midya2025experimental}. The inward entrainment leaves a signature in the in-plane vorticity. In the left subplot, the wall-normal vorticity $\omega_y^*$ shows a vertically elongated region of positive values along the lateral edge of the plane, consistent with fluid being drawn into the wake from the sides. In the right subplot, two horizontal bands of spanwise vorticity $\omega_z^*$ of opposite sign trace the upper and lower edges of the rear face of the model, corresponding to the upper and lower shear layers of the wake across which fluid is entrained from the freestream into the recirculation bubble. These vorticity signatures are most clearly attributed to the action of the corner vortices that develop downstream of the model. The wide gap between the two $\omega_z^*$ bands reflects the role of the bound vortex, whose induced flow opposes the inward inflow along the top and bottom edges and prevents the upper and lower shear layers from penetrating towards the centre, again as observed by \citet{midya2025experimental}. The spatial correspondence between the inward $C_p$ gradient and the vorticity features at the lateral and vertical edges of the plane therefore provides indirect support for the physical consistency of the assimilated pressure.

On Plane~$3$ (figure~\ref{fig:chap_4_cp_vort_plane3}), the assimilated $C_p$ field is again negative across most of the plane but with a substantially reduced magnitude compared to Plane~$2$. A small positive $C_p$ region appears at the centre of the plane, signalling that the downstream end of the primary recirculation bubble is near. The negative-$C_p$ region surrounds this central positive patch, producing an inward pressure gradient that is more uniformly directed inwards than on Plane~$2$, where the gradient was concentrated towards the lower half of the plane. This is consistent with the observation of \citet{midya2025experimental} that on Plane~$3$ the bound vortex core is fully contained within the plane and induces an inward flow in all directions. The in-plane vorticity fields are also weaker in magnitude than on Plane~$2$, consistent with the bubble approaching its downstream end and the entrainment moderating. The wall-normal vorticity $\omega_y^*$ retains a structure broadly similar to that on Plane~$2$, with a vertically elongated region of positive values along the lateral edge, indicating that the spanwise extent of the bubble has not changed appreciably between the two planes. The spanwise vorticity $\omega_z^*$, by contrast, shows a narrower gap between the upper and lower bands, indicating that the bubble has shrunk in the vertical direction. The spatial correspondence between the central $C_p$ recovery and the narrowed $\omega_z^*$ gap, alongside the persistent lateral $\omega_y^*$ signature provides indirect support for the physical consistency of the assimilated pressure on Plane~$3$.

\section{Data efficiency}
\label{section:data_efficiency}

Experimental campaigns are often constrained by cost, time, and instrumentation, resulting in sparse spatial coverage of observational data. This section examines the sensitivity of the variational DA framework to the number of measurement planes provided as input. Rather than prescribing a minimum data requirement (which would be specific to this configuration and unlikely to generalise), the aim is to provide physical reasoning for why certain plane selections are more effective than others from a DA perspective. Section~\ref{subsection:assimilated_variables_efficiency} presents the assimilated mean velocity fields for four cases of progressively reduced input data coverage and compares it to the case when all planes of data are provided, while \S \ref{subsection:data_efficiency_derived_variables_2} presents the Reynolds shear stress (which is a quantity that is not directly provided as input to the assimilation) as a metric for comparing the reduced data coverage cases with the case when all planes of data are provided as input. This allows us to examine the effect that reduced data coverage has on derived variables.

\subsection{Assimilated variables}
\label{subsection:assimilated_variables_efficiency}

Four cases are studied, using the first $6$, $5$, $4$, and $3$ planes of data (designated as $6$p, $5$p, $4$p and $3$p, respectively) measured immediately downstream of the model. The selection of these planes is motivated by two considerations. First, the flow in the near wake is dominated by the primary recirculation bubble. Due to the complex 3D nature of the bubble and the associated challenge of ascertaining the exact length of the bubble, providing between $3$ and $6$ planes therefore tests cases where data coverage spans the interior of the bubble, its closure region, and the region just beyond it. Second, the natural wake state begins approximately at Plane~$7$, beyond which the flow transitions to a momentum recovery regime. It was hypothesised that constraining the optimisation within the dynamically active near-wake region would be more effective than providing data further downstream where the flow is relatively smooth. Since the precise location of the bubble closure cannot be known \textit{a priori}, the range of cases studied is designed to bracket this region and reveal the sensitivity of the reconstruction to whether data coverage includes, terminates at, or falls short of the bubble closure.

\begin{figure}
    \centering
    \psfrag{P}[cc][cc]{$L_1$ norm}
    \psfrag{Q}[cc][cc]{Plane}
    \psfrag{a}[rc][rc]{$0.08$}
    \psfrag{b}[rc][rc]{$0.16$}
    \psfrag{c}[rc][rc]{$0.24$}
    \psfrag{n}[cc][cc]{$1$}
    \psfrag{o}[cc][cc]{$2$}
    \psfrag{p}[cc][cc]{$3$}
    \psfrag{q}[cc][cc]{$4$}
    \psfrag{r}[cc][cc]{$5$}
    \psfrag{s}[cc][cc]{$6$}
    \psfrag{t}[cc][cc]{$7$}
    \psfrag{u}[cc][cc]{$8$}
    \psfrag{v}[cc][cc]{$9$}
    \psfrag{w}[cc][cc]{$10$}
    \psfrag{x}[cc][cc]{$11$}
    \psfrag{y}[cc][cc]{$12$}
    \psfrag{A}[cl][cl]{$U_x$}
    \psfrag{B}[cl][cl]{$U_y$}
    \psfrag{C}[cl][cl]{$U_z$}
    \psfrag{Z}[cl][cl]{Baseline}
    \psfrag{X}[cl][cl]{$12$p}
    \psfrag{W}[cl][cl]{$6$p}
    \psfrag{V}[cl][cl]{$5$p}
    \psfrag{U}[cl][cl]{$4$p}
    \psfrag{T}[cl][cl]{$3$p}
    \includegraphics[width=\textwidth]{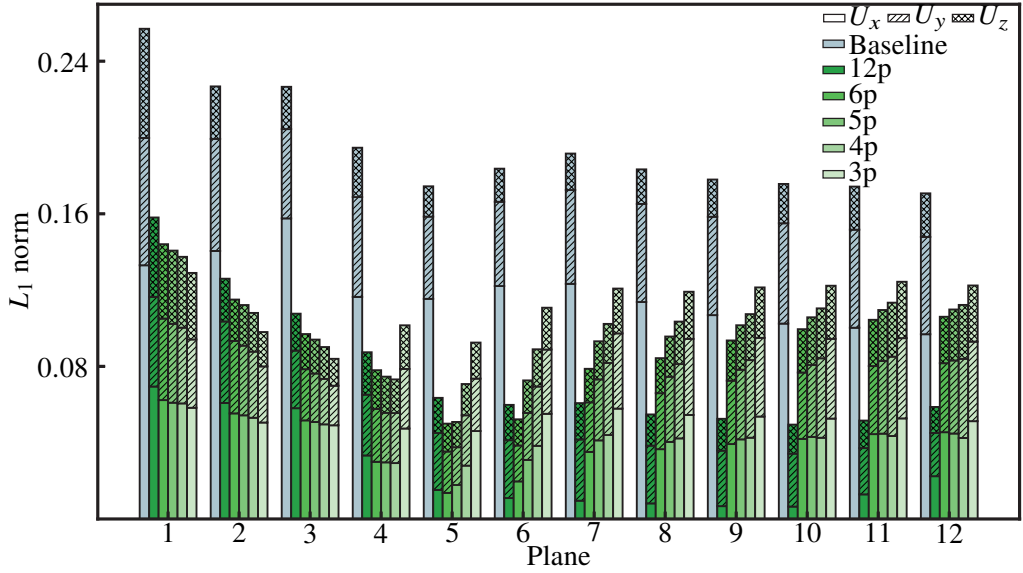}
    \caption{Plane-by-plane $L_1$ norm of the mean velocity field comparing the baseline SA model and the body-wake assimilated cases with progressively reduced measurement data. The stacked bars show the individual contributions of $U_x$, $U_y$, and $U_z$ to the total $L_1$ norm. The green gradient denotes the number of input planes used in the assimilation, with the darkest shade corresponding to the full 12-plane case and progressively lighter shades representing the 6-, 5-, 4-, and 3-plane cases respectively.}
    \label{fig:chap_4_l1_data_efficiency}
\end{figure}

The $L_1$ norm of the mean velocity field is presented in figure~\ref{fig:chap_4_l1_data_efficiency} for all cases across all 12 measurement planes, with the stacked bars showing contributions from $U_x$, $U_y$, and $U_z$. Two trends are evident. First, all sparse cases show improvement over the baseline SA model across all planes, confirming that even minimal data coverage produces a meaningful correction to the flow field. Second, within the data coverage region, sparse cases show lower $L_1$ norms than the $12$p case -- a consequence of the optimiser being able to apply a more focused correction when the misfit is evaluated on fewer planes. Beyond the last input plane of each sparse case, the $L_1$ norm increases sharply relative to the $12$p case, with the degradation becoming more pronounced as the number of input planes decreases. This is physically consistent since the control variable has no direct constraint in regions without data, and the ability of the adjoint to propagate corrections downstream diminishes as the data coverage recedes further upstream.

\begin{figure}
    \centering
    \psfrag{A}[bc][bc]{$3$p}
    \psfrag{B}[bc][bc]{$4$p}
    \psfrag{C}[bc][bc]{$5$p}
    \psfrag{D}[bc][bc]{$6$p}
    \psfrag{E}[bc][bc]{$12$p}
    \psfrag{P}[cc][cc]{$y^*$}
    \psfrag{Q}[cc][cc]{$z^*$}
    \psfrag{a}[rc][rc]{$-0.15$}
    \psfrag{b}[rc][rc]{$0.00$}
    \psfrag{c}[rc][rc]{$0.15$}
    \psfrag{e}[cc][cc]{$-0.2$}
    \psfrag{f}[cc][cc]{$0.0$}
    \psfrag{g}[cc][cc]{$0.2$}
    \psfrag{I}[lc][lc]{$0.6$}
    \psfrag{J}[lc][lc]{$0.9$}
    \psfrag{K}[lc][lc]{$1.3$}
    \psfrag{R}[cc][cc]{$U_x^*$}
    \includegraphics[width=\textwidth]{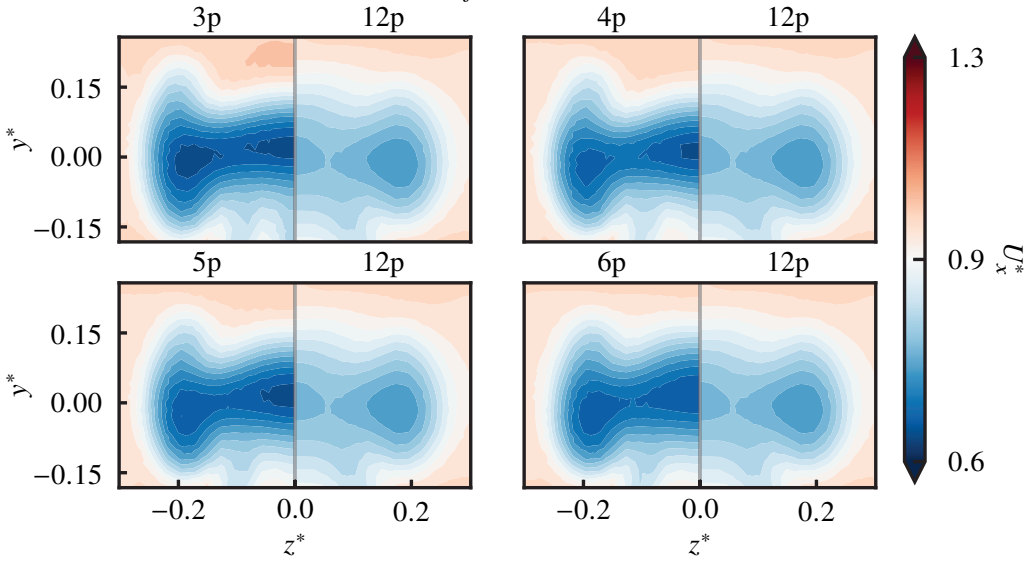}
    \caption{Mean streamwise velocity $U_x^*$ on Plane~$10$ ($x^* = 1.041$), which lies downstream of all input planes in every sparse case. Each panel compares a sparse assimilation case (left half) with the full 12-plane reference (right half) in a butterfly plot format. The panels show progressively increasing numbers of upstream input planes from top-left to bottom-right ($3$, $4$, $5$, and $6$ planes).}
    \label{fig:chap_4_ux_butterfly_plane9}
\end{figure}

While the $L_1$ norm provides a useful integrated measure of the mean velocity misfit, it does not capture the quality of derived quantities that are not directly provided as input to the assimilation. Figure~\ref{fig:chap_4_ux_butterfly_plane9} presents the mean streamwise velocity $U_x^*$ on Plane~$10$ ($x^* = 1.041$), which lies downstream of all input planes across all cases and therefore receives no direct data constraint in any of the sparse cases. Each panel compares a sparse case (left half) with the $12$p reference (right half). The sparse cases consistently show a lower momentum recovery than the $12$p case, evidenced by the deeper blue region near the wake centreline indicating a greater velocity deficit. This suggests that the reduced data coverage in the near wake produces a weaker correction to the eddy viscosity, which in turn limits the momentum recovery rate further downstream. Among the sparse cases, the $4$p, $5$p, and $6$p fields closely resemble each other, while the $3$p case shows a noticeably larger velocity deficit. The $3$p case provides data coverage only up to Plane~$3$, which lies upstream of the onset of pressure recovery identified in \S\ref{subsection:mean_flow_assimilation_derived_variables} as a marker for the end of the primary recirculating region. This suggests that data coverage extending at least to this region is necessary for the assimilation to adequately constrain the near-wake dynamics, and that falling short of it produces a qualitatively different reconstruction.

\subsection{Derived variables}
\label{subsection:data_efficiency_derived_variables_2}
\begin{figure}
    \centering
    \psfrag{A}[bc][bc]{$3$p}
    \psfrag{B}[bc][bc]{$4$p}
    \psfrag{C}[bc][bc]{$5$p}
    \psfrag{D}[bc][bc]{$6$p}
    \psfrag{E}[bc][bc]{$12$p}
    \psfrag{F}[bc][bc]{$12$p}
    \psfrag{G}[bc][bc]{$12$p}
    \psfrag{H}[bc][bc]{$12$p}
    \psfrag{I}[cc][cc]{Plane~$3$}
    \psfrag{J}[cc][cc]{Plane~$4$}
    \psfrag{K}[cc][cc]{Plane~$5$}
    \psfrag{L}[cc][cc]{Plane~$6$}
    \psfrag{P}[cc][cc]{$y^*$}
    \psfrag{Q}[cc][cc]{$z^*$}
    \psfrag{a}[rc][rc]{$-0.15$}
    \psfrag{b}[rc][rc]{$0.0$}
    \psfrag{c}[rc][rc]{$0.15$}
    \psfrag{e}[cc][cc]{$-0.25$}
    \psfrag{f}[cc][cc]{$0.0$}
    \psfrag{g}[cc][cc]{$0.25$}
    \psfrag{R}[cc][cc]{$\overline{u'v'}/U_\infty^2$}
    \psfrag{M}[lc][lc]{$-0.030$}
    \psfrag{N}[lc][lc]{$-0.004$}
    \psfrag{O}[lc][lc]{$0.025$}
    \psfrag{s}[lc][lc]{$-0.026$}
    \psfrag{q}[lc][lc]{$-0.003$}
    \psfrag{p}[lc][lc]{$0.019$}
    \psfrag{T}[lc][lc]{$-0.022$}
    \psfrag{U}[lc][lc]{$-0.004$}
    \psfrag{V}[lc][lc]{$0.013$}
    \psfrag{W}[lc][lc]{$-0.022$}
    \psfrag{X}[lc][lc]{$-0.006$}
    \psfrag{Y}[lc][lc]{$0.009$}
    \includegraphics[width=\textwidth]{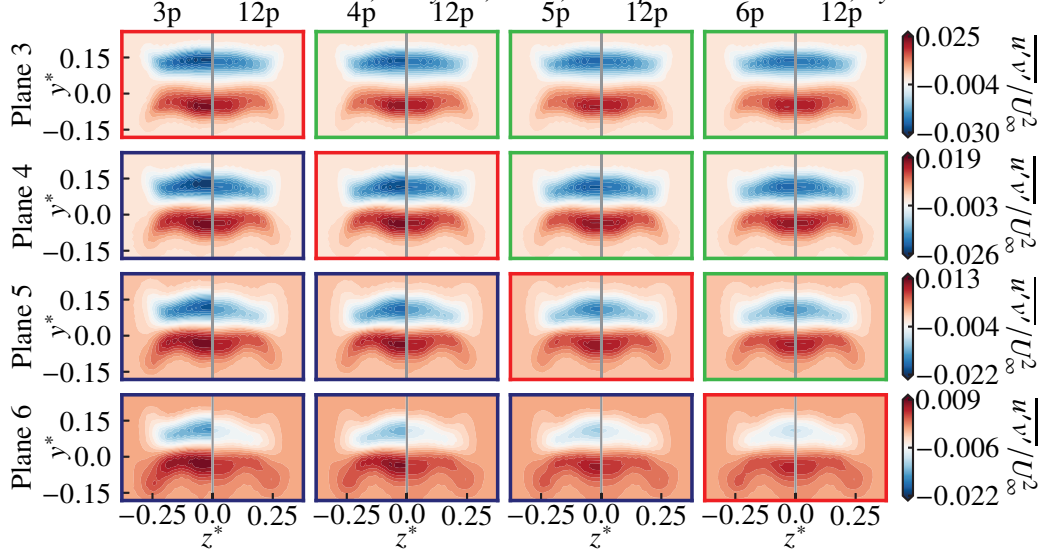}
    \caption{Reynolds shear stress $\overline{u'v'}/U_\infty^2$ on Plane~$3$--$6$ comparing sparse data assimilation cases with the full 12-plane reference in a butterfly plot format. In each panel, the left half shows the sparse case and the right half shows the 12-plane reference. The columns correspond to progressively increasing the number of input planes from left to right ($3$, $4$, $5$, and $6$ planes). The row labels indicate the evaluation plane. The coloured borders denote the relationship between the number of input planes and the evaluation plane: a \brightgreenline~border indicates that the evaluation plane lies upstream of the last input plane (data available beyond this plane), a \brightredline~border indicates that the evaluation plane coincides with the last input plane (this plane is directly constrained by data), and a \darkblueline~border indicates that the evaluation plane lies downstream of all input planes (pure extrapolation).}
    \label{fig:chap_4_uv_data_efficiency}
\end{figure}

The Reynolds shear stress $\overline{u'v'}/U_\infty^2$ is examined as a derived quantity that is not directly constrained by the assimilation. Its comparison across all four sparse cases is presented in figure~\ref{fig:chap_4_uv_data_efficiency}, arranged as a $4 \times 4$ matrix of panels. Each row corresponds to a measurement plane on which the stress is evaluated, with the topmost row showing Plane~$3$ and the bottommost Plane~$6$. Each column corresponds to the number of input planes used in the assimilation, with the leftmost column showing the $3$p case and the rightmost the $6$p case. Within each panel, the left half shows the sparse assimilation result and the right half the 12-plane reference in a butterfly format. The coloured borders encode the spatial relationship between the input data and the evaluation plane: a \brightgreenline~border indicates that the evaluation plane lies upstream of the last input plane, a \brightredline~border indicates that the evaluation plane coincides with the last input plane, and a \darkblueline~border indicates that the evaluation plane lies downstream of all input planes. Interpreting the figure as a matrix, the panels above the leading diagonal correspond to cases where data are provided beyond the evaluation plane, the diagonal panels to cases where data are provided exactly up to the evaluation plane, and the panels below the leading diagonal to cases where the evaluation plane lies downstream of all input data.

For panels in the upper triangle, where the evaluation plane lies upstream of the last input data plane, all sparse cases recover a Reynolds shear stress field that is in close agreement with the 12-plane reference in both shape and magnitude. This is physically consistent with the fact that the control variable is active throughout the assimilation domain, including at planes upstream of the last data plane. The optimiser therefore acts on these planes indirectly through the adjoint, and the resulting correction to the eddy viscosity is sufficiently well-constrained to reproduce the stress field with fidelity comparable to the fully-dense case.

For the diagonal panels, where the evaluation plane coincides with the last input data plane, the Reynolds shear stress field remains in close agreement with the 12-plane reference for the 4p, 5p, and 6p diagonals. The 3p diagonal shows a modest elevation in stress magnitude, consistent with the broader observation that this case is the marginal one across most diagnostics. The agreement on the 4p, 5p, and 6p diagonals is consistent with the observation that the $L_1$ norm is at its lowest on the diagonal plane for each sparse case, since the objective function is evaluated directly on that plane and the control variable is optimised to minimise the local velocity misfit. This suggests that a well-matched velocity field on the evaluation plane is sufficient to recover a physically consistent Reynolds shear stress, irrespective of how many planes precede it. Whether this reflects a genuine insensitivity of the stress to the upstream data history, or a coincidental cancellation of errors in the eddy viscosity field, is difficult to determine from the present results given the high dimensionality of the problem.

For panels in the lower triangle, where the evaluation plane lies downstream of all input data, the Reynolds shear stress field shows consistently higher magnitudes than the 12-plane reference, and this excess grows as the number of input planes decreases. This behaviour is directly attributable to an elevation of the eddy viscosity in the unconstrained downstream region. In the absence of observational data beyond the last input plane, the adjoint carries no gradient information to those locations and the SA transport equation evolves without correction. A comparison of the peak eddy viscosity between the baseline SA model and the 12-plane assimilated field (not shown) reveals that the baseline progressively overpredicts $\nu_t$ downstream of the primary recirculating region, with the ratio of baseline to assimilated peak values increasing from approximately $1.06$ on Plane~$5$ to $1.33$ on Plane~$12$. The sparse cases, receiving less adjoint correction in this region, inherit a greater proportion of this baseline overprediction in $\nu_t$, which propagates directly into the Reynolds shear stress through the Boussinesq relation. The degree of stress overprediction diminishes monotonically as more input planes are added. Notably, within the primary recirculating region (Planes~$2$--$4$), the baseline underpredicts $\nu_t$ relative to the assimilated field, consistent with the known tendency of the SA model to underestimate turbulent mixing in separated flows. The assimilation corrects this by increasing the eddy viscosity, which in turn produces the higher stress magnitudes seen in the upper triangle panels relative to the baseline.

\textcolor{black}{While the choice of planes that demonstrates the effect of limited data coverage in this study is purely illustrative, it may very well be possible to choose the planes in a smart way. The sensitivity of the objective function, which minimises the discrepancy between the model and the experiment, is spatially distributed and identifies those regions that are most sensitive to the optimisation of the control variable. As such, the discrepancy is very high in the near wake region, as observed in figure~\ref{fig:l1_barchart}, so it can be expected that the sensitivity would also take larger values in this region (not shown here). This sensitivity information can be used to designate regions that are more receptive to the optimisation of the control variable, and therefore, guide the selection of planes of data that will be used as input for the assimilation. In fact, the choice of adjoint localisation can be limited to those points only where the sensitivity is greater than a certain cut-off value. While these are interesting avenues to explore, they are out of the scope of the current study.}

\section{Conclusion}
\label{section:conclusion}
The full-field mean-flow reconstruction of a 3D separated flow past a multi-wake model, representative of a vehicle-like bluff body at a Reynolds number of $Re_L = 5.64 \times 10^5$ based on the streamwise body length, is performed using 3DVar by optimising a momentum forcing term in the RANS equations. Three-component mean velocity data measured on $12$ cross-stream planes using stereo PIV are used as reference data, the SA model as the baseline simulation. Using all $12$ planes of data results in a substantially improved mean velocity field, with the most notable improvements observed in the large streamwise momentum deficit in the near wake and the double-lobe structure in the far wake. To overcome the steep memory requirements associated with the discrete adjoint computation for 3D flows, an adjoint localisation is developed and tested, which restricts the space of control variables to a user-defined subdomain. It is shown that restricting the control variable space to approximately $12$~\% of the full mesh results in a peak memory reduction of at most $64$~\% relative to the global case. Different choices of localisation subdomain are tested, with the body-wake configuration, encompassing the model and its immediate wake, yielding a reconstruction that closely matches the full-domain adjoint case at a fraction of its peak memory cost.

The assimilated field is improved not only along the planes at which reference data are provided, but also in regions that receive no direct data constraint, demonstrating the ability of 3DVar to recover the correct asymmetric topology of a complex 3D recirculation bubble from sparse cross-stream measurements alone. Derived quantities not directly provided as inputs to the assimilation are also examined. The normalised Reynolds shear stress $\overline{u'v'}/U_\infty^2$ on Plane~$3$ ($x^* = 0.241$) shows good agreement with the experiment in terms of overall spatial structure, with peak magnitudes closer to the experimental values than the baseline. This enhancement is a direct consequence of the increase in eddy viscosity driven by the optimised velocity field, confirming that the momentum forcing acts on the turbulence representation indirectly through the SA transport equation, consistent with the mechanism documented in \cite{cadambi2026three}. While no experimental pressure measurements are available for direct validation, the assimilated mean pressure field on Plane~$2$ and~$3$ is shown to be physically consistent when correlated with the in-plane vorticity fields, $\omega_y^*$ and $\omega_z^*$.

A data efficiency study is performed by progressively reducing the number of input measurement planes from $12$ to $6$, $5$, $4$, and $3$, selecting planes sequentially from the rear face of the model. The assimilation yields an improvement over the baseline SA model in all sparse cases across all measurement planes, with the sparse cases consistently achieving lower $L_1$ norms than the 12-plane case within the data coverage region and higher norms beyond it. The $3$p case is found to be qualitatively distinct from the $4$p, $5$p, and $6$p cases. Its data coverage terminates upstream of the onset of pressure recovery, identified at Plane~$4$ as a marker for the end of the primary recirculating region, and the resulting reconstruction exhibits a markedly larger velocity deficit downstream of all input planes than the other sparse cases. This indicates that data coverage extending at least to the end of the primary recirculating region is required to adequately constrain the near-wake dynamics. The Reynolds shear stress, examined as a derived quantity, is recovered with fidelity broadly comparable to the 12-plane case on planes either upstream of or coincident with the last input plane of each sparse case. However, the planes downstream of all input data exhibit a systematic overprediction of the stress. This is attributable to the baseline SA tendency to overpredict the eddy viscosity in the wake recovery region, which the sparse cases partially inherit in the absence of adjoint correction there. These findings collectively highlight the importance of informed measurement plane placement relative to the dominant flow structures when designing experimental campaigns for variational DA of 3D separated flows.

Several future directions can be envisaged both methodologically and physics-based. The adjoint localisation strategy demonstrated in this paper is selected \textit{a priori} and is restricted to a fixed geometry. A physics-based localisation, in which features of the flow inform the selection procedure for the subspace of control variables, would extend the framework's reach. This direction would align with recent developments in zonally-augmented turbulence modelling, such as those proposed in~\cite{buchanan2025data} and~\cite{he2026field}, where physics-driven classification strategies are used to partition the correction field according to the underlying flow mechanism. The optimised correction fields could be used to construct symbolic models, with invariant flow quantities serving as inputs. Such a symbolic representation would enable prediction of flow phenomena across a range of flow conditions, generalising beyond the cases on which the correction was trained \cite{wu2025development, he2026field}. \cite{moldovan2021multigrid} demonstrated a multigrid strategy for sequential data assimilation, in which ensemble members of an EnKF are computed on a coarse grid while a single fine-grid simulation is corrected from the projected ensemble statistics. An analogous strategy for variational DA, in which the primal RANS equations are solved on a fine grid and the adjoint is computed on a coarser one, remains to be explored. Such a dual-mesh approach would reduce the size of the state Jacobian itself, and combined with adjoint localisation could substantially lower the barrier to adopting discrete adjoint methods for more complex problems.\\

\noindent \textbf{Acknowledgments}
The authors acknowledge the use of the IRIDIS High-Performance Computing Facility and its support services at the University of Southampton. We wish to acknowledge Dr. Renan Soares for his assistance in setting up the experiments. We gratefully acknowledge funding from EPSRC (Grant Ref: EP/W009935/1) and the School of Engineering at the University of Southampton for UCP's PhD studentship.\\

\noindent \textbf{Declaration of Interest}
The authors report no conflict of interest.\\

\noindent \textbf{Declaration of AI assistance}
During the preparation of this manuscript, large-language-model-based AI tools were used to assist with language editing and stylistic refinement. All scientific content and interpretations remain the responsibility of the authors.\\

\begin{appen}
\section{}\label{appA}
To determine the optimal value of $\lambda$, an L-curve analysis is performed, as shown in figure~\ref{fig:lcurve}. The L-curve plots the velocity objective function $J^u$ (the data misfit) against the regularisation term $J^f$ as $\lambda$ is varied, and is a standard tool for identifying the hyperparameter value that achieves a balance between the two competing terms. Each point on the curve corresponds to a fully converged assimilation at a given $\lambda$. The magnitude of $\lambda$ is varied over the range $10^{-6}$ to $10^{-3}$ and is shown on a logarithmic scale, such that a more negative value of $\log_{10}(\lambda)$ corresponds to a smaller regularisation weight. At the right extreme of the curve (large $\lambda$), the regularisation term is heavily weighted, producing a smooth forcing field but at the cost of a higher data misfit, since the optimiser cannot freely adjust $\mathbf{f}_c$ to fit the reference data. At the left extreme (small $\lambda$), the regularisation contributes negligibly to the total objective, allowing the optimiser to drive $J^u$ to a lower value; however, the resulting forcing field develops sharp, localised gradients that are physically undesirable and indicative of an ill-posed solution. The optimal $\lambda$ in a classical sense lies at the corner of the L-curve, where a marginal decrease in $\lambda$ begins to yield rapidly diminishing returns in misfit reduction while the regularisation term grows sharply. In the present case, however, the selected value, indicated by the red circle in figure~\ref{fig:lcurve}, corresponds to $\log_{10}(\lambda) \approx -4.3$, one step above the classical corner. Upon inspection of the forcing fields at these two values, it was found that increasing $\lambda$ to the corner point produced no discernible improvement in the spatial smoothness of $\mathbf{f}_c$, while the data misfit increased. The selected value therefore achieves a lower misfit without any meaningful loss of smoothness in the correction field.

\begin{figure}
    \centering
    \psfrag{P}[cc][cc]{$J^f/J_0^u \times 10^2$}
    \psfrag{Q}[cc][cc]{$J^u/J_0^u \times 10^{-1}$}
    \psfrag{a}[tc][tc]{$1.0$}
    \psfrag{b}[tc][tc]{$3.0$}
    \psfrag{c}[tc][tc]{$4.5$}
    \psfrag{e}[rc][rc]{$8$}
    \psfrag{f}[rc][rc]{$16$}
    \psfrag{g}[rc][rc]{$24$}
\psfrag{T}[cc][cc]{$\log_{10}(\lambda)$}
\psfrag{n}[lc][lc]{$-3.3$}
\psfrag{o}[lc][lc]{$-4.0$}
\psfrag{p}[lc][lc]{$-4.3$}
\psfrag{q}[lc][lc]{$-5.0$}
\psfrag{r}[lc][lc]{$-5.3$}
    \includegraphics[width=0.6\textwidth]{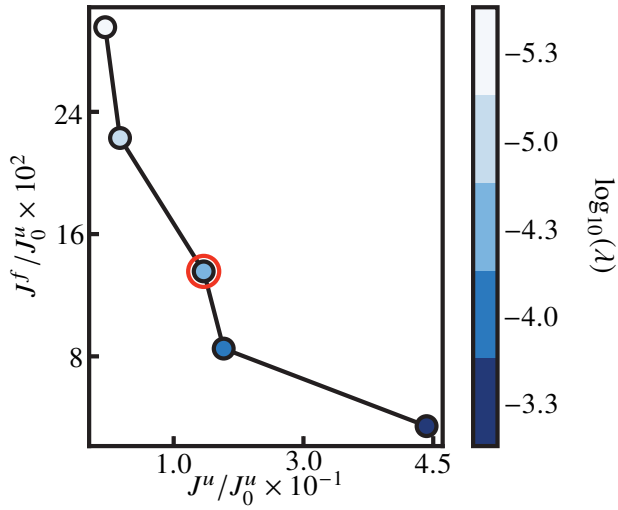}
    \caption{L-curve analysis showing the trade-off between the residual
             and regularisation norms. $J_0^u$ is the magnitude of the baseline misfit or the velocity objective function computed at the first optimisation iteration. The selected value $\lambda = 5\times10^{-5}$
             is highlighted.}
    \label{fig:lcurve}
\end{figure}

\begin{figure}
    \centering
    \psfrag{P}[cc][cc]{$y^*$}
    \psfrag{Q}[cc][cc]{$x^*$}
    \psfrag{A}[cc][cc]{Streamwise forcing}
    \psfrag{B}[cc][cc]{Wall-normal forcing}
    \psfrag{C}[cc][cc]{No penalty}
    \psfrag{D}[cc][cc]{Penalised}
    \psfrag{a}[rc][rc]{$-0.25$}
    \psfrag{b}[rc][rc]{$0$}
    \psfrag{c}[rc][rc]{$0.25$}
    \psfrag{d}[rc][rc]{$0.5$}
    \psfrag{e}[cc][cc]{$-0.8$}
    \psfrag{f}[cc][cc]{$0$}
    \psfrag{g}[cc][cc]{$0.8$}
    \psfrag{h}[cc][cc]{$1.6$}
    \psfrag{I}[cc][cc]{$-0.6$}
    \psfrag{J}[cc][cc]{$0.0$}
    \psfrag{K}[cc][cc]{$0.6$}
    \psfrag{L}[cc][cc]{$-0.6$}
    \psfrag{M}[cc][cc]{$0.0$}
    \psfrag{N}[cc][cc]{$0.6$}
    \psfrag{R}[cc][cc]{$f_{c,x}L_x/U_\infty^2$}
        \psfrag{S}[cc][cc]{$f_{c,y}L_x/U_\infty^2$}
    \includegraphics[width=\textwidth]{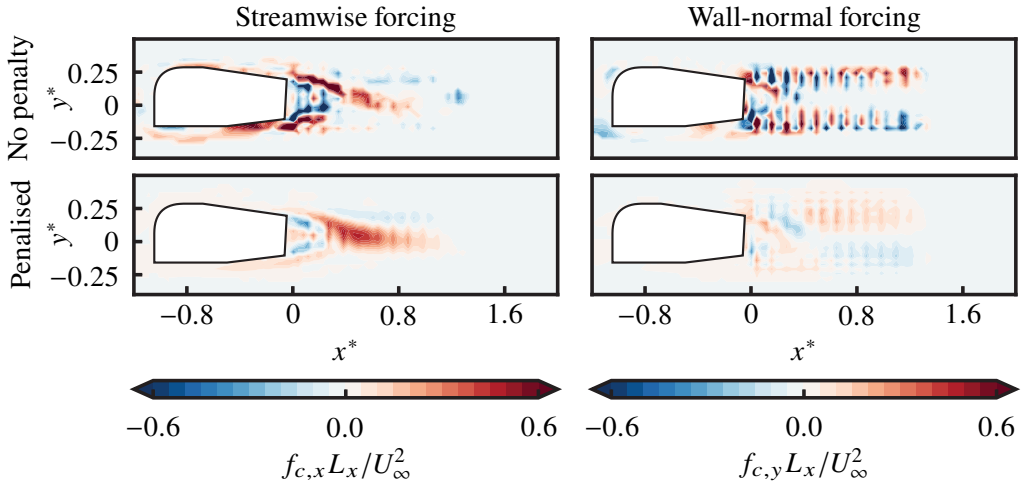}
    \caption{Effect of regularisation on the assimilated forcing field for the global control variable, shown on the symmetry plane $z^* = 0$. Left column: streamwise component $f_{c,x} L_x / U_\infty^2$. Right column: wall-normal component $f_{c,y} L_x / U_\infty^2$. Top row: assimilation without regularisation ($\lambda = 0$). Bottom row: assimilation with regularisation ($\lambda = 5 \times 10^{-5}$).}
    \label{fig:forcing_reg_vs_unreg_global}
\end{figure}

Figure~\ref{fig:forcing_reg_vs_unreg_global} presents the streamwise ($f_{c,x}L_x/U_\infty^2$) and wall-normal components ($f_{c,y}L_x/U_\infty^2$) of the optimised forcing comparing the case without regularisation ($\lambda = 0$) and one with regularisation using $\lambda = 5\times 10^{-5}$ obtained from figure~\ref{fig:lcurve}. It is quite evident that the absence of regularisation demonstrates a high-amplitude dirac forcing (which has been discussed in \cite{symon2017data} and \cite{franceschini2020mean}), especially in the wall-normal component. Applying regularisation smooths out the forcing fields and also reduces its magnitude. 

\end{appen}

\bibliographystyle{jfm}
\bibliography{jfm}

\end{document}